\documentclass[preprint,floatfix,nofootinbib]{revtex4}
\usepackage{color}
\usepackage{amsmath,empheq}
\usepackage{amssymb}
\usepackage[subnum]{cases}
\usepackage{slashed}
\usepackage{isotope}
\usepackage{graphicx}
\usepackage{mathtools}
\usepackage[utf8]{inputenc}
\usepackage{multirow}
\usepackage{hyperref}
\usepackage{soul}
\hypersetup{colorlinks, linkcolor = [rgb]{0,0.0,0.75}, citecolor = [rgb]{0,0.0,0.75}, urlcolor = [rgb]{0,0.0,0.75}}

\usepackage{amsmath,amssymb}
\newcommand\bef{\begin{figure}}
\newcommand\eef[1]{\label{fg:#1}\end{figure}}
\newcommand\beq{\begin{equation}}
\newcommand\eeq[1]{\label{#1}\end{equation}}
\newcommand\bet{\begin{table}}
\newcommand\eet[1]{\label{tb:#1}\end{table}}

\makeatletter
\g@addto@macro\bfseries{\boldmath}
\makeatother

\begin{document}
\allowdisplaybreaks

\title{An empirical model on the dynamics of Covid-19 spread \\
  in human population}

\author{Nilmani Mathur}
\email{nilmani@theory.tifr.res.in}
\affiliation{Department of Theoretical Physics, \\
Tata Institute of Fundamental Research, 1 Homi Bhabha Road, Mumbai 400005, India}
\author{Gargi Shaw}
\email{gargishaw@gmail.com}
\affiliation{Department of Astronomy and Astrophysics, \\
Tata Institute of Fundamental Research, 1 Homi Bhabha Road, Mumbai 400005, India}



\begin{abstract}
We propose a mathematical model to analyze the time evolution of the total 
number of infected population with Covid-19 disease at a region in the
ongoing pandemic. Using the available data of Covid-19 infected
population on various countries we formulate a model which can successfully
track the time evolution from early days to the saturation period in a given
wave of this infectious disease. It involves a set of effective
parameters which can be extracted from the available data. Using those
parameters the future trajectories of the disease spread can also be
projected. A set of differential equations is also proposed whose solutions
 are these time evolution trajectories. Using such a formalism we project the future time evolution trajectories of infection spread for a number of countries where the Covid-19 infection is still rapidly rising.
\end{abstract}

\maketitle

\section{Introduction}
Currently a pandemic is ongoing throughout the world caused by a contagious
respiratory disease, called Covid-19. The pathogen of this respiratory disease is a novel coronavirus, named SARS-CoV-2 \cite{who}. It
started from China and subsequently spreads to most of the countries, and as of
August 12, 2020, it has infected more than 20.2 million human population worldwide causing more than 740 thousand deaths \cite{who, global_data, covid_wiki}.
Though the spread of infection
has substantially reduced in several countries, particularly in China
and Europe, in many countries with a large population, such as USA,
Brazil and India, the pandemic is surging prominently at this time. It
is also not clear whether this respiratory disease will be seasonal
and a second wave will come later. There is no clear consensus in the
scientific community on the possible future evolution of this disease
and naturally there is no consensus on the ideal intervention strategy
by a government in minimizing the number of fatalities while allowing
economic and social activities. Many mathematical models have been put
forward \cite{adam, wu2020_1, imperial_2, tang2020_3, tang2020_4, li2020_5,
  wang2020_6, kucharski2020_7, lauer2020_8, backer2020_9,
  linton2020_10, chinazzi2020_11, kraemer2020_12, prem2020_13,
  li2020_14, yang2020_15, boldog2020_16, anderson2020_17,
  Kisslereabb5793_18, adhikari, model_gsonnino, stat_model1, stat_model2, stat_model3, erf_fun1, erf_fun2, fodorintegral, adhar, gardner} {\footnote{References are
    not at all exhaustive. There are many other similar articles in
    medRxiv and
    bioRxiv:\url{https://connect.biorxiv.org/relate/content/181}}} to
track the time evolution of the disease spread, to understand its
dynamics, as well as to provide a feasible guidance to governments
around the world to control this pandemic.

As in ecology and population growth, in epidemiology too the time evolution
of virus growth in a population is of fundamental importance.  In
general, one builds a model of disease transmission using a system of
differential equations with an assumption of initial exponential
growth \cite{math_model1, math_model2, math_model3, math_model4, math_model5, math_model6, math_model7, math_model8, math_model9, math_model10, math_model11}.  A standard way for mathematical
analysis of the dynamics of infection spread is to adopt a variety of compartmental models originated
primarily from the so-called SIR model \cite{sir1, sir2,sir3}, incorporating
susceptible (S), infectious (I) and Recovered (R) population. For the
Covid-19 disease growth, one of these models, the so-called SEIR model, and its extensions have been utilized extensively \cite{wu2020_1, imperial_2, tang2020_3, tang2020_4, li2020_5,
  wang2020_6, kucharski2020_7, lauer2020_8, backer2020_9,
  linton2020_10, chinazzi2020_11, kraemer2020_12, prem2020_13,
  li2020_14, yang2020_15, boldog2020_16, anderson2020_17,
  Kisslereabb5793_18, adhikari, adhar}. In this scheme of models, one divides the total
population, $N$, at a infected region (e.g., city, country), into
susceptible ($S$), exposed ($E$), Infected ($I$) and Recovered ($R$) population
with a constraint $S+E+I+R=N$. A set of differential equations incorporating these correlated compartments then provides the time evolution of disease spread.  The success and problem of this model
have been discussed over last few months in detail and there is no clear consensus  whether this type of models can predict the spread of Covid-19 with reasonable accuracy \cite{model_prob, adam1}.
There is an alternate view that the integral equations based models could be
more effective than the models based on differential equations
mentioned above to describe the dynamics of epidemics
\cite{fodorintegral, hethcote1980, wearing2005}.

Another approach in studying the dynamics of an epidemic is to employ
data-driven phenomenological statistical models \cite{model_gsonnino, stat_model1, stat_model2, stat_model3, erf_fun1, erf_fun2} where one constructs a mathematical description utilizing the existing data
on the epidemics. For example, the numbers of total infections, day to
day infections, fatalities etc. can be
utilized to construct a model with a number of parameters and then
constrain those parameters with the data.  Of course, these models
do not include any microscopic parameters but can describe the data
in an effective way. Once the model parameters are fixed, in principle,
it is possible to project\footnote{Here projection means prediction within a set of assumptions \cite{projection_ref}.} the future dynamics of the infection spread. 

In this work, adopting the phenomenological approach of the statistical models, we propose a mathematical 
model for the time evolution of the number of infected
population. Rather than proposing a microscopic model, by analyzing
the available data on various countries and cities, we develop an
effective time evolution trajectory on the number of total infection
which can be employed at Covid-19-affected regions. The reason behind adopting such an approach is that since the Covid-19 disease is contagious, assuming a region as a closed system, the number of today's infected population is directly correlated to the number of infected population in the past, and moreover, today's number will also determine how many people will be infected in the near future.
The parameters of the
model can be thought of as an effective mean-field type parameters, which
can be constrained with the available data, and later those can be
employed for the future time evolution of the disease spread. We find that
there is a clear common pattern in the initial growth, mitigation and
saturation periods of the time evolution of the number of infected
population at various Covid-19-affected regions.
The only difference in
describing the Covid-19 virus spread between various affected
population is the difference between parameter sets of these regions.
However,
they are all confined within a smaller subspace of the parameter
space. The differences between parameters of different regions are
possibly due to their differences in total number of population, density, mobility, age
and gender distributions, testing facility, lockdown effect, social
distancing etc. microscopic factors.  Of course, it will be
interesting to find mathematical correlations between the effective
parameters and the microscopic parameters.

Using the proposed model, we also formulate a set of differential
equations which can equally well describe the Covid-19 spread at an
affected population from the initial days to the saturation period. The
time evolution of these differential equations can be performed with
any good first-order differential equation solver with matching the
boundary conditions at different periods of infection. We find that
by solving these differential equations it is possible to find a common
trajectory that can describe the available data on the total number of
infected population for all periods of Covid-19 spread.  The parameters
inside these differential equations can be tuned with the available
data. 
Instead of finding a
single trajectory, we use the available data set with error-bars which
yield a set of parameters and hence a set of trajectories within the
allowed errorbars. The use of errorbars on the available data is
justified as the exact number of infection is unknown due to the lack
of adequate testing facility and often for social as well as political reasons.
We use a
larger error (mostly within 10\%, and in a few cases maximum up to 20\%) at the first 14 days (considering the
incubation period of SARS-CoV-2 virus) and later reduced it to less than a percent level in the saturation period. This is a  reasonable assumption, as in
the initial days the testing facilities are in general severely
inadequate, the symptoms are not well recognized in the population and
hence under-reported and so the reported infected numbers could be well below the true numbers. In the later days, these bottlenecks in general
reduce substantially and the number of reported cases become less
erroneous. We adopt a $\chi^2$ minimization procedure to incorporate
these errors and this also helps to get a band of trajectories around
its mean values (reported numbers). In this way, the onset of
saturation period of infection can also be span over a few days (or
weeks), and not on a single day which is also what we observe at
various affected regions.

In this work, our main objective is to build a model, utilizing the
available data on Covid-19 spread of various regions, which can track
the time evolution of the number of infected population, and also to
project the possible future trajectory.
As it is data-driven, the
dynamics of our model and its predictive power is dependent on
the correct source of data, and hence the model parameters and
projection can change if the data is erroneous. We use data mainly
from the coronavirus resources of Wikipedia \cite{covid_wiki}, W.H.O. \cite{who} as well as of local governments. We assume each region as a closed system and all conditions during the disease spread more or less remain to be the same so that a time correlation of infection can be built. If
the prevailing measures against the disease spread,
such as effectiveness of lockdown, social
distancing, contact tracing and quarantine,
preventive mask-wearing, forbidding large public gatherings
etc., under which the data were available, change substantially
then the parameters and hence the trajectories will also
change. However, this model can be progressively improved with more
data, and projection for a few weeks to a month or more can also be made.

We organize the article as below. In section II, we detail our model
and also elaborate on the set of differential equations corresponding
to this model. In section III, we provide results with numerical
details. First, we validate the model by analyzing data on various
European countries and New York City. Then to demonstrate the
predictive ability of this model we show how a subset of data can
help to predict the future time evolution trajectory of the infection
at a region.  Next, we proceed to analyze data on Russia, Brazil, India
and the USA where Covid-19 infections are still increasing rapidly.
For India we separately analyze the data for its two biggest cities: Mumbai and
Delhi. For each of the regions, within this model, we show the
projections of the future time evolution of the infection with the
most probable time-scale for the onset of saturation along with the cumulative number of infection. At the end, we
discuss our results and conclude.


\section{\label{sec:model} A Dynamical Model for Covid-19 spread}
The basis of building a phenomenological data-driven model is to
analyze the available data on the targeted problem and formulate a
mathematical framework to represent the data, with a minimal set of
parameters, in a consistent plausible way. This mathematical model can
also predict (project) the dynamics in a domain where data is not
available. With this in mind we analyze the number of cumulative
infected population of Covid-19 disease for a number of countries
and cities. In Fig. \ref{fig:general_plot} we represent the available data
in a few possible ways. In the top
plots we show the cumulative number of infected populations ($N(t)$)
as a function of time (Days(t)), in linear and log scales
respectively. The bottom left plot shows the same data when we
normalize the infected population of a region by its total population,
whereas the bottom
right plot shows the number of infection per day as a function of day
(log-scale is used to show all data in one plot so that they can be
compared together).
\bef[h]
\centering
\includegraphics[scale=0.63]{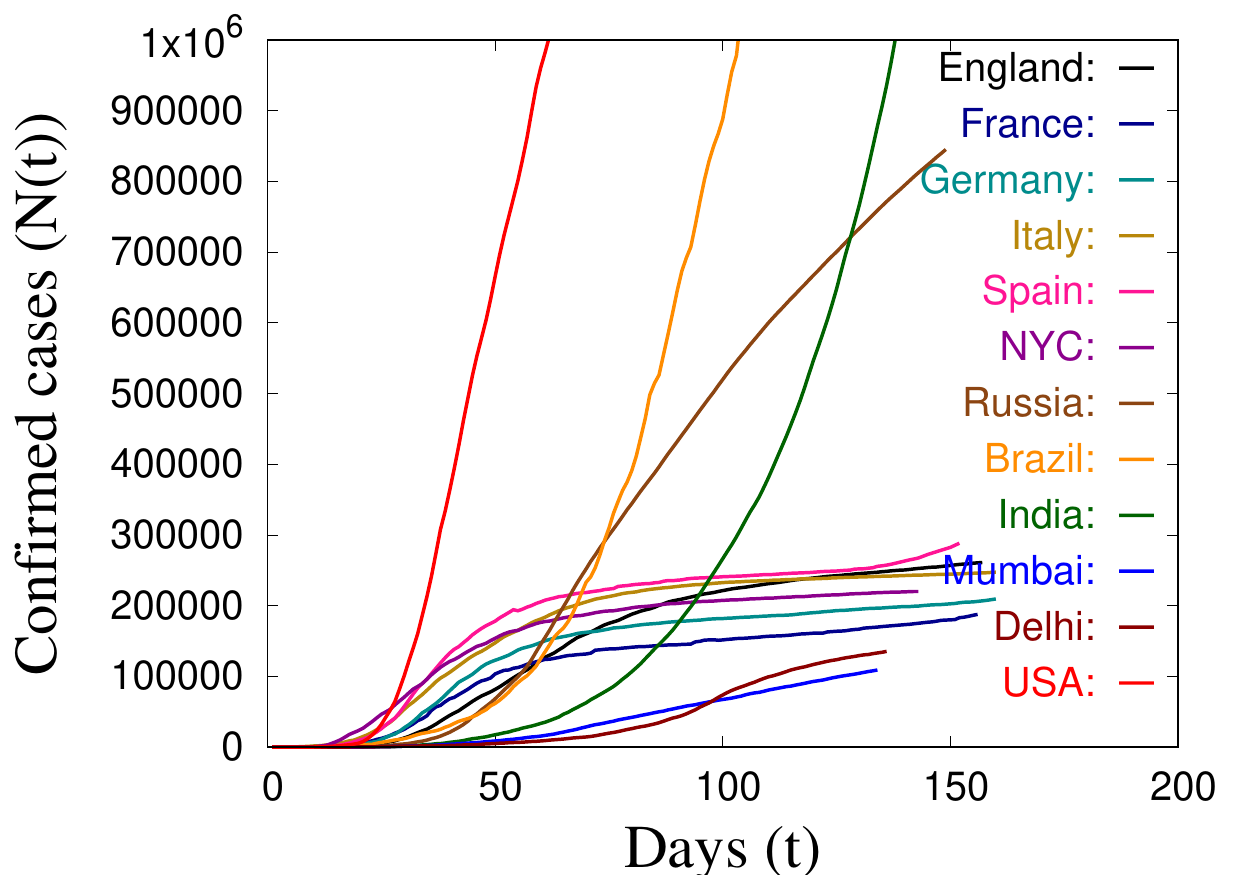}
\includegraphics[scale=0.63]{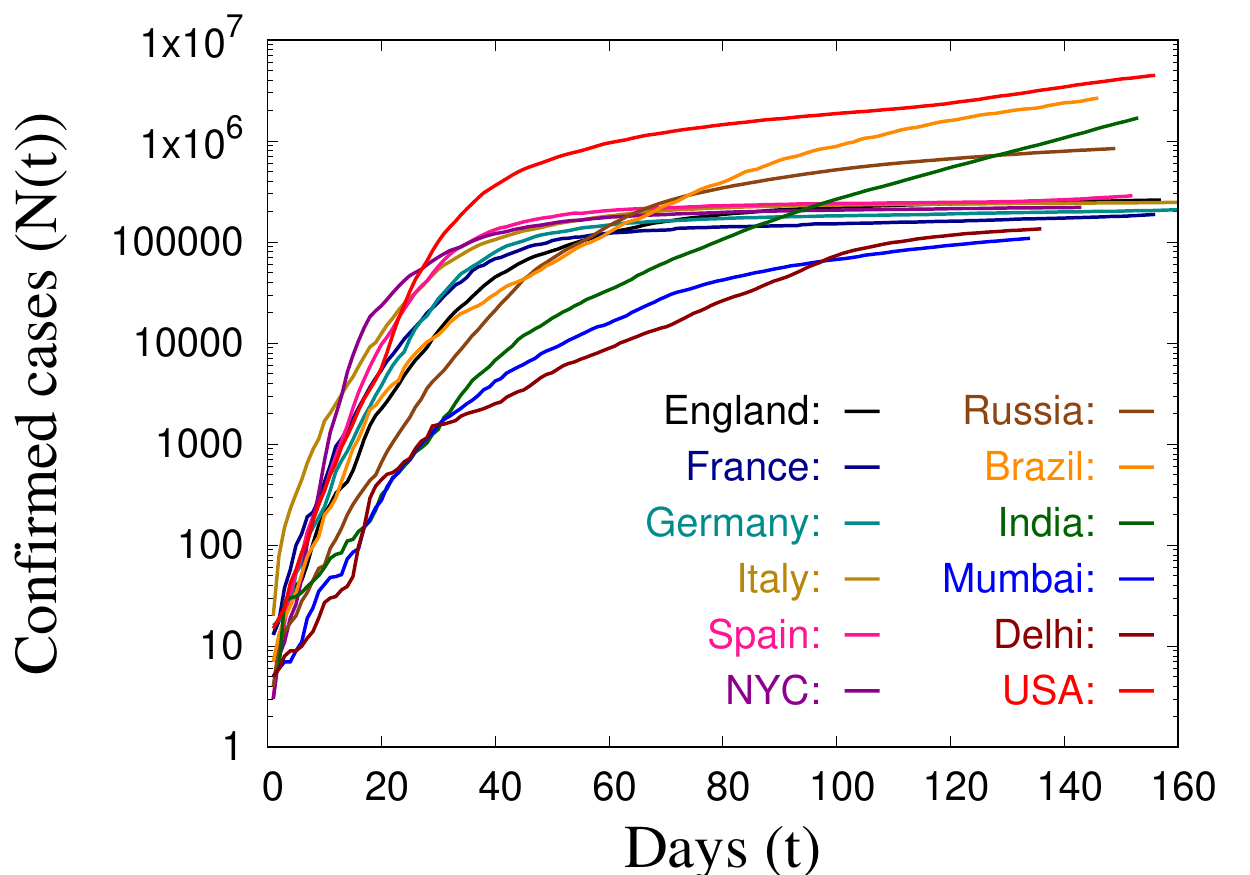}\\
\includegraphics[scale=0.63]{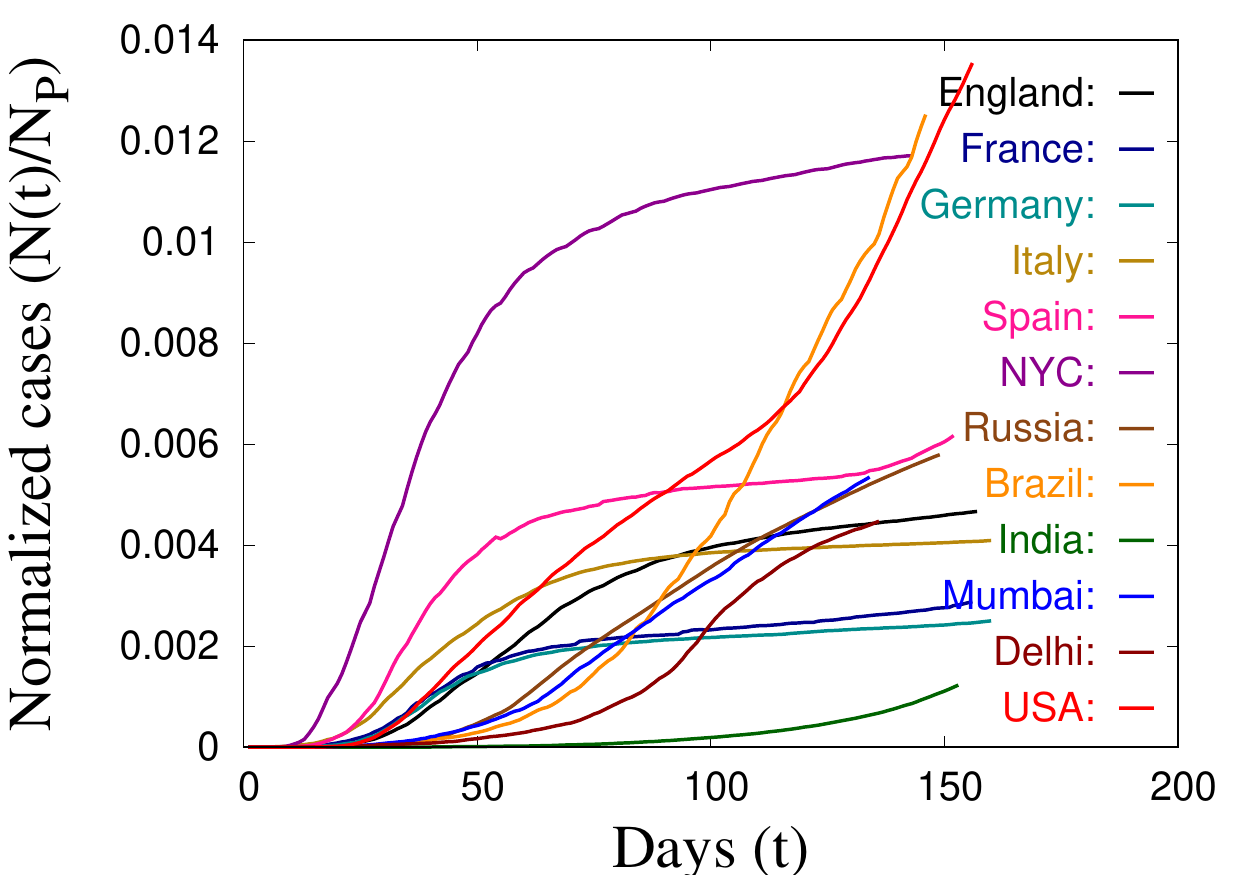}
\includegraphics[scale=0.63]{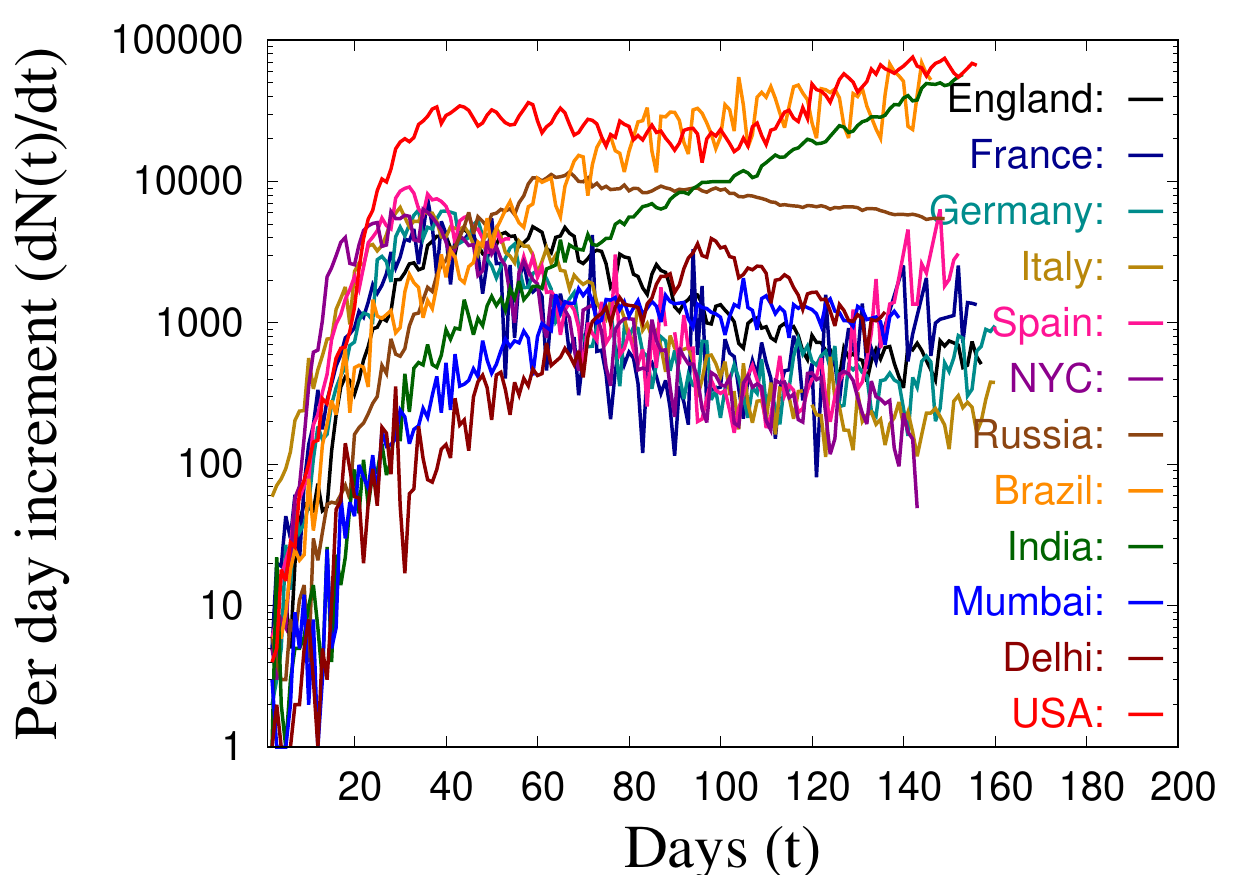}
\caption{\label{fig:general_plot} The time-evolution trajectories
  of Covid-19-infected population for various countries and
cities are shown.
  The top left plot shows the cumulative number of infected population ($N(t)$) in linear-scale while the top left one represents the same in log-scale. The bottom left plot represents data normalized by the total population ($N_p$) of the affected regions. The bottom right plot shows the number of infected population per day. Data (till July 31,2020) are obtained from Refs. \cite{covid_wiki, who, england_data, covid_wiki_france, covid_wiki_germany, covid_wiki_italy, covid_wiki_spain, covid_wiki_nyc, covid_wiki_rus, covid_wiki_brazil, covid_wiki_india, covid_india1, covid_india2, mumbai_data,delhi_data, covid_wiki_usa}.}
\eef{}

By observing the progress of Covid-19 disease spread at
various countries, as shown in Fig. \ref{fig:general_plot}, we find that
the time evolution of cumulative infected population
shows more or less a common pattern, irrespective of the inherent different conditions prevailing at the affected regions.
We find that the total infection time can be divided broadly within
the following three periods:
\begin{itemize}
\item an early increment period ($t_0 \, \le \,t\, \le \,t_{m}$)
\item a mitigation period  ($t_m \, < \,t\, < \,t_{s_{i}}$)
\item a decline or saturation period ($t_{s_{i}} \, \le \,t\, \le \,t_{s_{f}}$).
\end{itemize}
Here $t_0, t_m, t_{s_{i}}$ and $t_{s_{f}}$ are the starting times of
infection, mitigation, saturation and the ending time of the infection,
respectively. Of course, the transitions from one period to the other
are not sharps, rather those are cross-over between two regions
and can span over a couple of days or weeks. For
example, for the latent factor, $t_0$ may not be a particular
day. Similarly $t_m, t_{s_{i}},t_{s_{f}}$ may also span over a couple
of days or weeks. Hence, if there are different mathematical forms for describing the time evolution at different periods, there should be a continuity of the
functional forms from one period to the others
as will be explained later.  The transition times from one period to other can be determined through the data, for example, by minimizing the $\chi^2$ of a particular model against the data. Naturally, the transition times will be different for different countries since the conditions responsible for
early infection rate,
decrement of that rate in the mitigation period, onset of the
saturation period are different for different countries.

To show the above findings more conclusively, as a representative case, in Fig. \ref{fig:ref_england_plot}, we show
the time evolution trajectory of the infected population of England
with top plot showing the cumulative infected population and the
bottom plot representing the per day increment of infection as a
function of days. The above-mentioned three periods are differentiated
with different colors mentioning the transition times. These
transition times are calculated by fitting our model with available
data with a minimum acceptable $\chi^2$ which we will discuss later in
detail. For England, with fixing the initial days of infection ($t_0$)
as February 24, 2020,  we find that the transition time from early rise to the
mitigation period started at around $t = 47$, and the onset of saturation period
started at around $t = 93$ days. We will discuss on this in more detail in
section II when all other results will be presented.

\bef[h]
\centering
\hspace*{-0.3in}
\includegraphics[scale=0.80]{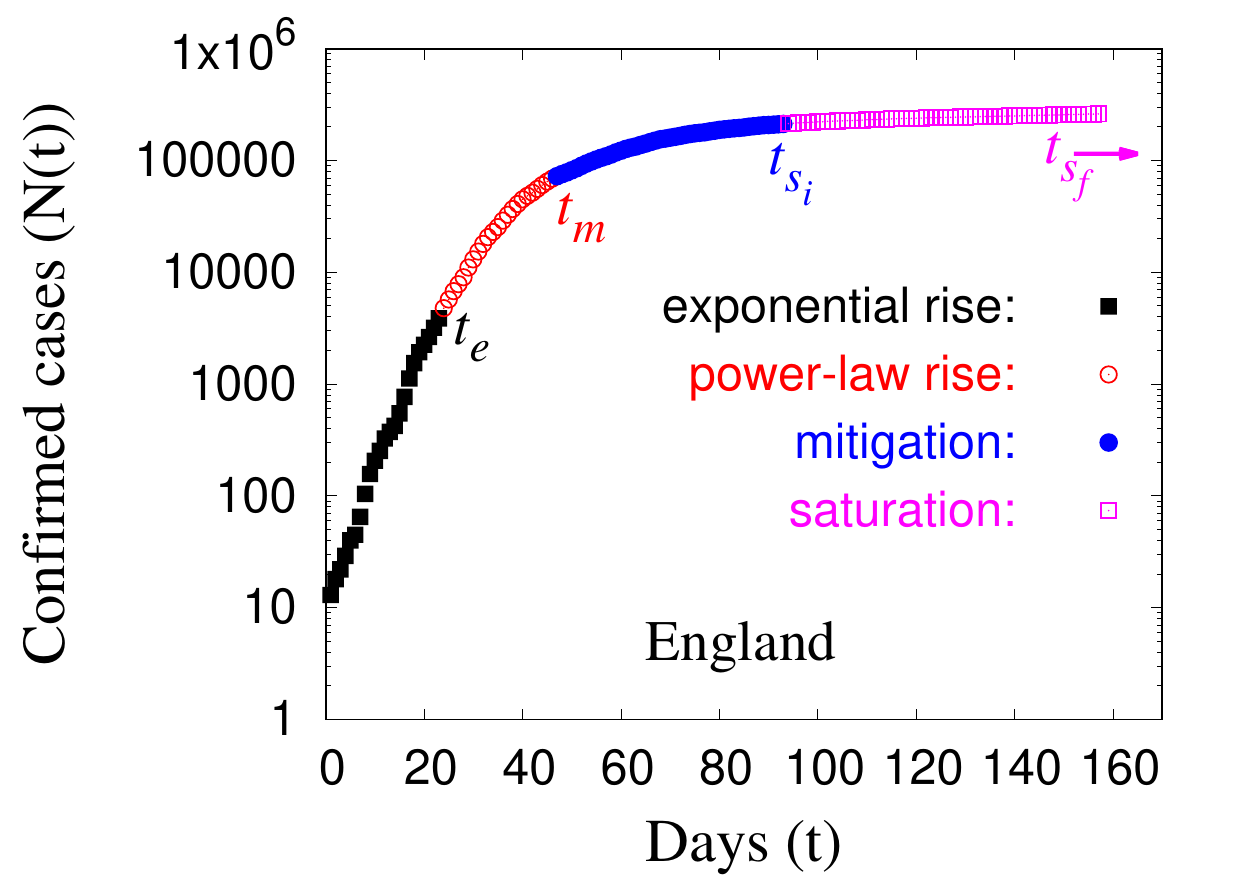}
\includegraphics[scale=0.75]{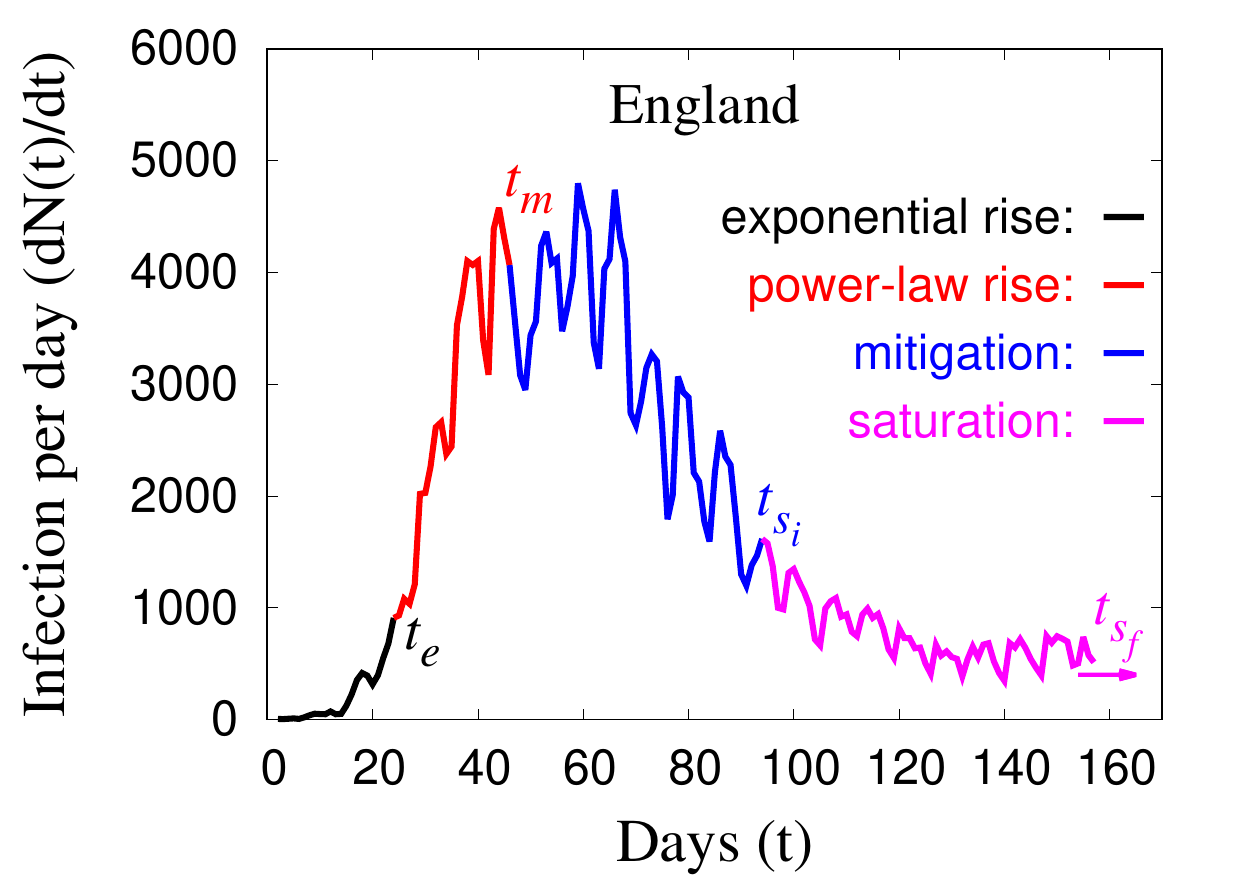}
\caption{\label{fig:ref_england_plot} The time-evolution trajectory 
  of Covid-19 infected population of England. The top plot represents the cumulative number of infection as a function of days while the bottom plot shows the infection per day. Data are taken from Ref. \cite{england_data}.}
\eef{}
By analyzing the available data for various countries, we find
empirically that a possible way to track the trajectory of the number
of infected population, within the above-mentioned three periods, starting from the
early days ($t_0$) to the saturation period ($t_{s_{f}}$), is through
the following equations:
\begin{subequations}
  \begin{empheq}[left =  {N(t) = \empheqlbrace}]{alignat = 3}
  \quad f_i(t) &\qquad t_0 \, \le \,t\, \le \,t_{m}\,, \label{eq:model_eq1}\\
  \quad f_m(t) & \qquad t_m \, < \,t\, < \,t_{s_{i}}\,,\label{eq:model_eq2}\\      
  \quad f_s(t) & \qquad t_{s_{i}} \, \le \,t\, \le \,t_{s_{f}}\,,\label{eq:model_eq3}
  \end{empheq}
  \label{eqn:model_eq}
\end{subequations}
where the functional form of the fast increment period could be an
exponential rise followed by a power-law-rise of the form:
\begin{equation}
f_{i}(t) = \left[A e^{\alpha_{i}^{e}(t-t_0)}\right]_{t_0 \, \le \,t\, \le \,t_{e}} + \left[B + Ct^{\alpha_i}\right]_{t_e \, < \,t\, \le \,t_{m}},
  \end{equation}
and in some cases, simply be the following power-rise form
\begin{equation}
f_{i}(t) = \left[B + Ct^{\alpha_i}\right]_{t_0 \, \le \,t\, \le \,t_{m}}.
\end{equation}
Here $A,B, C$ are constants and mainly depend on the latent population
and the density of population, while $\alpha^{e}_{i}$ and $\alpha_i$
govern the rate of infection in a given population. In reality,
$\alpha^{e}_{i}$ and $\alpha_i$ are time-dependent variables and
depend on many factors, such as the density of population, the total
population, effectiveness of the preventive measures against the disease spread.  In this study we
assume that these variables can be taken as constants in the sense of
an effective mean-field approximation: $\alpha^{e}_{i}(t) =
{\tilde{\alpha^{e}_i}}$ and $\alpha_{i}(t) =
{\tilde{\alpha_i}}$. However, if there are rapid mutations of the
pathogen, migration between different population in a short interval
as well as rapid change in environmental factors, these assumptions may
not be valid.  For a given population, depending on its density,
immunity in population, social distancing or any other preventing
factor against the disease spread,
one of the above forms enhances the infection to a large
number in a short period of time. By tracking the infected population
in the early rise period one can fix one of these forms to explain the
rise of infected of population. Such power-law-rise form was also
observed in Refs. \cite{model_gsonnino}.

In the second period (so-called mitigation period with $t_m \le t < t_{s_{i}}$),
we find that the time evolution trajectory for the total
number of infected population can be tracked with a combination of rising and damping factors through the following
equation:
\begin{equation}
f_m(t) = \left[Dt^{\alpha_m} e^{-\lambda(t-t_m)^{\gamma}}\right]_{t_m \,< \,t \,  < \,t_{s_{i}}}
  \end{equation}
where, $\alpha_m$ is related to the rate of infection in this period,
while the parameters $\lambda$ and $\gamma$ determine how fast the
saturation period can be reached.
The decrease of early infection rate may happen for various reasons, for example for government imposed preventive measures, development of immunity in
the community, availability of drugs etc. Eventually the infection rate
saturates and then decreases depending on the effectiveness of these
external factors.

For a given wave of infection, we observe that the increment of
infection in the saturation/decline period ($t_{s_{i}} \le t \le
t_{s_{f}}$) can be tracked through the following equation:
\begin{equation}
  f_s(t) = N(t_{s_{i}}) + E \left[1 - e^{-\alpha_s(t - t_{s_{i}})}\right],
  \end{equation}
where $E$ is a constant, $\alpha_s$ is the infection rate in this period,
 and these constants depend on the population density, herd
immunity factor, availability of remedy through drugs etc. However, in
this period, at any time, a second wave may start if there is a large
unaffected population and the restrictions to contain the virus become
much softer. If the virus is seasonal, on which there is no consensus
yet, it can also come back to unaffected population. In the case of USA,
such a second rise of infection by this virus is quite prominent (and probably also for Spain, albeit slowly) which we will discuss later.

To be noted that rather than using a single mathematical formula for
tracking the infection throughout the contamination period, from the early rise to the
saturation period, as used in \cite{model_gsonnino, erf_fun1, erf_fun2},
we use three different functional
forms. The reason behind such a formulation is that the dynamics of disease spread in three periods are different due to change in various external conditions as infection progresses (this is quite apparent in Fig. \ref{fig:ref_england_plot}). A single functional form, as used in Refs. \cite{model_gsonnino, erf_fun1, erf_fun2} is thus not suitable to follow the time evolution dynamics throughout the contamination period.  As mentioned earlier, in our case, at the boundaries the model-evaluated trajectories obtained from different functional forms are matched between two periods so that no discontinuity arises.

Though these data-driven mathematical trajectories can be effective to
project the future evolution of the disease, it is important to find
out a set of differential equations whose solution are these
equations. That may help to understand the dynamics of infection in a
better way as well as the origin of such equations and their parameters. To
achieve such a formulation it is worthwhile to look towards the growth
equations in ecology and epidemic studies. In fact in epidemic studies,
to estimate the infection rate, cumulative number of cases, the  peak
number of infected population, future dynamics of the epidemics etc.,
one can use the growth equations. One such approach is the standard
logistic model \cite{logistic_model1} where population growth is determined through an
exponential growth, constrained to the number of population, as below:
\begin{equation}
{dN\over dt} = \alpha N(t) {\left(1 - {N(t)\over K} \right)}, 
\end{equation}
where the first term $\alpha N (\alpha > 0)$ implies an initial
exponential growth with an exponent $\alpha$, while the second term,
known as the {\it bottleneck factor} with the parameter $K$, the {\it
  carrying capacity}, adjusts the growth with critical resources in
the population. A situation with maturity of population arrives as the
competition between two terms reduces the combined growth rate, until
the infected population saturates. Since the environmental
conditions influence the carrying capacity ($K$), as a consequence it
can be time-varying ($K(t) > 0$), leading to the following
mathematical model of growth:
\begin{equation}
{dN\over dt} = \alpha N(t) {\left(1 - {N(t)\over K(t)} \right)}.
\end{equation}
This can be associated with a more general growth model, the so-called
generalized Richards Model \cite {richards_model} which was originally introduced
in the context of ecological population growth, as below:
\begin{equation}
  {dN\over dt} =
  \alpha N (t) {\left[1 - {\left({N(t)\over K(t)}\right)}^{\nu} \right]},
  \label{eqn:richard_eq}  
\end{equation}
where the additional parameter $\nu > 0$ determines the asymptote of
maximum growth. This model has a solution (when $K$ is independent of time)
\begin{equation}
N(t)  = {K\over{{\left(1+\nu e^{-\alpha\nu(t-t_{tp})}\right)}^{1/\nu}}}
\end{equation}
with $t_{tp}$ as the turning point where the growth rate becomes
maximum. This model has already been employed
for real-time prediction in epidemiology, for example in Refs. \cite{grm_appl1, grm_appl2}.

At this point, we would like to correlate our observed empirical growth model
(Eq.\eqref{eqn:model_eq}), with Richards Model
(Eq.\eqref{eqn:richard_eq}). However, once a government introduces a
measure, such as lockdown and/or introduction of preventive drugs, to
reduce the disease spread, the mitigation period also reduces. If one
considers a time-dependent {\it log-kill} factor, $c(t)N(t)$, as in
reference \cite{model_gsonnino}, the above equation modifies to
\begin{equation}
  {dN(t)\over dt} = \alpha N(t) {\left[1 - {\left({N(t)\over K(t)}\right)}^{\nu} \right]} -c(t)N(t).
    \label{eqn:richard_eq1}  
\end{equation}
If Eq.\eqref{eq:model_eq2} is a theoretical model for infectious
growth with preventive measures against infection spread, and
Eq.\eqref{eqn:richard_eq1} can also explain the same growth, then
Eq.\eqref{eq:model_eq2} could well be a solution of
Eq.\eqref{eqn:richard_eq1}. Following the similar strategy as in
Ref. \cite{model_gsonnino}, we find that for $t > t_0$,
Eq.\eqref{eq:model_eq2} is a solution for Eq.\eqref{eqn:richard_eq1}
with the following set of parameters:
\begin{eqnarray}
  \nu = 1; &&\, \,\alpha = {\gamma(\gamma-1)\lambda t_m};\, \,K(t) = {2D\over t_m}t^{\alpha_m+1}; \nonumber\\
 && \mathrm{and}, \quad c(t) = {{\lambda \gamma t^\gamma -\alpha_m}\over t} - {{\lambda (\gamma-1)\, t_m^2\, e^{-\lambda(t-t_m)^{\gamma}}}\over t} +\lambda \gamma (\gamma-1)\, t_m\, (1-t^{\gamma-2}).
\end{eqnarray}
With this parameterization, we arrive at the following dynamical
evolution equation of the infection growth for Covid-19 with
preventive measures (such as lockdown, social distancing, introduction
of preventive drugs etc.)
\begin{eqnarray}
 {dN\over dt} &=&  \alpha {N(t)\left(1-{N(t)\over {\beta t^{\alpha_m+1}}}\right)}\nonumber\\
 && -\left[{{\lambda \gamma t^\gamma -\alpha_m}\over t} - {{\lambda (\gamma-1)\, t_m^2\, e^{-\lambda(t-t_m)^{\gamma}}}\over t} +\lambda \gamma (\gamma-1)\, t_m\, (1-t^{\gamma-2})\right]N(t)\,.
 \label{eq:mitigation_deq}
\end{eqnarray}
The above equation has five parameters: $\alpha, \alpha_m, \beta,
\lambda$ and $\gamma$, and those are dependent on the particular
virus that is spreading, the infection rate in a particular community
as well as on the externally imposed conditions such as the lockdown
measures to which the population is subjected to ($t_m$ will be determined by varying it dynamically and requiring minimum $\chi^2$). With the available
data these parameters can be constrained and then
Eq. \eqref{eq:mitigation_deq} can be utilized for future time
evolution.  To be noted
that with $\alpha_m = 1,$ and $\gamma = 2$, Eq.\eqref{eq:model_eq2}
turns to Covid-19 model of Ref. \cite{model_gsonnino}, which can be thought of as a special case
of Eq.\eqref{eq:model_eq2}.

Combining all the terms, the final compilation of a set of
equations that we propose for the time evolution of the cumulative
infected population at different period is the following: 
  \begin{numcases}
       {{dN\over dt} = }  \alpha^{e}_i N(t) + {\alpha_i N(t)\over t} \left[1-{A\over N(t)}\right] &\nonumber\\
       \qquad \qquad  {\hbox{or}}  \hspace*{3.15in} t_0 \, \le \,t\, \le \,t_{m} & \label{eq:model_deq1}\\
      \quad  {\alpha_i N(t)\over t} \left[1-{A\over N(t)}\right], & \nonumber\\
      \quad &\nonumber\\
        \alpha {N(t)\left[1-{N(t)\over {\beta t^{\alpha_m+1}}}\right]} \hspace*{2.5in} t_m \, < \,t\, < \,t_{s_{i}} & \nonumber\\
       -\left[{{\lambda \gamma t^\gamma -\alpha_m}\over t}
         - {{\lambda (\gamma-1)\, t_m^2\, e^{-\lambda(t-t_m)^{\gamma}}}\over t} +\lambda \gamma (\gamma-1)\, t_m\, (1-t^{\gamma-2})\right]N(t), \label{eq:model_deq2}\\
      \quad &\nonumber\\
      \quad -m_sN(t)\left[1-{F\over N(t)}\right] \hspace*{2.2in}  \,t_{s_{i}} \le \,t\, \le \,t_{s_{f}} & \label{eq:model_deq3}     
\end{numcases}

These equations can be solved numerically for a given set of parameters, and by matching the boundary conditions at $t_m$ and $t_{s_i}$ one can get a time evolution trajectory for the whole infection period. The parameters of Eq.(13) can be fixed through a $\chi^2$-fitting of the numerical trajectory against the available data. The transition periods can be evaluated dynamically by requiring the minimum $\chi^2$. Here one can introduce an error $\sigma_i$'s on each days data which will then generate a set of trajectories allowed within that errors.
  
The onset of saturation (plateauing), that is $t_{s_{i}}$, can be
determined by finding the maximum of $N(t)$ at the end of the mitigation
period. Taking the first derivative one arrives at
\begin{equation}
\lambda \gamma (t-t_m)^{\gamma-1} - n = 0,
\end{equation}
a positive-time solution of which provides $t_{s_{i}}$. One can also
find the inflection point, that is the onset of approach towards
$t_{s_{i}}$, by taking the second derivative which yields
\begin{equation}
\lambda\gamma t^2[1+\gamma\{\lambda(t-t_m)^\gamma-1\}](t - t_m)^{m-2} -n [1+2\lambda\gamma t (t-t_m)^{m-1}] + n^2 = 0,
\end{equation}
and then finding a solution of this equation.

We use both Eqs. \eqref{eqn:model_eq} and (13) to generate the time evolution trajectories and the parameters are fixed using the available data. The full trajectories, covering the whole contamination period, are then generated which also project the disease spread at the future times. Results obtained for various affected regions are presented in the next section.



\section{Results}\label{sec:results}
In this section we present the results obtained through our proposed
model. First, we will check whether the proposed model is able to successfully
track the 
available data on the cumulative number of infection at various Covid-19-affected
regions. Then we will probe the predictive ability of our model using a
subset of data and reproducing the later trajectory.
Finally, we will project the
future time evolution of the number of infected population
for a few countries where the infection is still rising rapidly.

To validate our model we use the available data on
Covid-19-infected population of the following countries: England, Germany,
Italy, France, Spain and also for New York City. These data are taken
from Refs. \cite{england_data, covid_wiki_france, covid_wiki_germany, covid_wiki_italy, covid_wiki_spain, covid_wiki_nyc}. We choose these countries and city as they
have already at the inside of the saturation period (after $t_{s_{i}}$) in the ongoing
wave of  Covid-19 disease. This will enable us to
test our model from the early rise to the saturation period.
After demonstrating the usefulness of the proposed model we proceed to
show the predictive ability of our model by analyzing the subset of data for
Italy and New York City. As will be evident later that this model can
predict the future time evolution trajectory for a few weeks to
months. Then we will show our results on the total number of infected population of USA, Brazil, Russia and India (separately on Mumbai and Delhi also), where the number of cases are still
rapidly increasing. This will enable us to project the time evolution
trajectories of these countries in the mitigation period, till they
reach the beginning of their respective plateau positions (starting at
$t_{s_{i}}$).
\subsection{Validation of the proposed model}
As mentioned earlier, to validate our proposed model, as in
Eqs. \eqref{eqn:model_eq} and (13), we first analyze
the time evolution of the number of Covid-19 positive population for a
number of countries for which the mitigation periods ended and the
saturation periods have started. For a given country, using the
available data on the cumulative number of infected population, obtained from Refs. \cite{covid_wiki, who, england_data, covid_wiki_france, covid_wiki_germany, covid_wiki_italy, covid_wiki_spain, covid_wiki_nyc, covid_wiki_rus, covid_wiki_brazil, covid_wiki_india, covid_india1, covid_india2, mumbai_data,delhi_data, covid_wiki_usa}, we perform a non-linear
$\chi^2$-fitting of Eq. \eqref{eqn:model_eq} (details of data analysis is given in the Appendix).  The data are fitted over
the entire time range simultaneously with three sub-equations in
Eq. \eqref{eqn:model_eq}.  We vary the extent
of different time-periods 
dynamically and choose $t_m$ and $t_{s_{i}}$ that minimize the total $\chi^2$, 
which is a sum over individual terms, {\it i.e.},
$\chi^2 = \chi^2_i + \chi^2_m + \chi^2_s$. Here the definition of $\chi^2$ is the usual one $\chi^2 = \sum{(f(t_i)-N(t_i))^2/\sigma_{N(t_i)}^2}$, where $f(t_i)$ is Eq. \eqref{eqn:model_eq}, $N(t_i)$ is the cumulative number of infected population on {\it i-th} day, and $\sigma_{N(t_i)}$ is the error on that.  We introduce this error ($\sigma_{N(t_i)}$) with each day's number with a
maximum of 20\% error for the first 14 days (incubation period, and most cases it is 10\%) and
less than a percent level error as the fit approaches to the saturation period.
An error of about 5-1\%
is imposed in between with gradual decrement with the number of
days. We have already mentioned it and elaborating it further here
with following arguments: at the onset of
 infection, the number of initial cases is not well determined as the
number of tests performed could well be too low and there is a good
probability for underestimation.  We keep the maximum error for the
first 14 days which is typically the incubation period for this
respiratory disease. The testing procedure, capacity as well as
reliability improves over time and hence the reported number for the
Covid-positive cases becomes less more erroneous over time. The
timing of a transition from one period to other, as mentioned at the
beginning of section II is chosen dynamically so that the total
$\chi^2$ is minimum. To avoid a sharp transition between
two periods (for example, early rise to mitigation, that is points
just before and after $t_m$) we interpolate the results between the
adjacent points obtained from two different fitting forms. This is justifiable
as we mentioned earlier that change form one period to others does not
happen in a single day and can happen with a smooth cross-over over several days. With this
procedure we fit the data for the above-mentioned countries uniformly. We
first show our results for the first two time-periods (initial and
mitigation, till the onset of saturation) with fit to
Eqs. \eqref{eq:model_eq1} and \eqref{eq:model_eq2}. This will show the
validity of our model at these two time-periods. Later we also show
our results of the entire time-periods (initial mitigation, and
saturation).  We show it separately as Eq.\eqref{eq:model_eq3} is the
ideal representation of the saturation period with the same
environmental factors as in the mitigation period. However, in this period
due to possible ease of various environmental factors and with a large unaffected
population a second wave of infection may resurface which needs to be
dealt separately. That we will discuss for the case of USA later.

In Fig. \ref{fig:time_traj_valid1} we show the cumulative
Covid-positive cases for the above-mentioned countries for the first
two time periods (initial to the onset of saturation, before
$t_{s_{i}}$), as determined by our fit to Eqs. \eqref{eq:model_eq1}
and \eqref{eq:model_eq2} dynamically with a minimum $\chi^{2} =
\chi^{2}_i + \chi^{2}_m$.
The red points
are the actual data points as obtained from Refs. \cite{england_data, covid_wiki_france, covid_wiki_germany, covid_wiki_italy, covid_wiki_spain, covid_wiki_nyc}, while the black lines are the fitted trajectories
with the time evolution form as in Eqs. \eqref{eq:model_eq1} and
\eqref{eq:model_eq2}. In Table \ref{tab:pars} we show the mean values of the fitted parameters.
As one can observe that the fitted
trajectories obtained through Eqs. \eqref{eq:model_eq1} and
\eqref{eq:model_eq2} describe the actual trajectories of
the time evolution, quite well, starting from the onset of disease to the onset of
saturation period as it approaches the plateau region. It is
interesting to see that the onset of saturation period for all
countries happened when the power $\gamma$ reaches a value closer to 1.
\bef[h]
\centering
\includegraphics[scale=0.63]{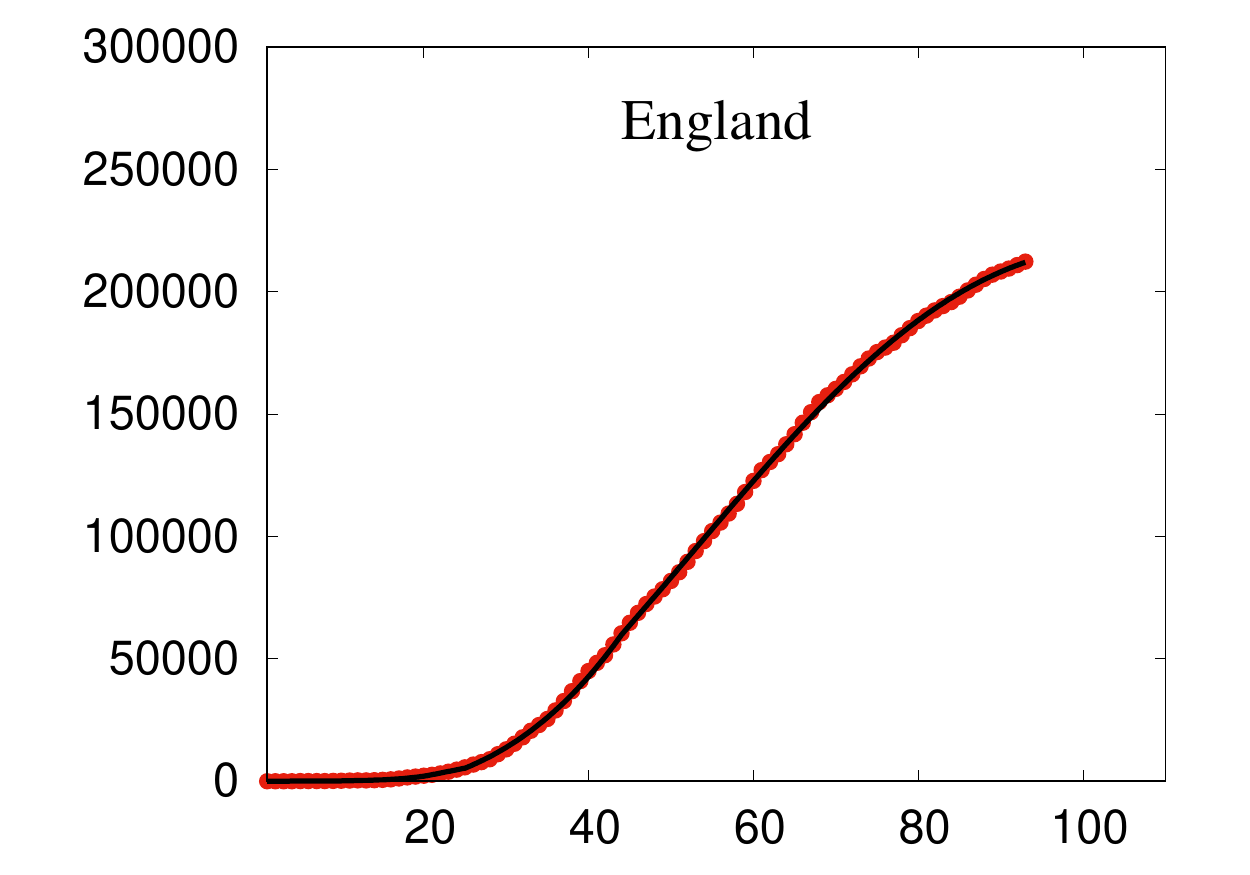}
\includegraphics[scale=0.63]{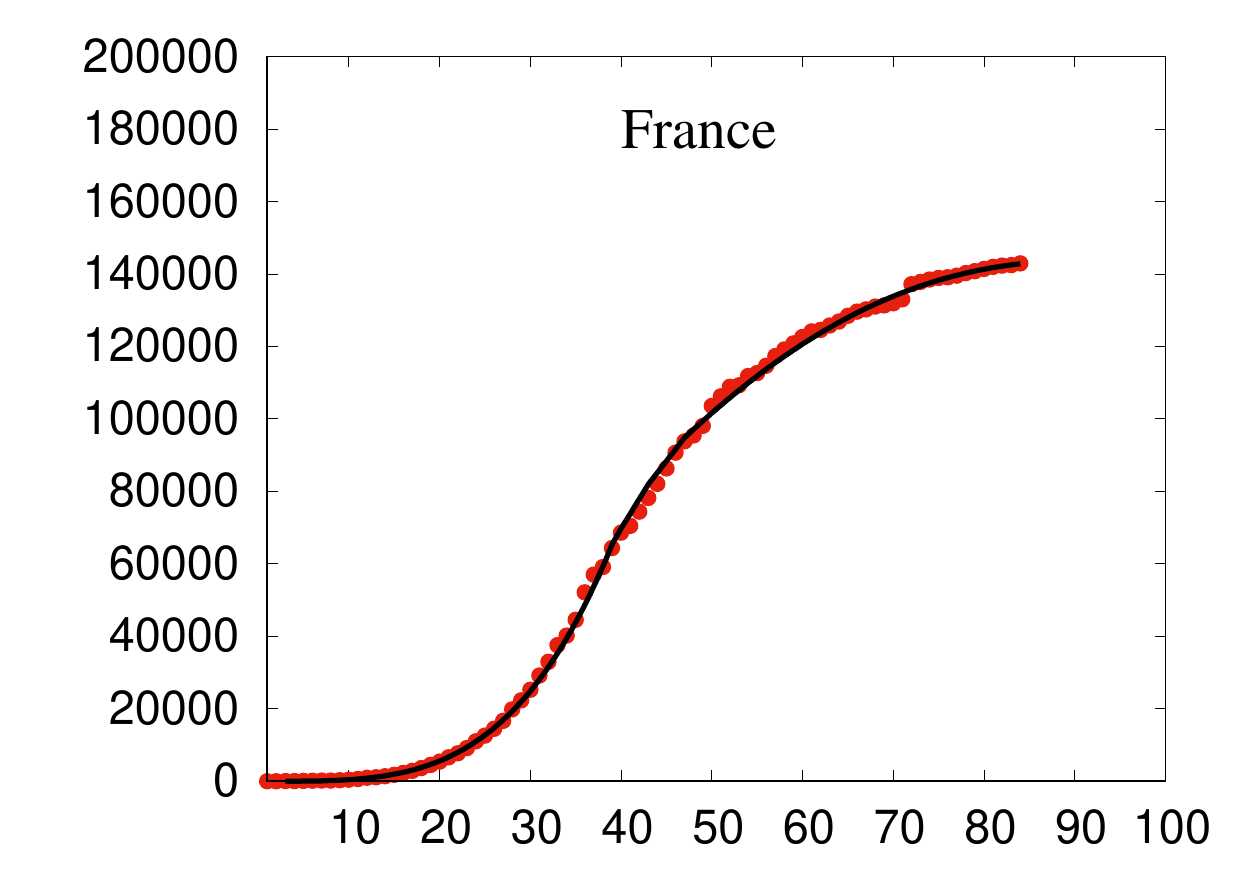}\\
\includegraphics[scale=0.63]{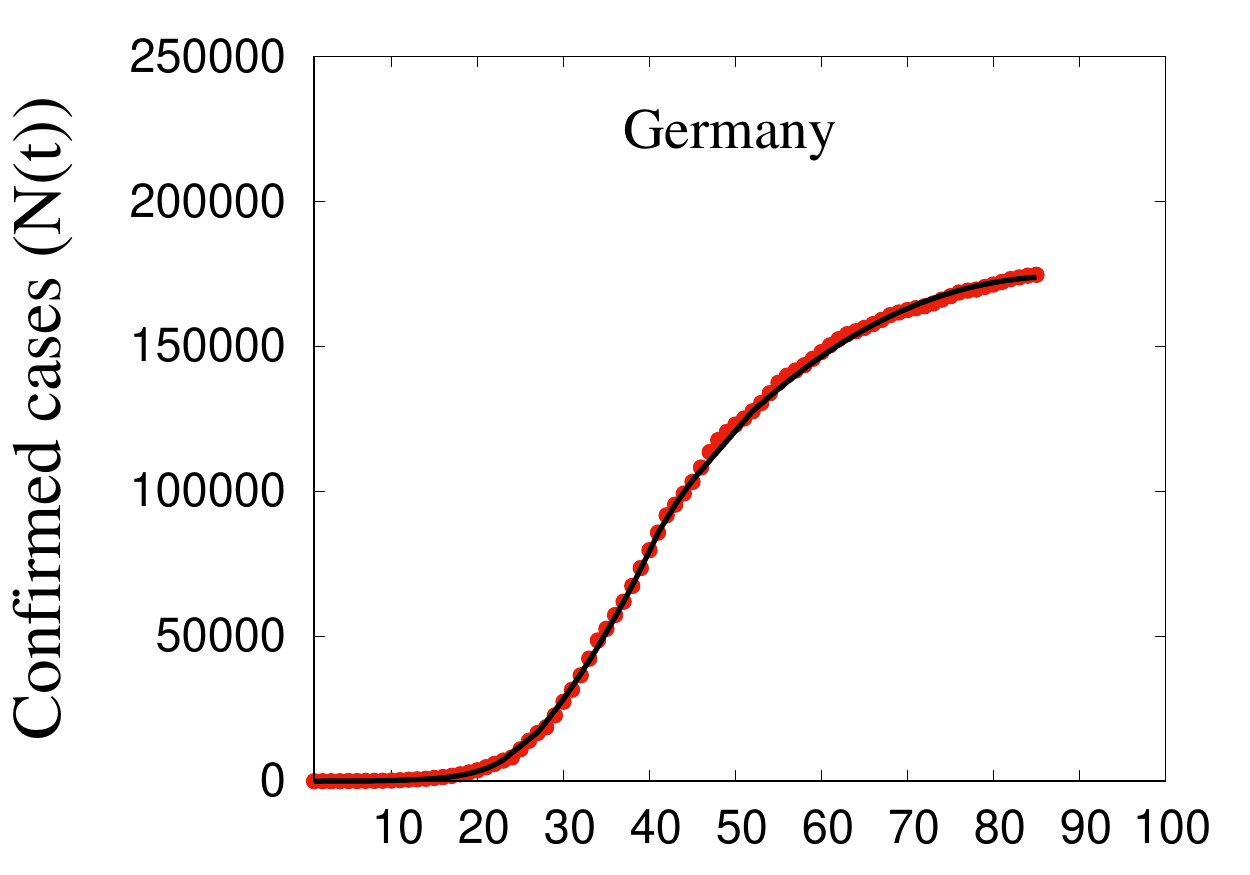}
\includegraphics[scale=0.63]{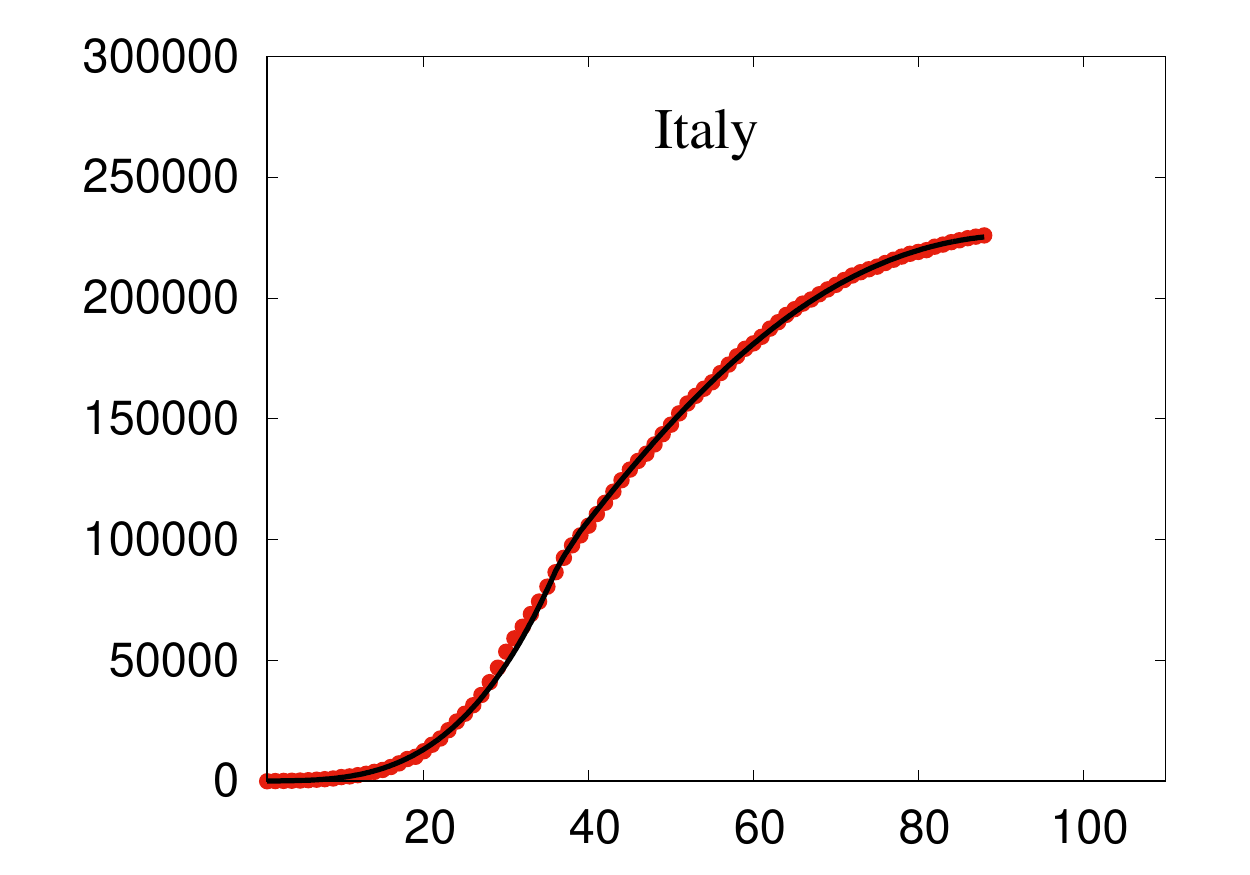}\\
\includegraphics[scale=0.63]{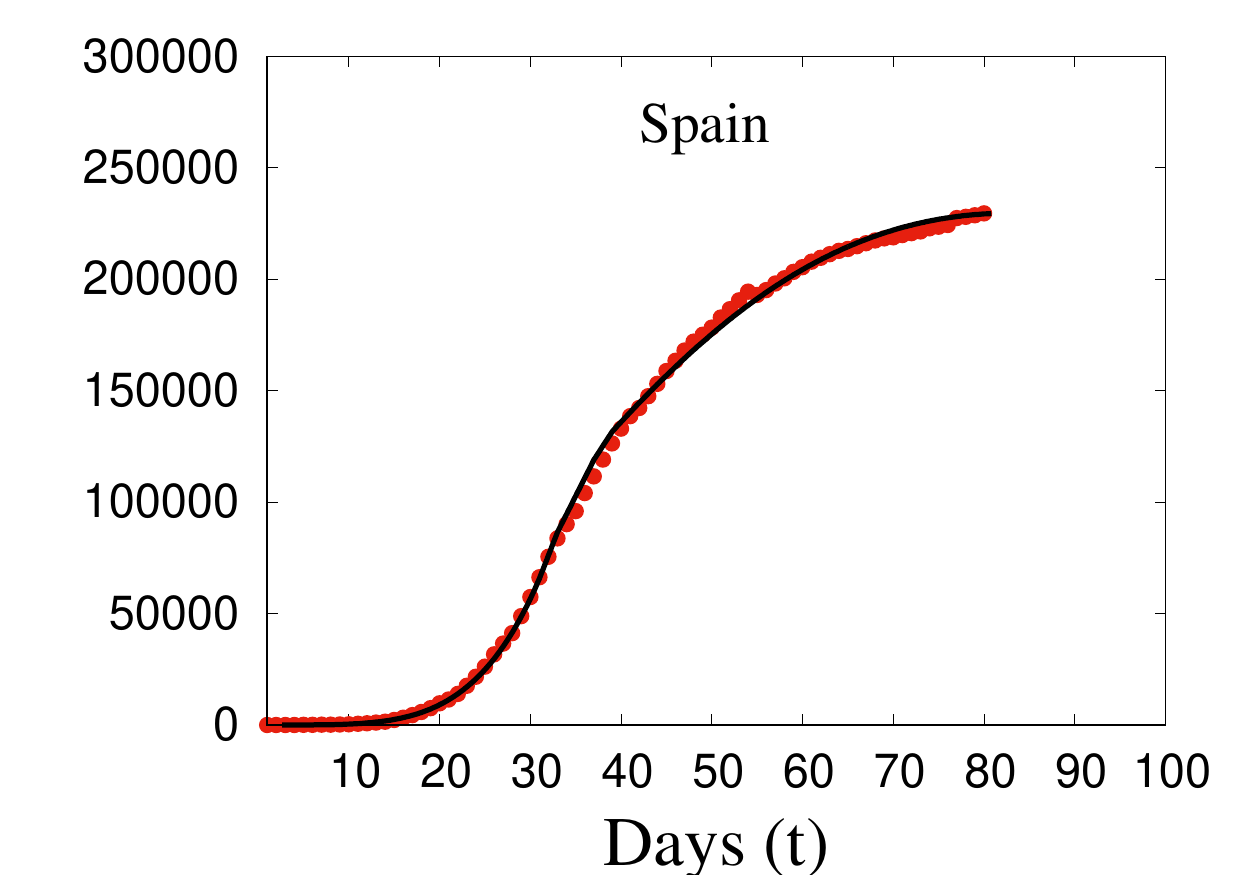}
\includegraphics[scale=0.63]{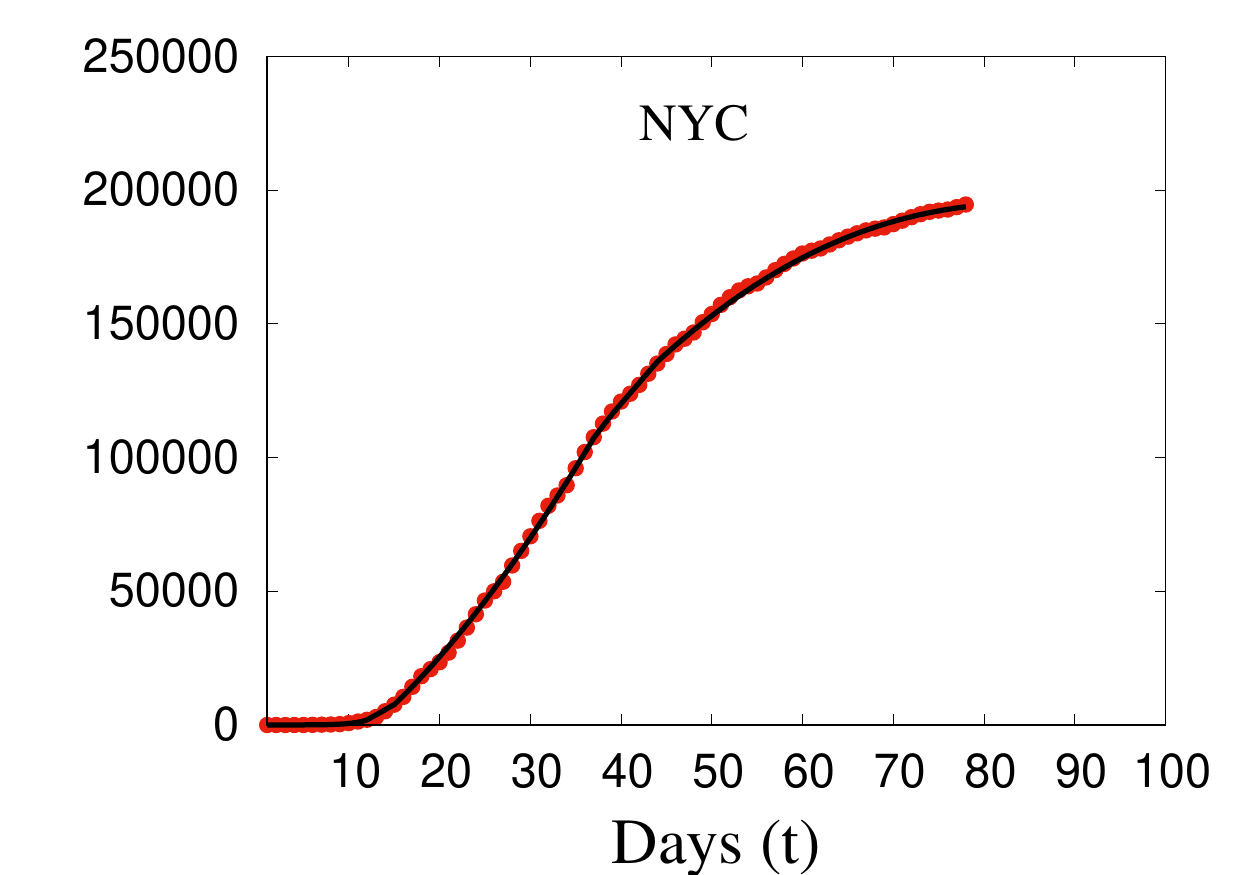}
\caption{\label{fig:time_traj_valid1} The time-evolution trajectories
  of infected population for various countries and for the New York
  City are shown.  The red filled circles are actual data from
  Refs. \cite{england_data, covid_wiki_france, covid_wiki_germany, covid_wiki_italy, covid_wiki_spain, covid_wiki_nyc}.
  The black solid lines are the trajectories
  obtained by fitting the data with Eqs. \eqref{eq:model_eq1} and
  \eqref{eq:model_eq2}. The data are plotted up to a time just before
  the onset of the saturation period as determined by the fit. The
  parameters obtained from these fittings are tabulated in Table \ref{tab:pars}.}
\eef{}
In Fig. \ref{fig:d_time_traj_valid1}, we show the per day infection as
a function of days for the above-mentioned countries. The red points
are the actual data points as obtained from Refs.
\cite{england_data, covid_wiki_france, covid_wiki_germany, covid_wiki_italy, covid_wiki_spain, covid_wiki_nyc}, while the black lines are results from Eq. \eqref{eqn:model_eq}.
\bef[h]
\centering
\includegraphics[scale=0.63]{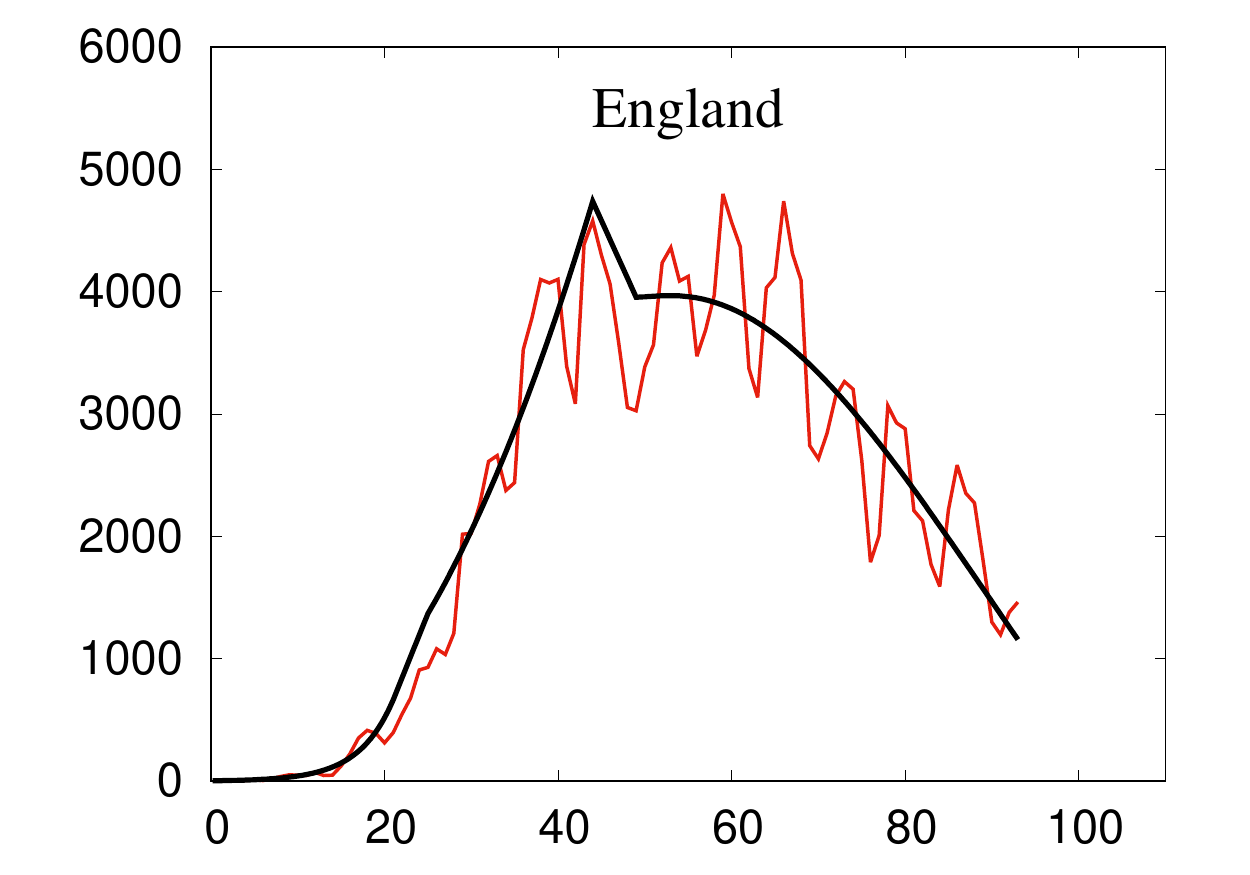}
\includegraphics[scale=0.63]{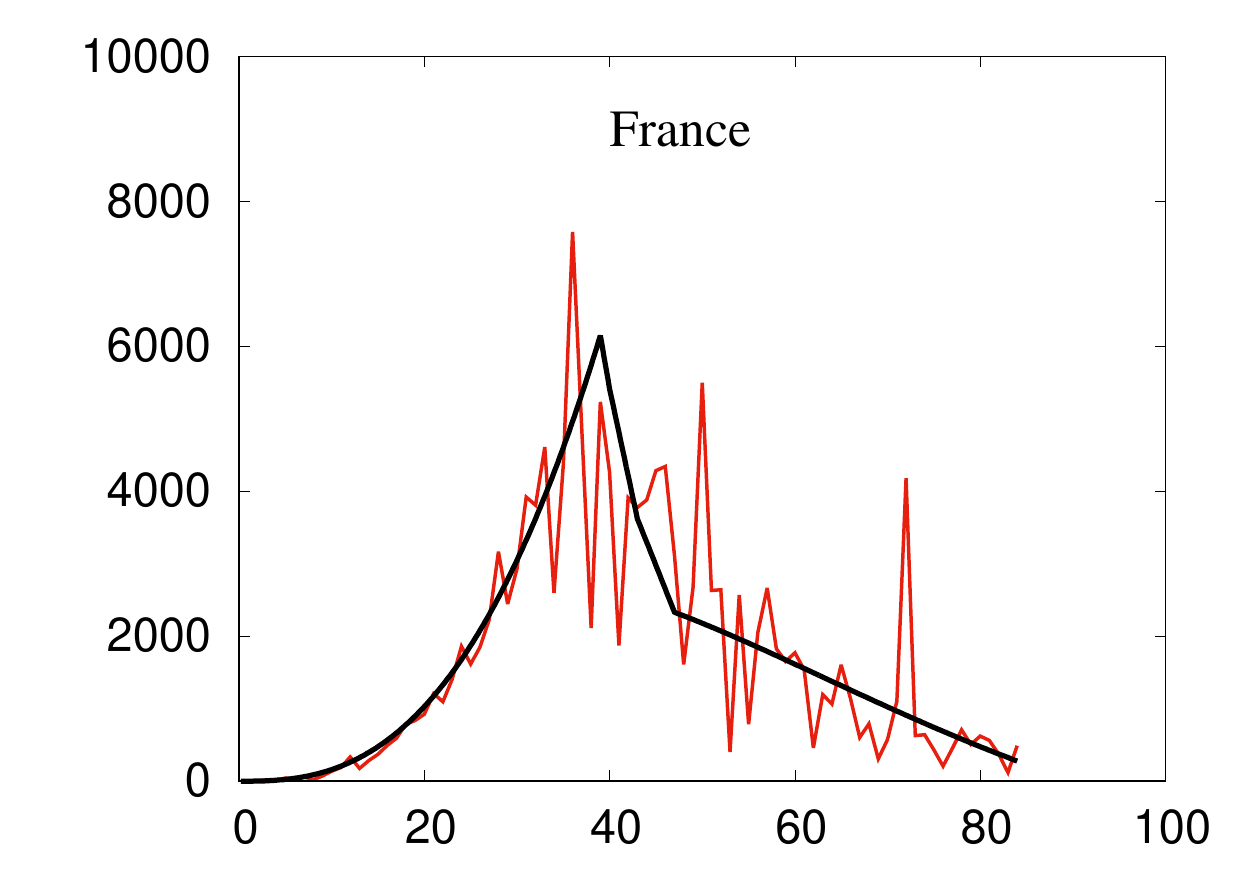}\\
\includegraphics[scale=0.63]{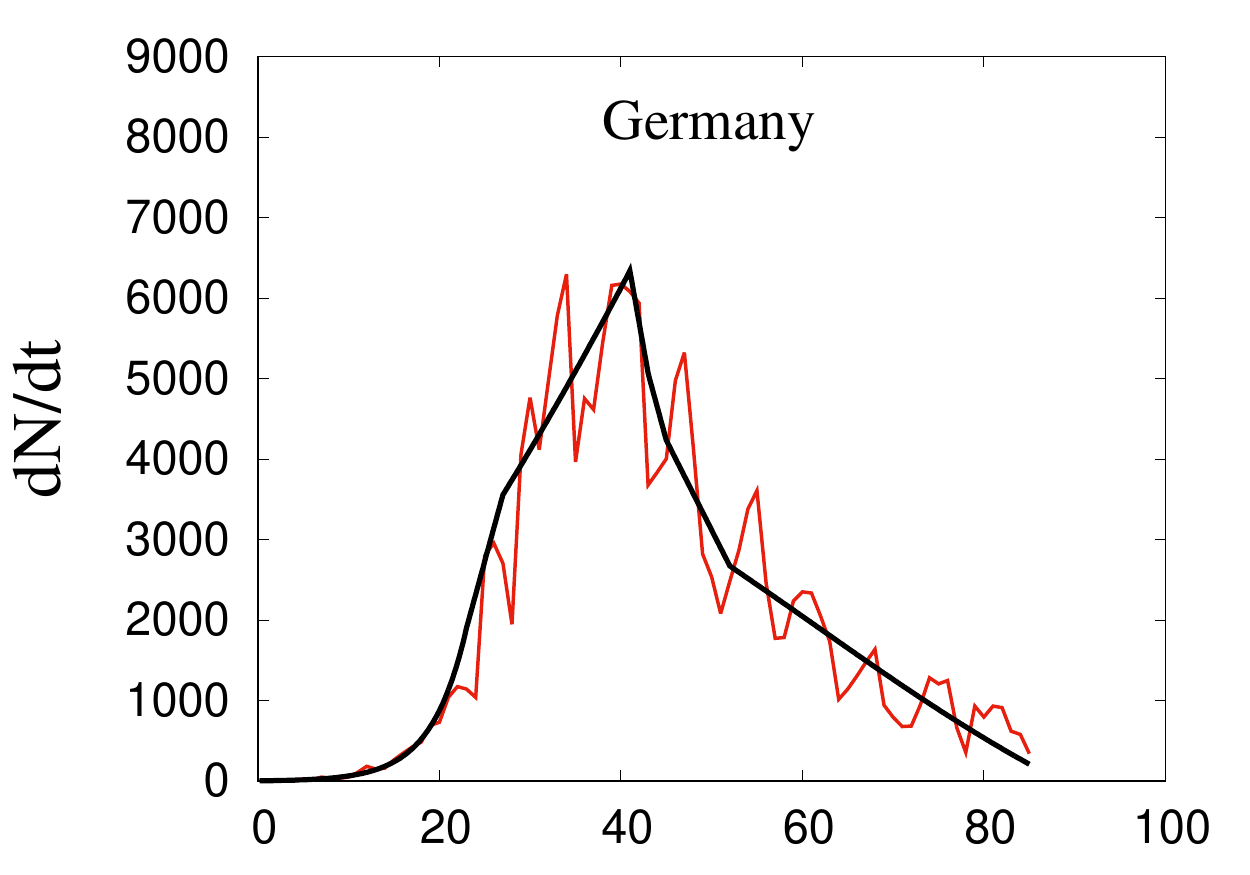}
\includegraphics[scale=0.63]{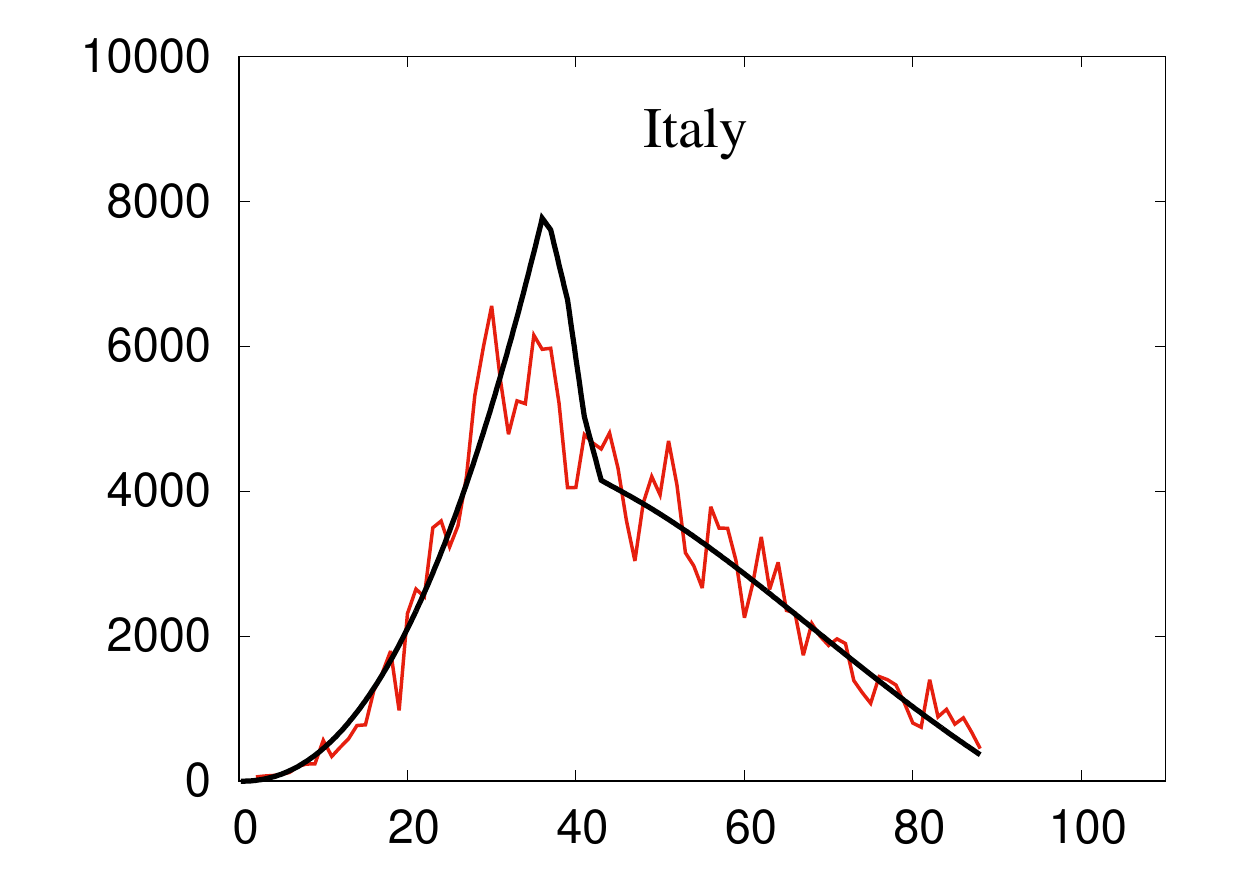}\\
\includegraphics[scale=0.63]{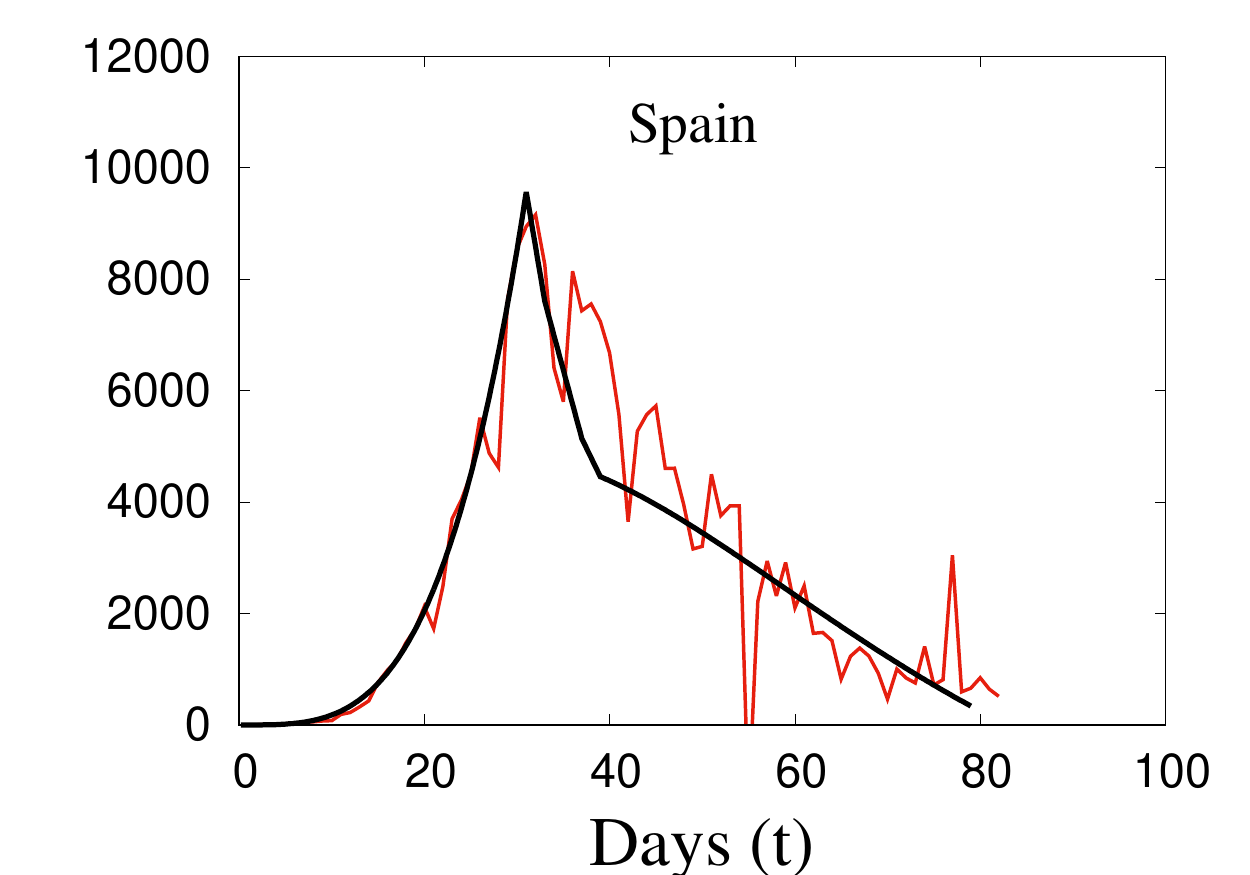}
\includegraphics[scale=0.63]{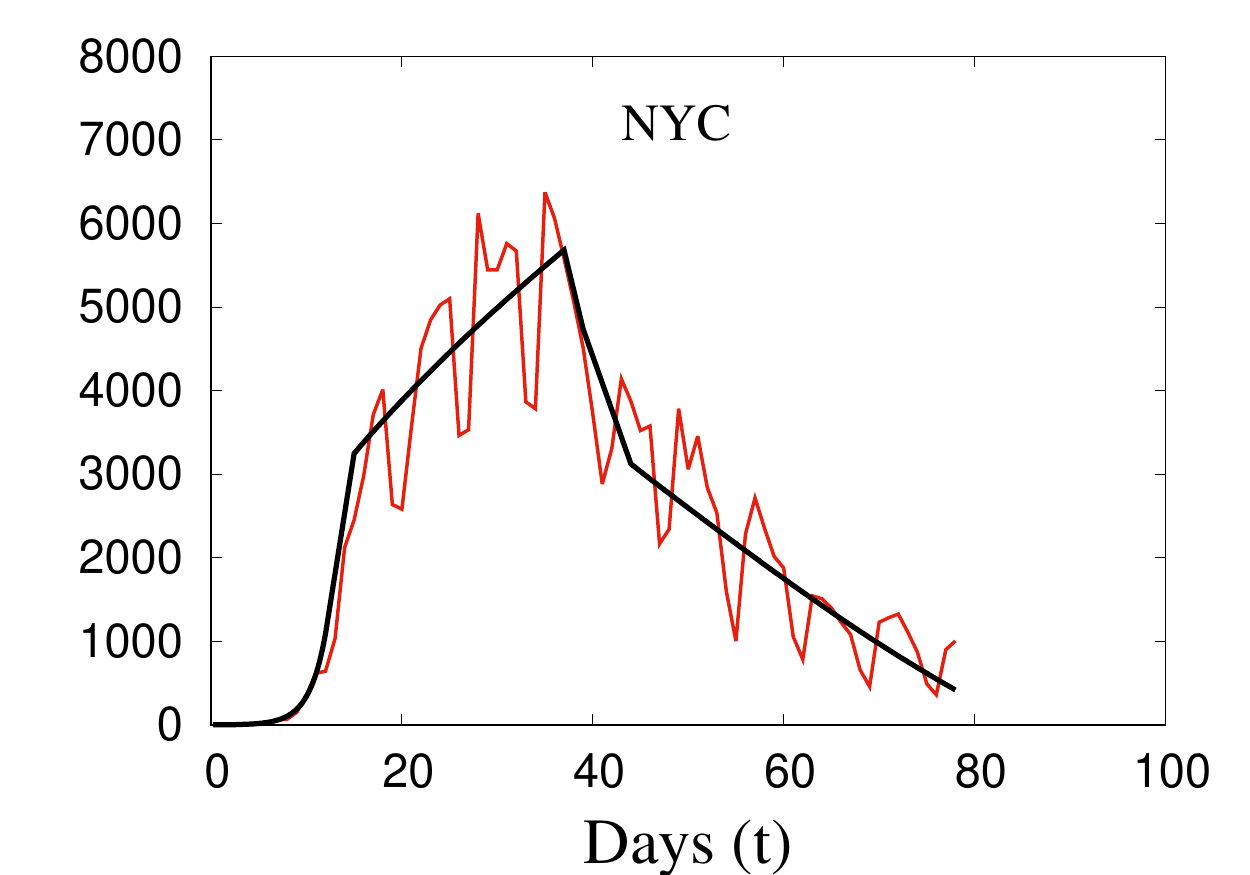}
\caption{\label{fig:d_time_traj_valid1} The number of infected population
  per day are shown for various countries and New York City . The red solid
  lines are the actual data \cite{england_data, covid_wiki_france, covid_wiki_germany, covid_wiki_italy, covid_wiki_spain, covid_wiki_nyc} while the black solid lines are
  obtained from Eq. \eqref{eqn:model_eq}.}
\eef{}
\begingroup
\begin{table}[ht]
	\centering
	\begin{tabular}{|c|c|c|c|c|c|c|c|c|c|}
          \hline
        {Countries} & {Initial} &
\multicolumn{8}{c|} {Parameters}\\
    \cline{3-10}
 and City   &time ($t_0 = 1$) &$\alpha^{e}_{i}$ & $\alpha_{i}$ & $\alpha_m$ & $\lambda$ & $\gamma$ & $t_m$ & $t_{s_i}$ &$\chi^2/dgf$\\
 \hline
England & Feb 24, '20&0.2493       &3.194   &3.107   &0.008   &1.2633   &46   &93& 1.1\\
 France &Feb 25, '20&0       &3.674   &2.340   &0.0242   &1.0141   &41   &80&3.2\\
 Germany &Feb 25, '20&0.2585  &2.383   &2.056   &0.0121   &1.1371   &43   &80&1.4\\
 Italy   &Feb 25, '20&0       &3.220   &2.282   &0.01457  &1.1056   &39   &85&1.6\\
 Spain   &Feb 24, '20&0       &4.515   &2.345   &0.0238   &1.0352   &33   &80&3.7\\
 NYC     &Mar 03, '20&0.5810  &1.619   &1.770   &0.0138   &1.0853   &39   &76 &0.8\\  
    \hline
        \end{tabular}
	\caption{\label{tab:pars}{Mean values of the parameters of Eq. \eqref{eqn:model_eq} as determined by the data \cite{england_data, covid_wiki_france, covid_wiki_germany, covid_wiki_italy, covid_wiki_spain, covid_wiki_nyc} for several countries and New York City.}}
\end{table} 
\endgroup


\bef[h]
\centering
\includegraphics[scale=0.63]{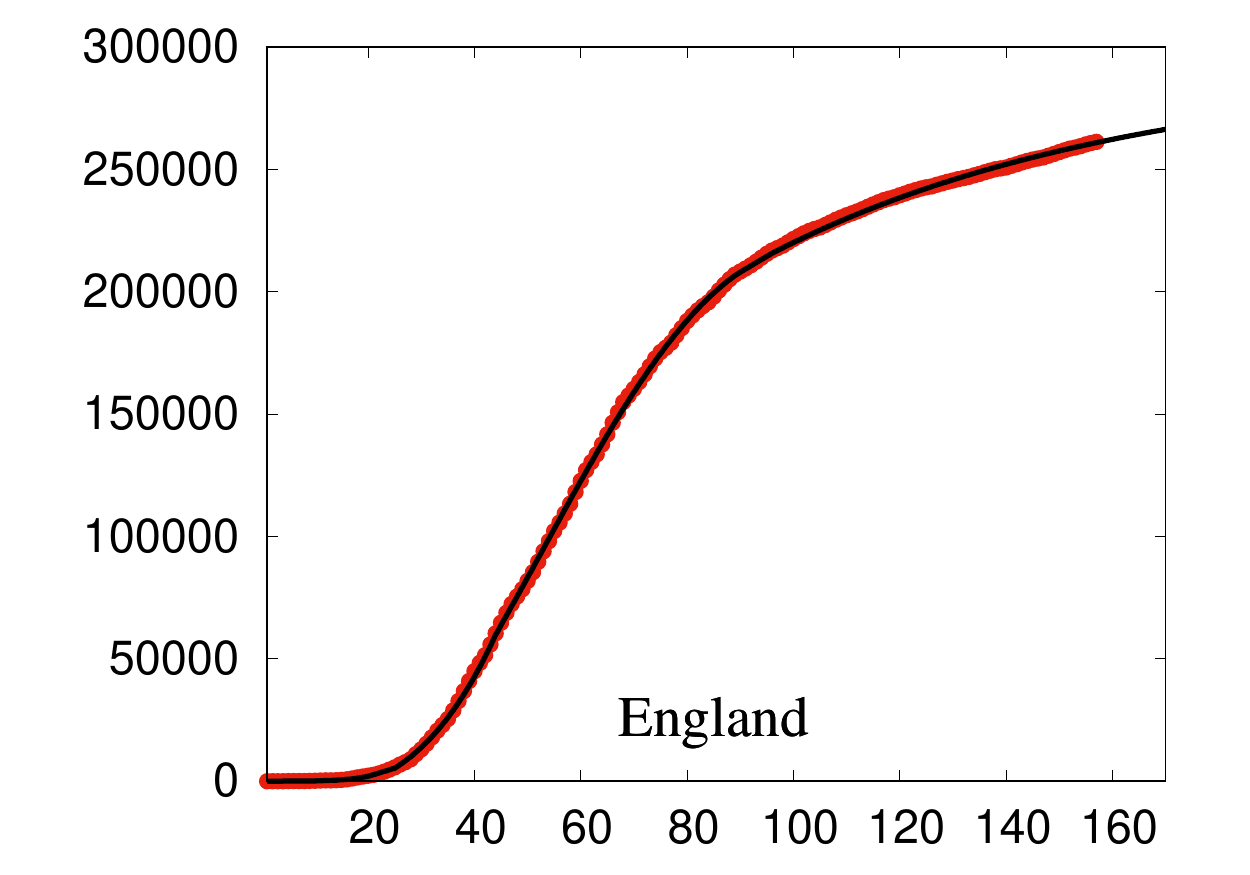}
\includegraphics[scale=0.63]{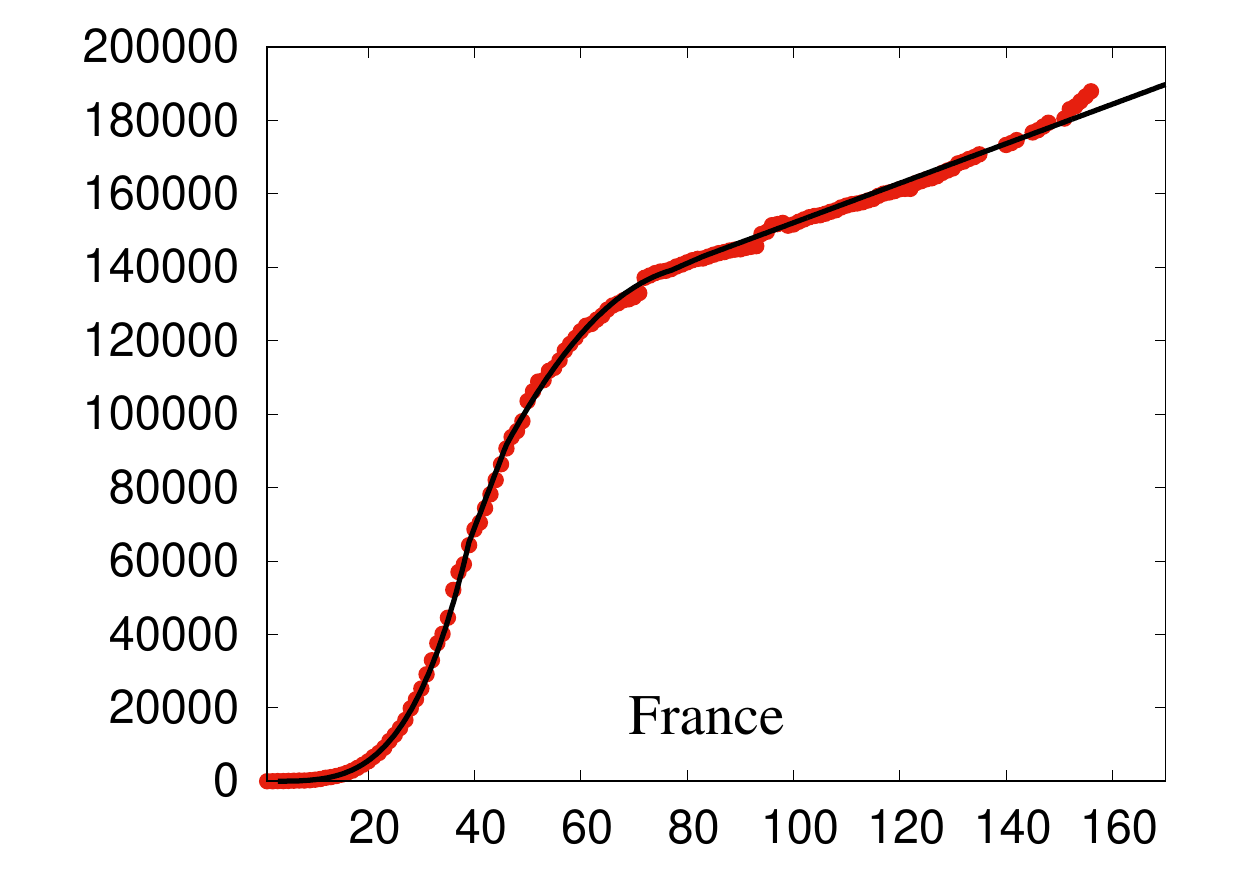}\\
\includegraphics[scale=0.63]{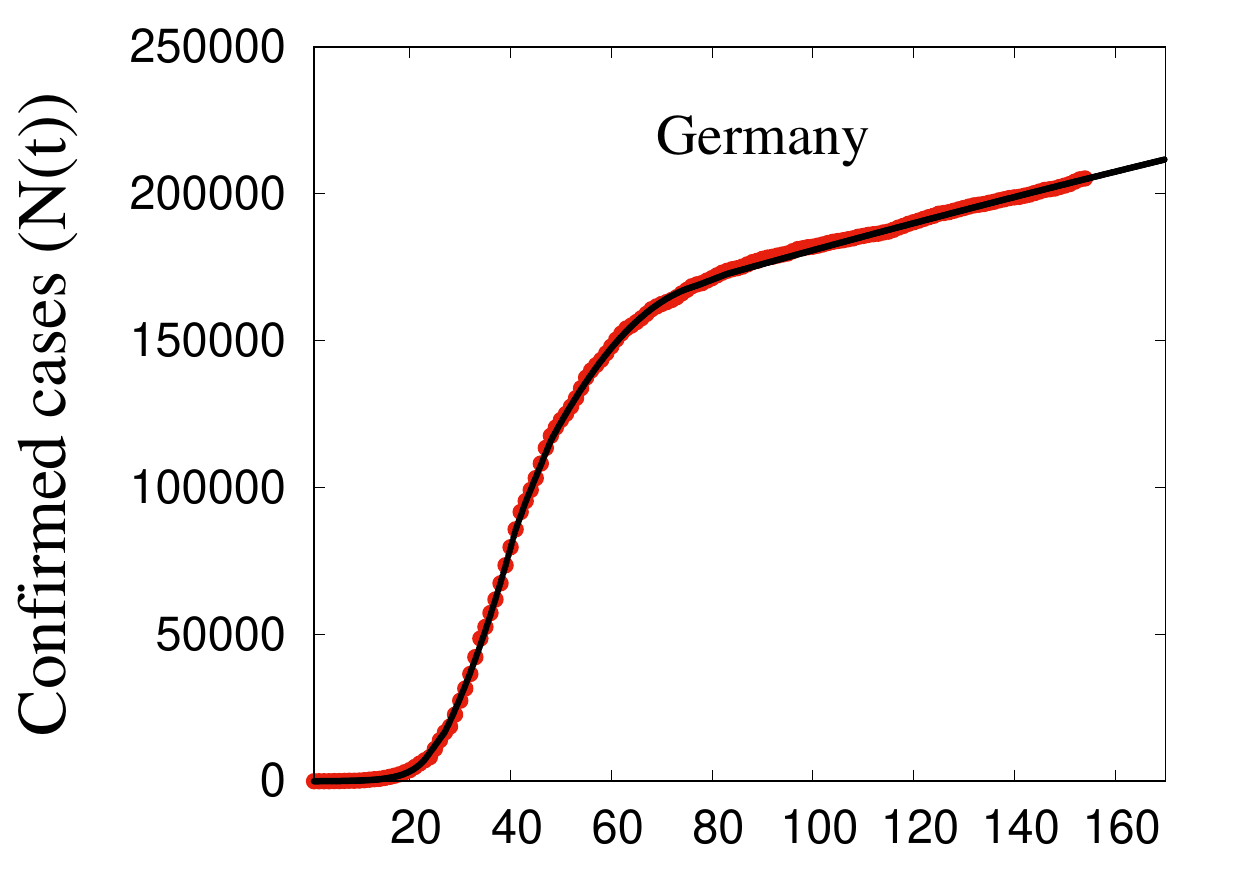}
\includegraphics[scale=0.63]{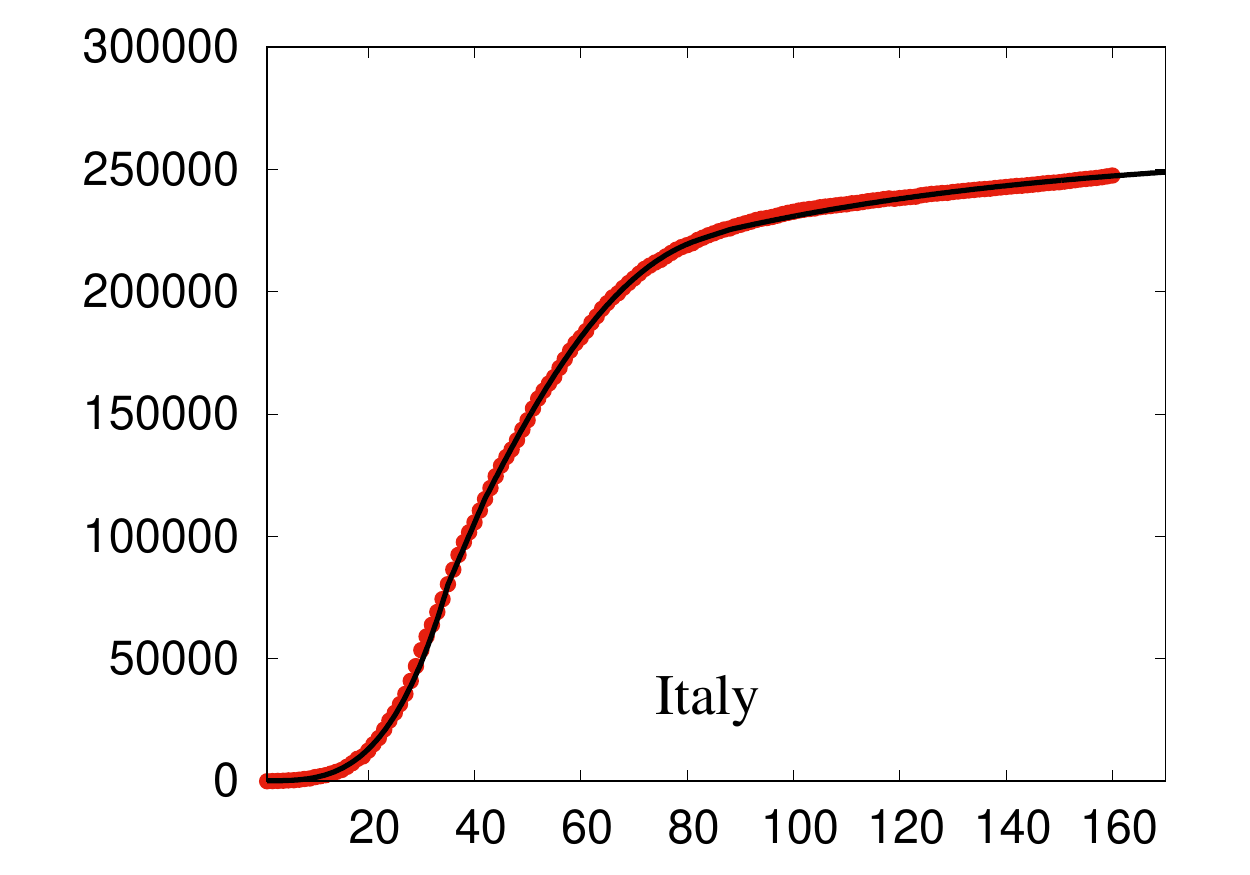}
\includegraphics[scale=0.63]{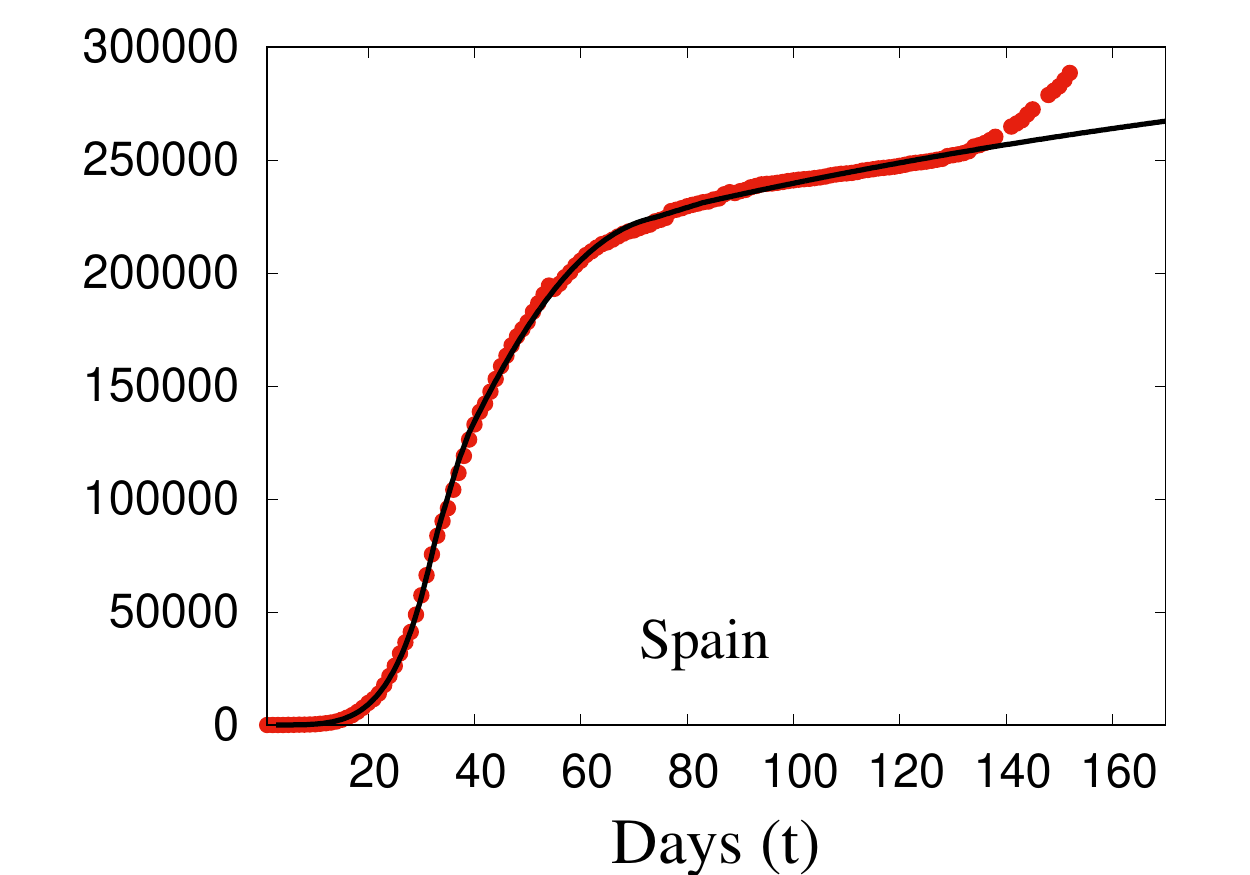}
\includegraphics[scale=0.63]{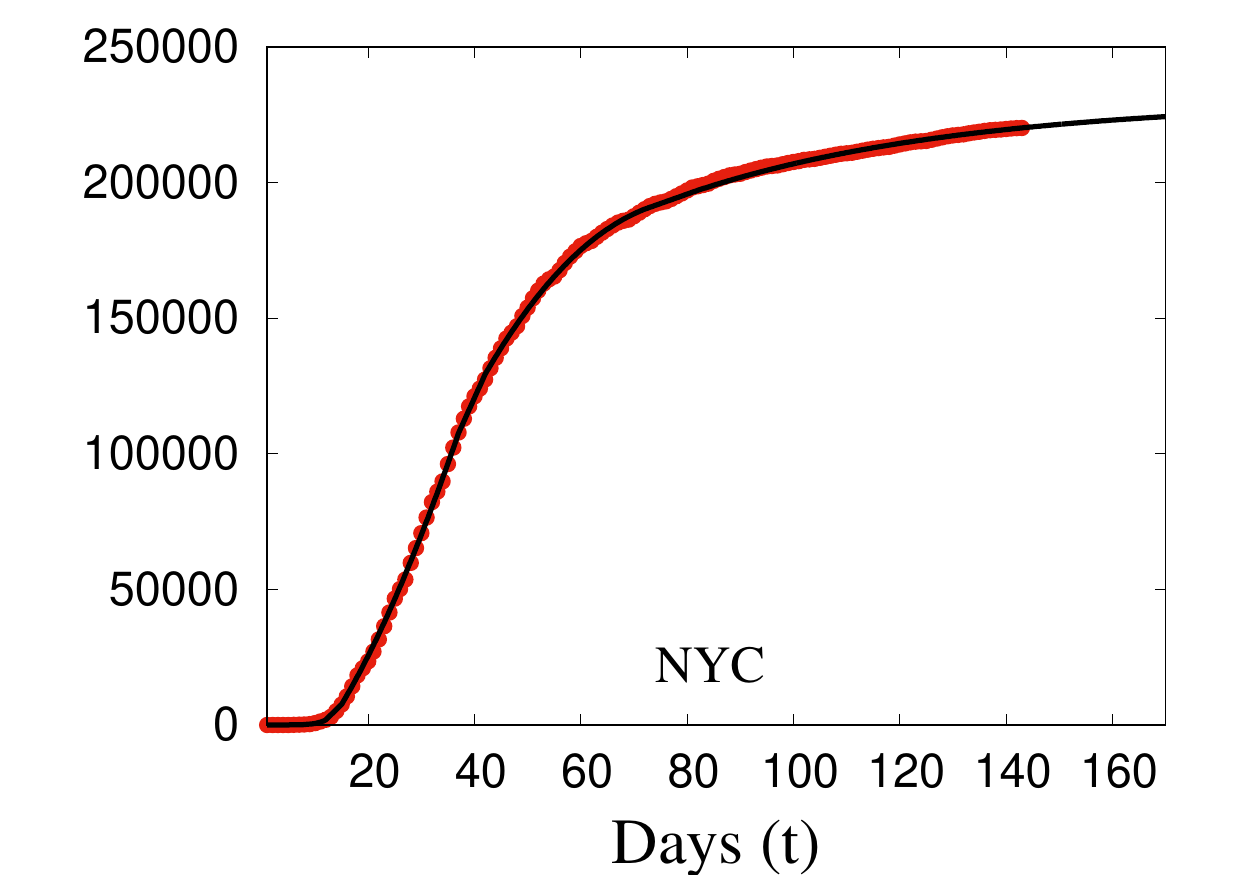}
\caption{\label{fig:time_traj_valid2} The time-evolution trajectories of
  infected population for various countries and for the New York City
  are shown with full set of data (till July 31) including the saturation period. The
  red filled circles are actual data from Refs. \cite{england_data, covid_wiki_france, covid_wiki_germany, covid_wiki_italy, covid_wiki_spain, covid_wiki_nyc}. The
  black solid lines are the trajectories obtained through
  Eq. \eqref{eqn:model_eq}. The thick black lines also include 95\% confidence intervals of the fitted parameters of the proposed model. The extended time trajectories with no data are not the projection since a second wave of infection can start in this period (see text for details).}
\eef{}
\bef[h]
\centering
\includegraphics[scale=0.63]{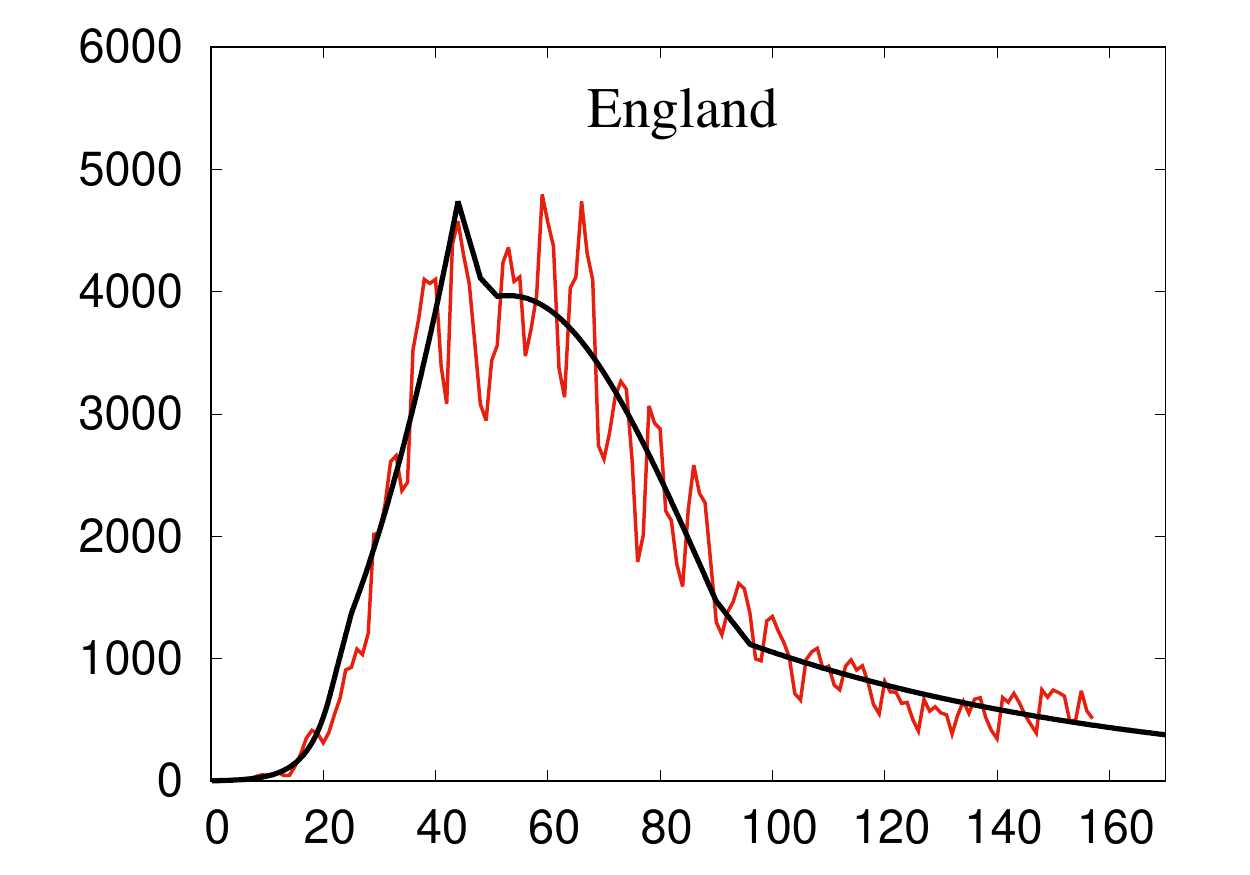}
\includegraphics[scale=0.63]{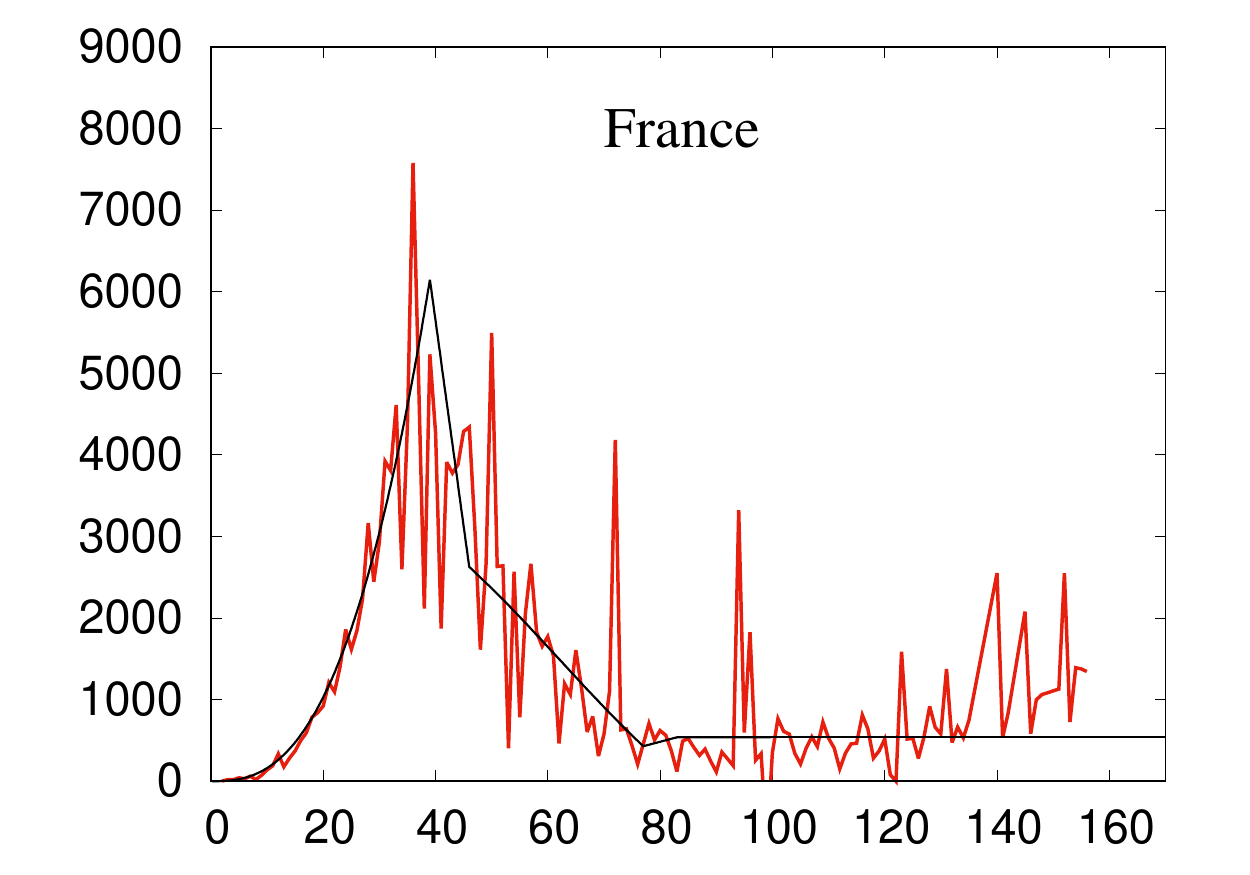}\\
\includegraphics[scale=0.63]{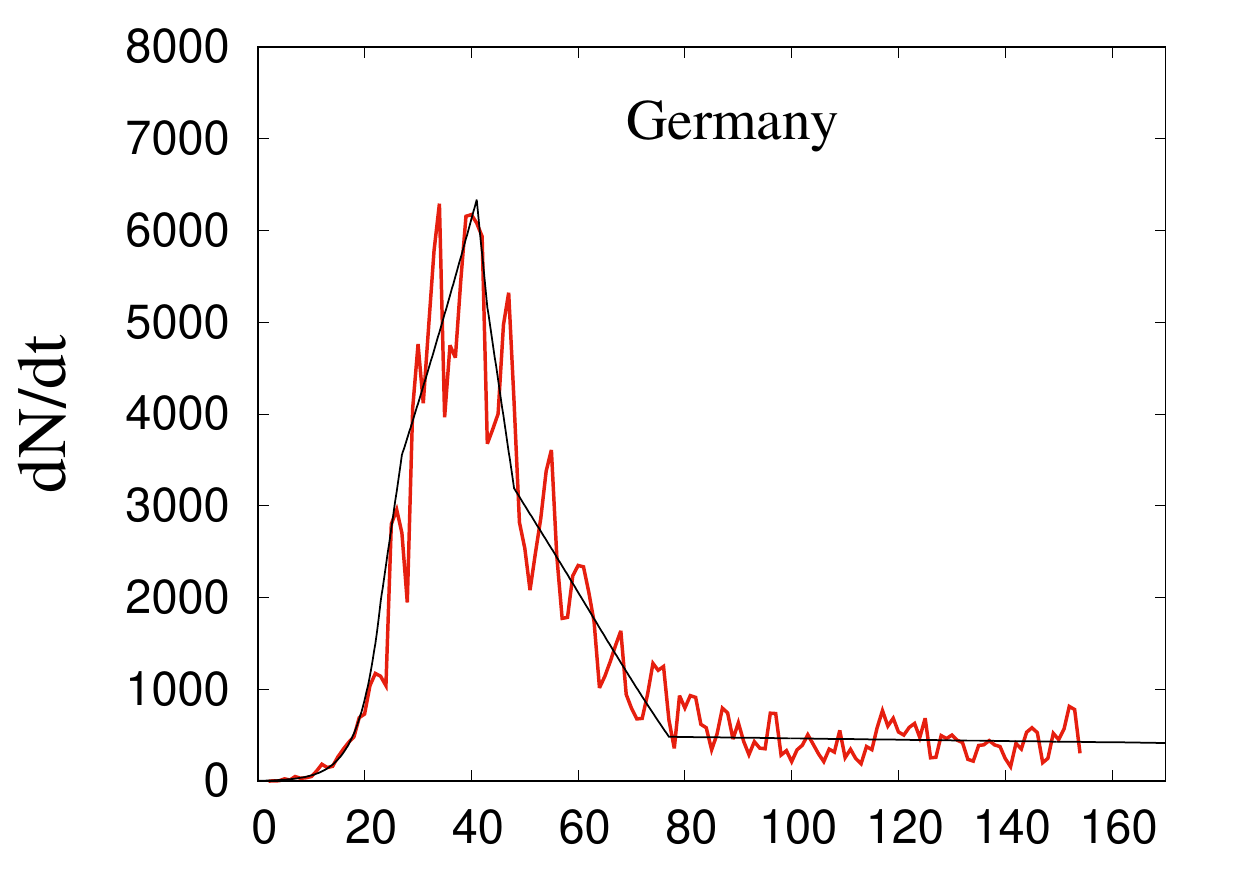}
\includegraphics[scale=0.63]{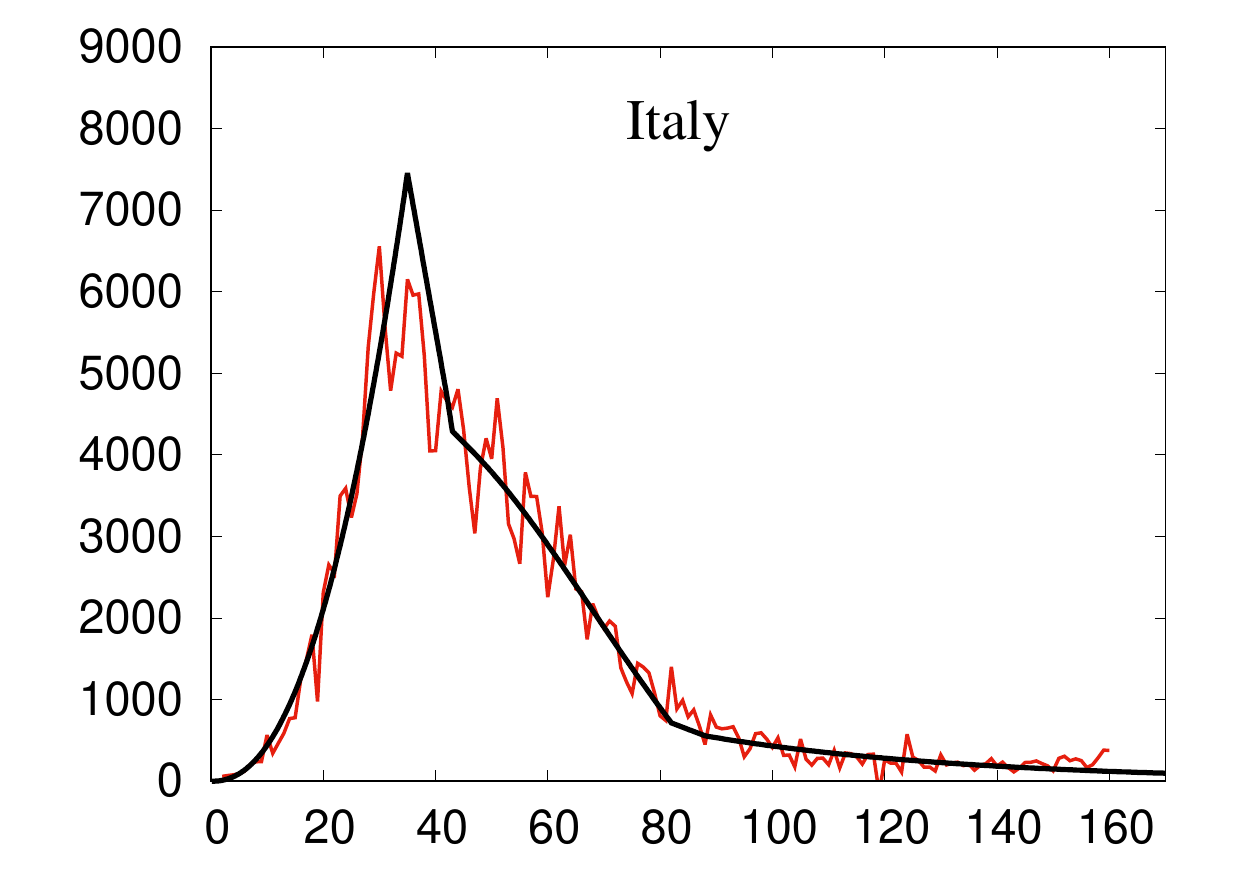}
\includegraphics[scale=0.63]{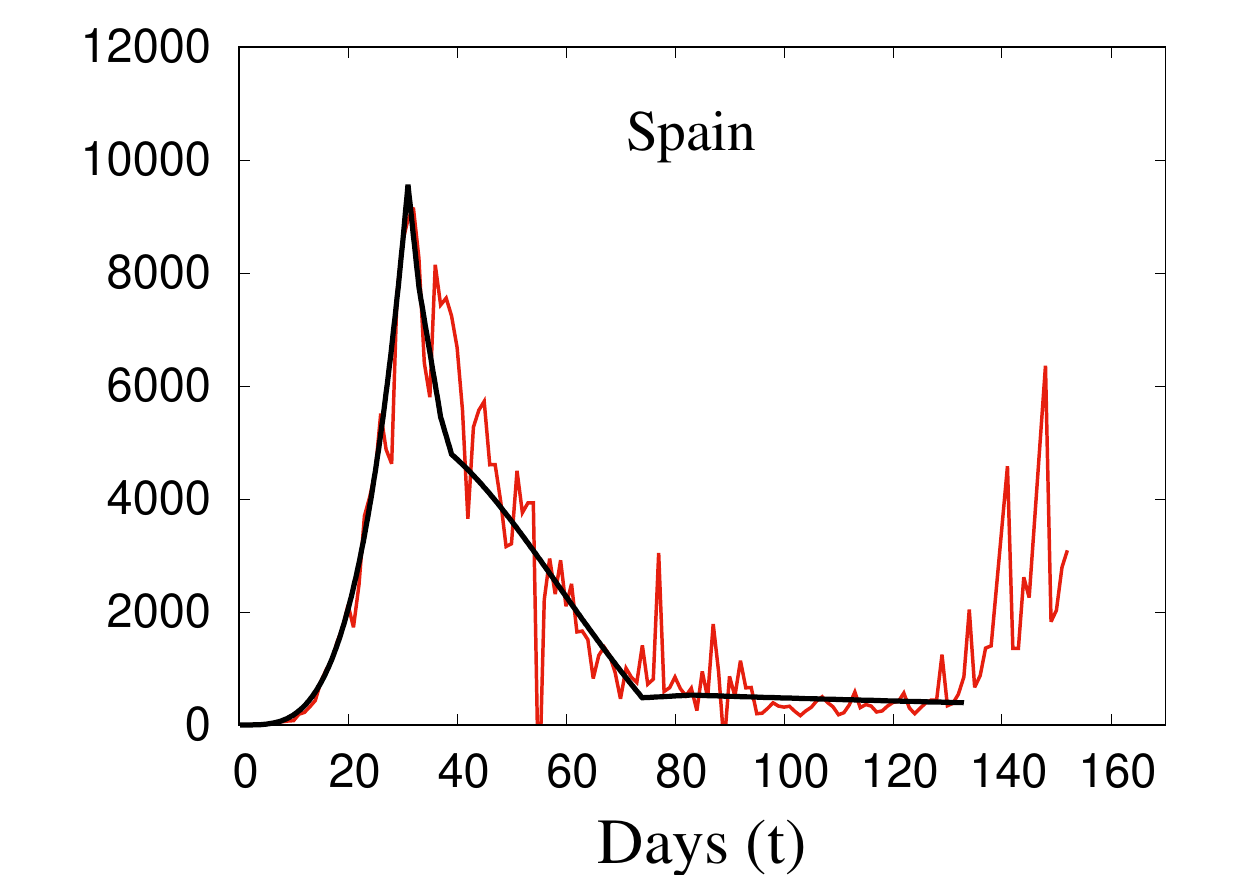}
\includegraphics[scale=0.63]{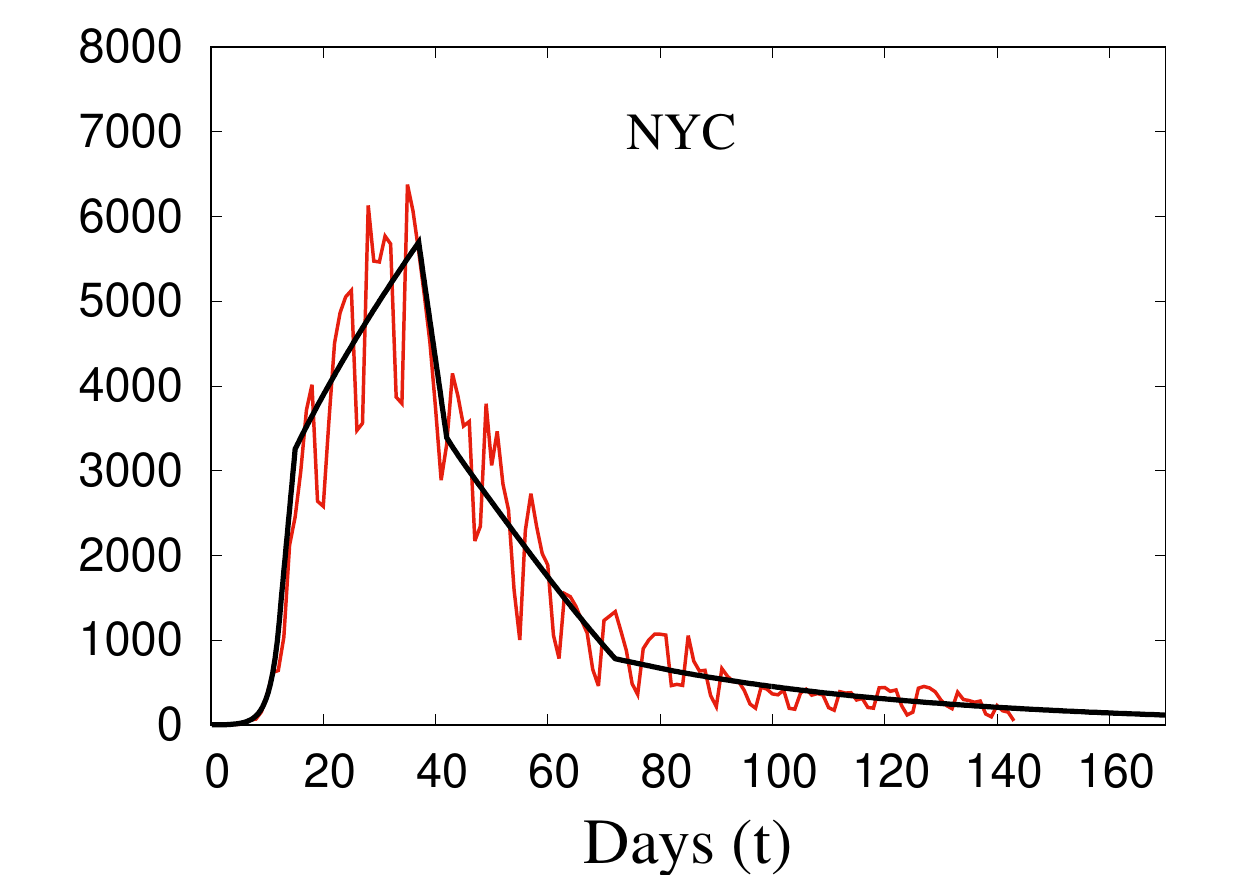}
\caption{\label{fig:d_time_traj_valid2} The number of infected
  population per day are shown for various countries and for New York
  City with full set of data (till July 31). The red solid lines are the actual data
  \cite{england_data, covid_wiki_france, covid_wiki_germany, covid_wiki_italy, covid_wiki_spain, covid_wiki_nyc}, while the black solid lines are obtained using Eq. \eqref{eqn:model_eq}, as in Fig. \ref{fig:time_traj_valid2}.}
\eef{}

It is evident from Figs. \ref{fig:time_traj_valid1} and \ref{fig:d_time_traj_valid1} that the Eqs. \eqref{eq:model_eq1} and
\eqref{eq:model_eq2} are able to track the time evolution of the
cumulative infected population very well, from the beginning to the onset
of saturation period. One would expect that Eq. \eqref{eq:model_eq2}
terminates at a point where the rate of growth vanishes. However, that
point is never reached due to multiple reasons: partial or full
un-lockdown when the rate of growth reduces, huge carrying capacity (when
 the total number to reach herd immunity is very high). From the onset of
saturation, thus Eq. \eqref{eq:model_eq2} is unsuitable to track the
progression of the disease. At that point the third phase of time
evolution with Eq. \eqref{eq:model_eq3} becomes effective. We choose
the onset of $t_{s_{i}}$ dynamically with the condition such that
$\chi^2 = \chi^2_i + \chi^2_m + \chi^2_s$ together provide the best
acceptable $\chi^2$/dgf with $t_{s_{i}}$ as high as possible. With
that method we fit the available full data set till July 31,
2020. Results for the total number of Covid-19 positive cases, ($N(t))$,
and the per day increments ($dN(t)/dt$), for the full data set along
with fitted trajectories are shown in Figs. \ref{fig:time_traj_valid2}
and \ref{fig:d_time_traj_valid2}, respectively, for various
countries and New York City. It is quite satisfying to see that the
full data set for various countries and cities as well can be tracked very
well through the model proposed in Eq. \eqref{eqn:model_eq}. It is to be mentioned here that in the saturation period our main purpose is to show the validity of the proposed model. We do not project the time evolution trajectory in the future in this period as is shown in the figure. Though the extended trajectories (beyond data point) are obtained with 95\% confidence intervals of the fitted parameters, a projection in this period may not hold as a second wave of infection can start at any point of time in this period. That is probably what is currently happening, albeit with less extent, in all the regions mentioned in this plot, except for Italy and New York city. Most notable one is on Spain's data where from the middle of July, the number for per day infection has started to increase again beyond what we observe throughout inside a large part of saturation period ($t = 81-135$). Considering that Covid-19 is a contagious disease, this second rise could be due to i) relaxation of measures against the
disease spread which was implemented earlier, ii) a large unaffected population, and iii) no availability of an effective vaccine. This will be elaborated further when later we discuss the infection of the USA.


Since data for an individual day fluctuates more and sometime extra data for a day gets reported on the next day (since the official announcement comes at a certain time) it may be worthwhile to study the same data taking an average over a few days. That will also check  the consistency of fits to extract the parameters of Eq. \eqref{eqn:model_eq}. We perform a study on that for England's data taking an average over 5 days (one can also choose a different bin size, but it should not be too large). We fit the data with the same errorbar as in each day's data with the corresponding periods. The results are shown in Fig. \ref{fig:avg5_eng}, both corresponding to per day data and average over 5 days. As one can see that  there is no significant difference between results obtained when we use each days data (magenta line) and average data (blue line).

\bef[h]
\centering
\includegraphics[scale=0.63]{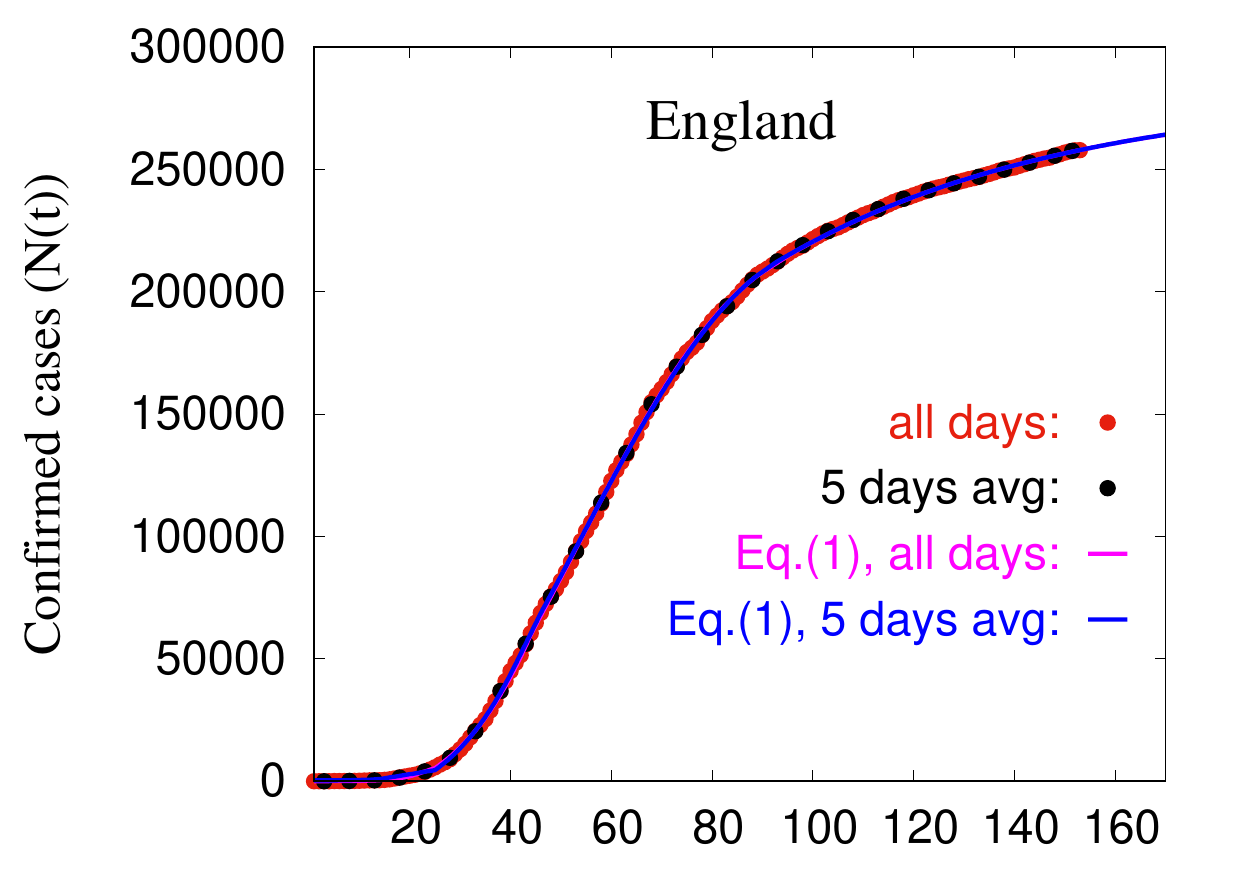}
\includegraphics[scale=0.63]{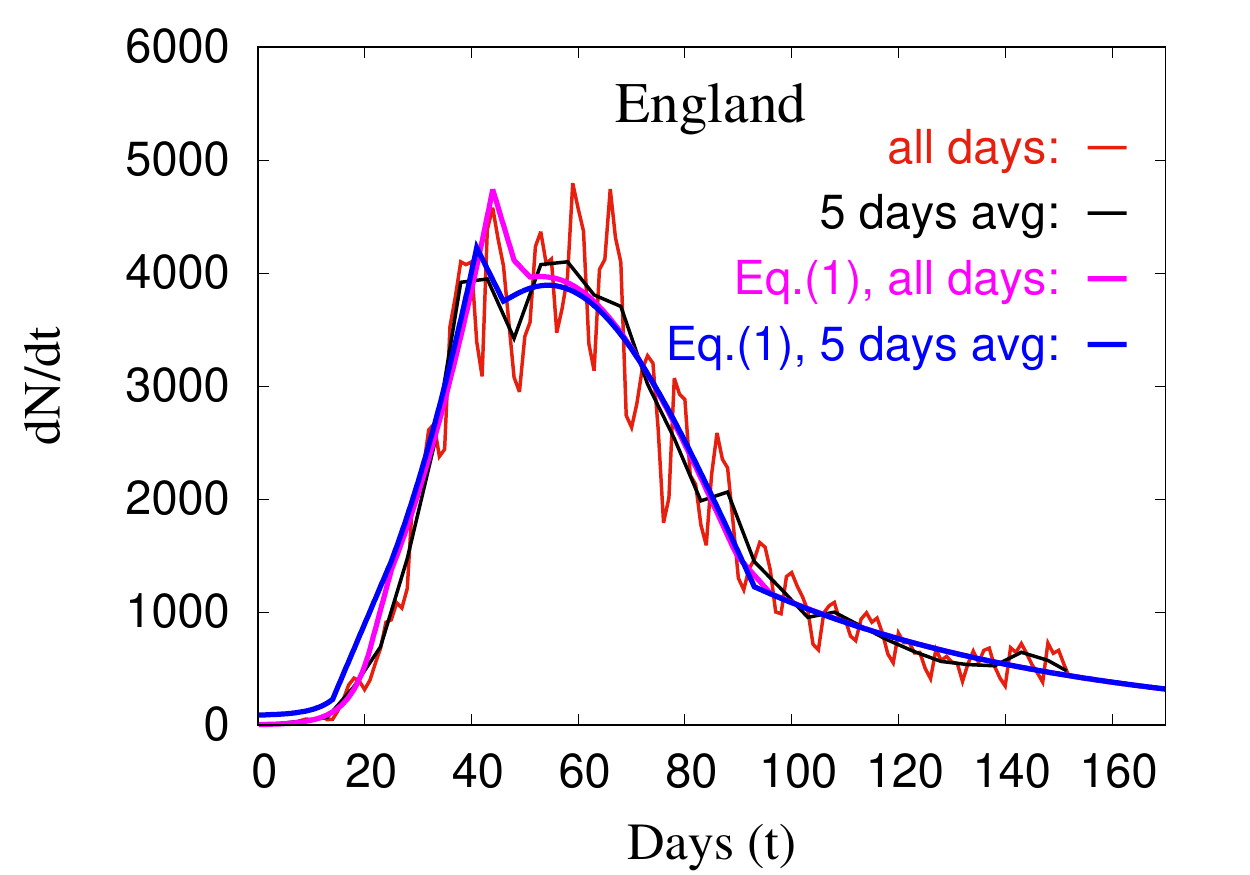}\\
\caption{\label{fig:avg5_eng} The time evolution data for the number of infected population is shown for England corresponding to data for each day and also for average over 5 days. The results obtained using Eq. \eqref{eqn:model_eq} are shown by magenta and blue lines respectively.}
\eef{}

\bef[h]
\centering
\includegraphics[scale=0.63]{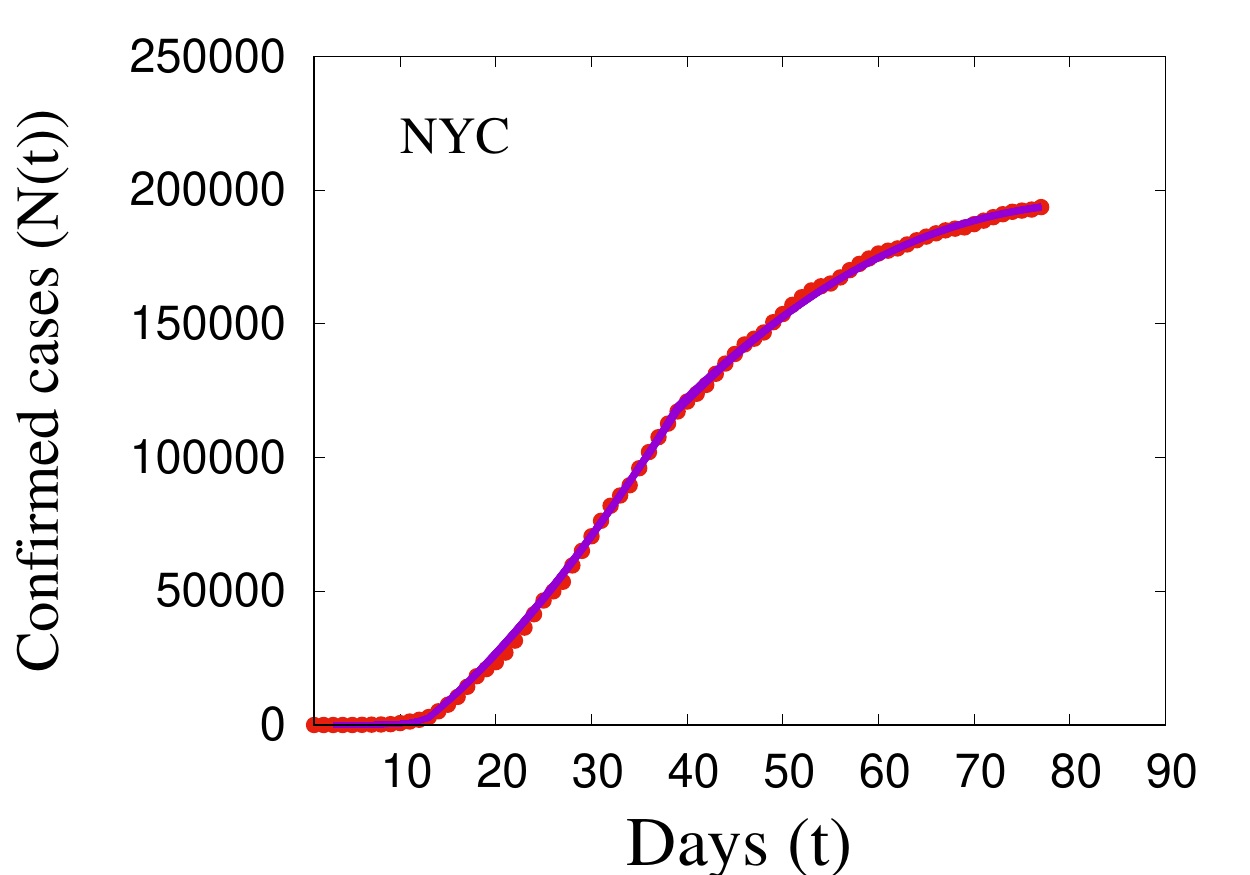}
\includegraphics[scale=0.63]{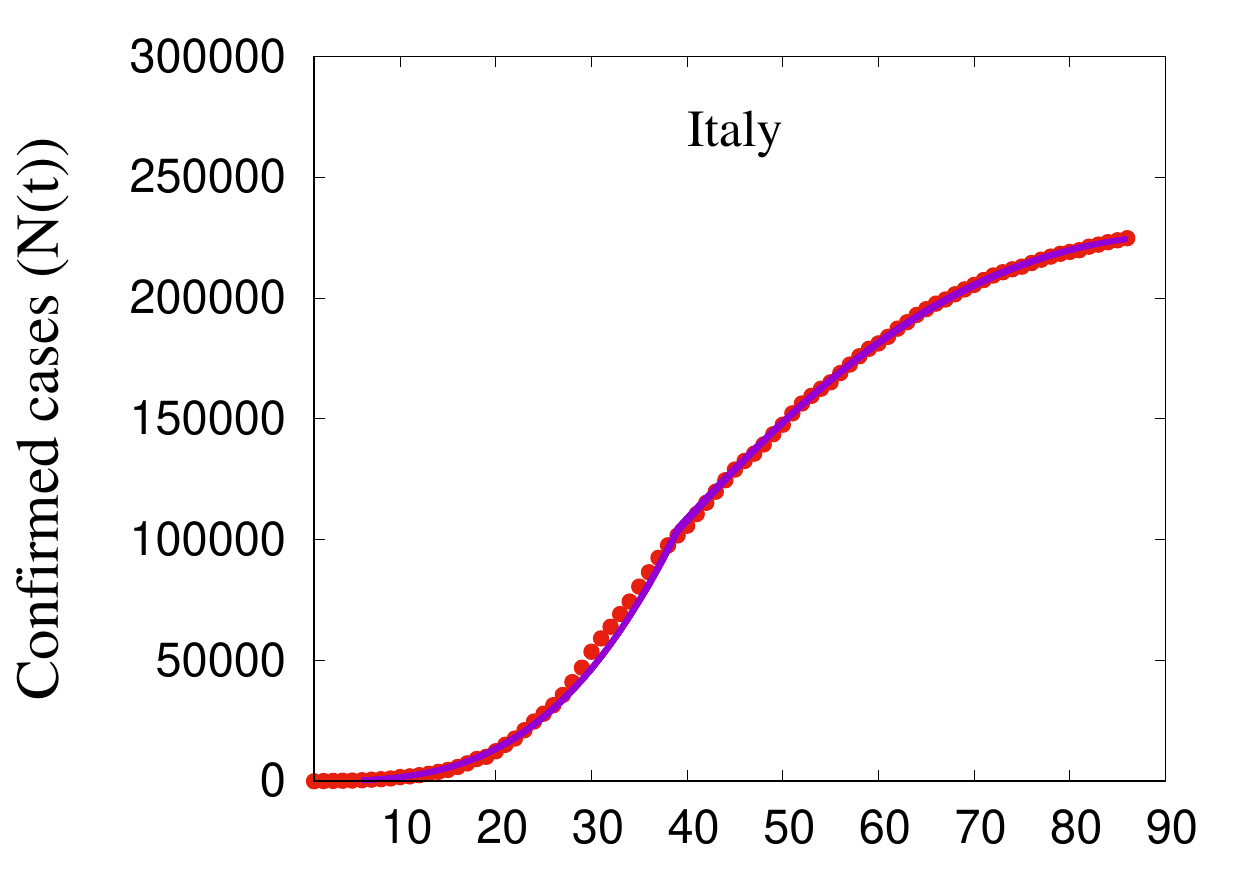}\\
\includegraphics[scale=0.63]{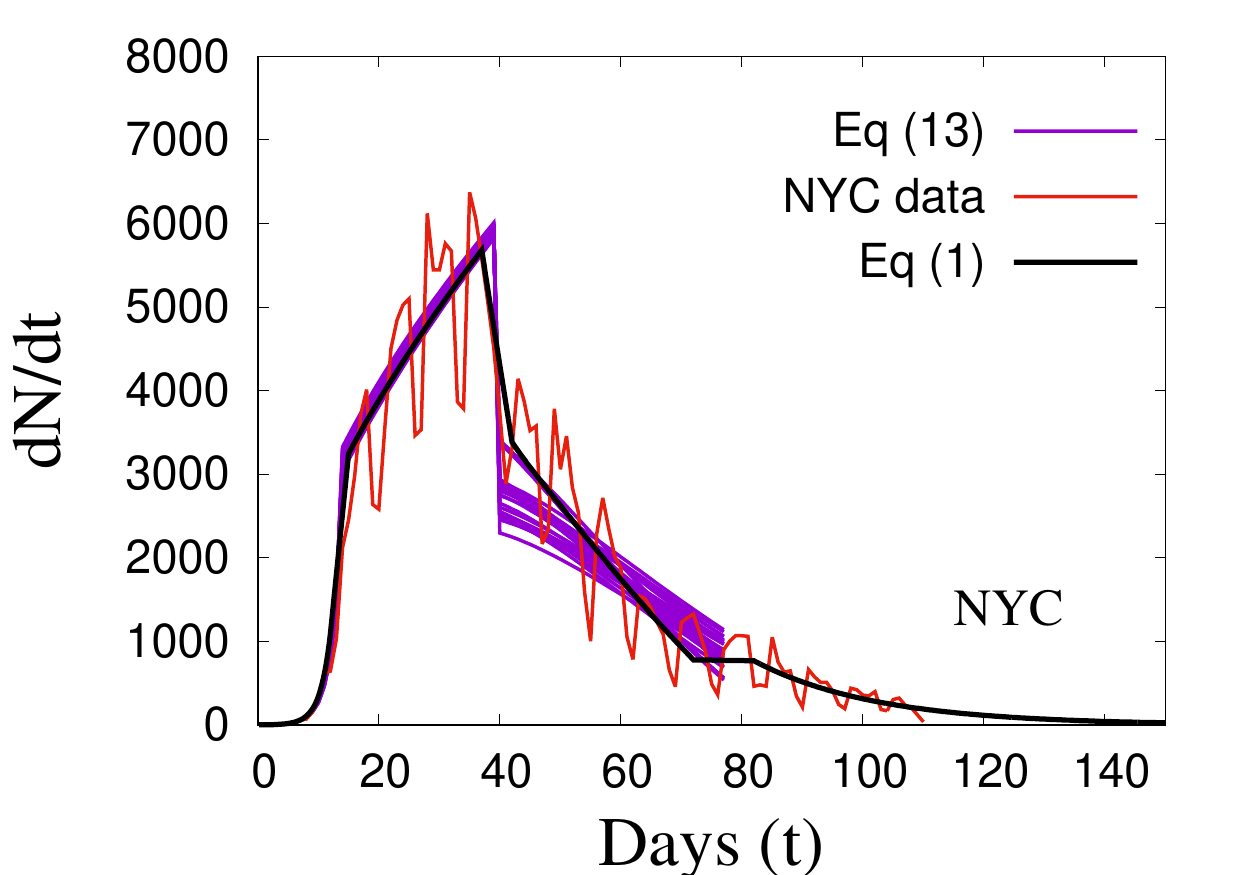}
\includegraphics[scale=0.63]{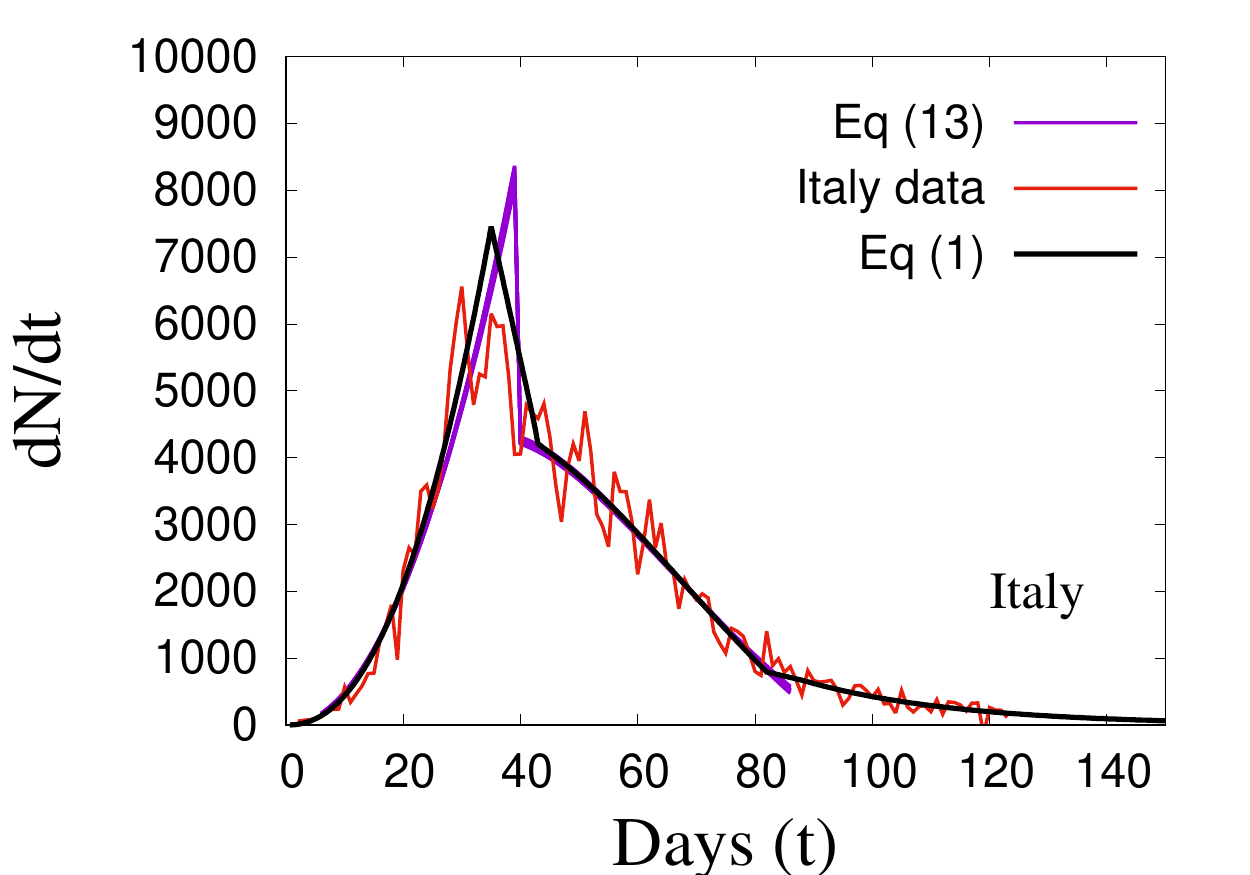}
\caption{\label{fig:time_traj_d} The time evolution trajectories as
  obtained from the proposed differential equation
  (Eq. (13)) for Covid-19 growth are plotted along
  with the data for New York City and Italy. The red solid lines are
  actual data from Refs. \cite{covid_wiki_italy, covid_wiki_nyc} while the blue lines are obtained from
  the solutions of the differential equations (Eq. (13)). The upper two figures are for the
  total number of infection while the bottom two figures represent
  corresponding per day infection.}
\eef{}

Next, we proceed to solve Eq. (13) to find the time
evolution trajectory governed by the dynamics of such differential
equations.  Eqs. \eqref{eq:model_deq1}, \eqref{eq:model_deq2} and
\eqref{eq:model_deq3} are first-order differential equations with a
number of parameters.  These equations can be solved numerically with
a given set of parameters leading to a time-evolution trajectory of
the total infected population.  We vary the parameter set over a wide
range and choose a parameter space which can describe the
time-dependent data with minimum $\chi^2$ with a given errorbar.
Here, again we take the usual definition of
$\chi^2 = \sum{(f(t_i)-N(t_i))^2/\sigma_{N(t_i)}^2}$, where $f(t_i)$'s are obtained
from the solution of Eq. (13) after matching the boundary conditions, whereas $N(t_i)$'s are the actual data with imposed errors $\sigma_i$, which accounts for the uncertainty in the reported numbers as mentioned earlier. To be noted that solutions of these equations
with a given errorbar will result in a band of trajectories
rather than just one. However, that is more realistic since a different set of parameters, with minimum $\chi^2$, generates a band of trajectories
which can spread over a few days which is also what we observe in reality.
At the boundary, we match the solution without any discontinuity that is
solutions are continuously progressed. For
example, the initial value of $N(t)$ at $t = t_m+1$ with
\eqref{eq:model_deq2} is generated from the solution of
\eqref{eq:model_deq1} at $t = t_m$. In this way, we match the
boundary conditions of solutions between different regions.  A
trajectory generated by these equations is accepted or rejected based
on the minimum $\chi^2$ within the errorbars.  The trajectories obtained through these differential equations for
Italy and New York City are shown in Fig \ref{fig:time_traj_d}.
Top two plots show the time evolution trajectory for the cumulative number of
infection while the bottom two plots show its rate of change (infected number per day).
Here again, it is satisfying to see that the differential equations
\eqref{eq:model_deq1} and \eqref{eq:model_deq2} can reproduce the
time evolution of disease spread from the initial days to the onset of saturation.
Similarly, the solutions of Eq. \eqref{eq:model_deq3} can also track the disease spread in
the saturation period assuming the same environmental condition prevails throughout this period.
Though we show results for only two regions, data for others regions can also be tracked with equal success
through the proposed differential equations (Eq. (13)).

\subsection{Predictive ability of the proposed model}
The success of a mathematical model not only depends on the validation
but crucially also on its predictive power.  After successfully
validating the model through reproducing the time evolution of the
number of infections, hence, we proceed to find the predictive
ability of our proposed model. Since the model is non-linear, a
small change in parameter space may result in a different time
evolution trajectory. However, if there is a parameter space that is
unique to disease spread and if that can be constrained with a subset
of data then the model can be used for future evolution. With this in
mind, we find that it is not possible to constrain the parameter
space until it reaches the beginning of the mitigation period ($t >
t_m$).
 That is apparent as the parameters of a region cannot be
 obtained from other regions since the dynamics of disease spread
 in two regions are
different. We use a subset of data till the 55th days (for NYC)
and 60th days (for Italy) and use those to constrain the respective
parameter spaces. We then employ those to obtain the respective future
trajectories till the onset of saturation period for the total number
of infection. To be mentioned that the onset of saturation period is
determined dynamically by minimizing the total $\chi^2$.  Results are
shown in Fig. \ref{fig:projection}, where the top two plots are for
the cumulative number of infection and the bottom plots are for per
day infection. The solid red lines represent the actual data
\cite{covid_wiki_italy, covid_wiki_nyc} while the blue solid lines are obtained using
differential equations (Eq. (13)).  To compare the effectiveness of
projection from the model results, in the bottom two figures we also
plot the actual data for the whole range along with its fitted results
through Eq. \eqref{eqn:model_eq}. As can be observed from this figure
that the proposed differential equations (Eq. (13)) can project the
future time evolution trajectories of the number of infection with a
correct trend for next few weeks quite reliably.
For New York City the projection is for 21 days $t_{s_{i}}(=76-55)$ and the trend of the projected trajectory goes even beyond that. For Italy this projection is for 25 days $t_{s_{i}}(=85-60)$. We find that this
projection can progressively be improved by gradually including more
data. It would be interesting to see how far in future time-scale this model can
make a projection, and that we will discuss in the next subsection while
considering the disease spread for countries where the infection is
still increasing substantially.
\bef[h]
\centering
\includegraphics[scale=0.63]{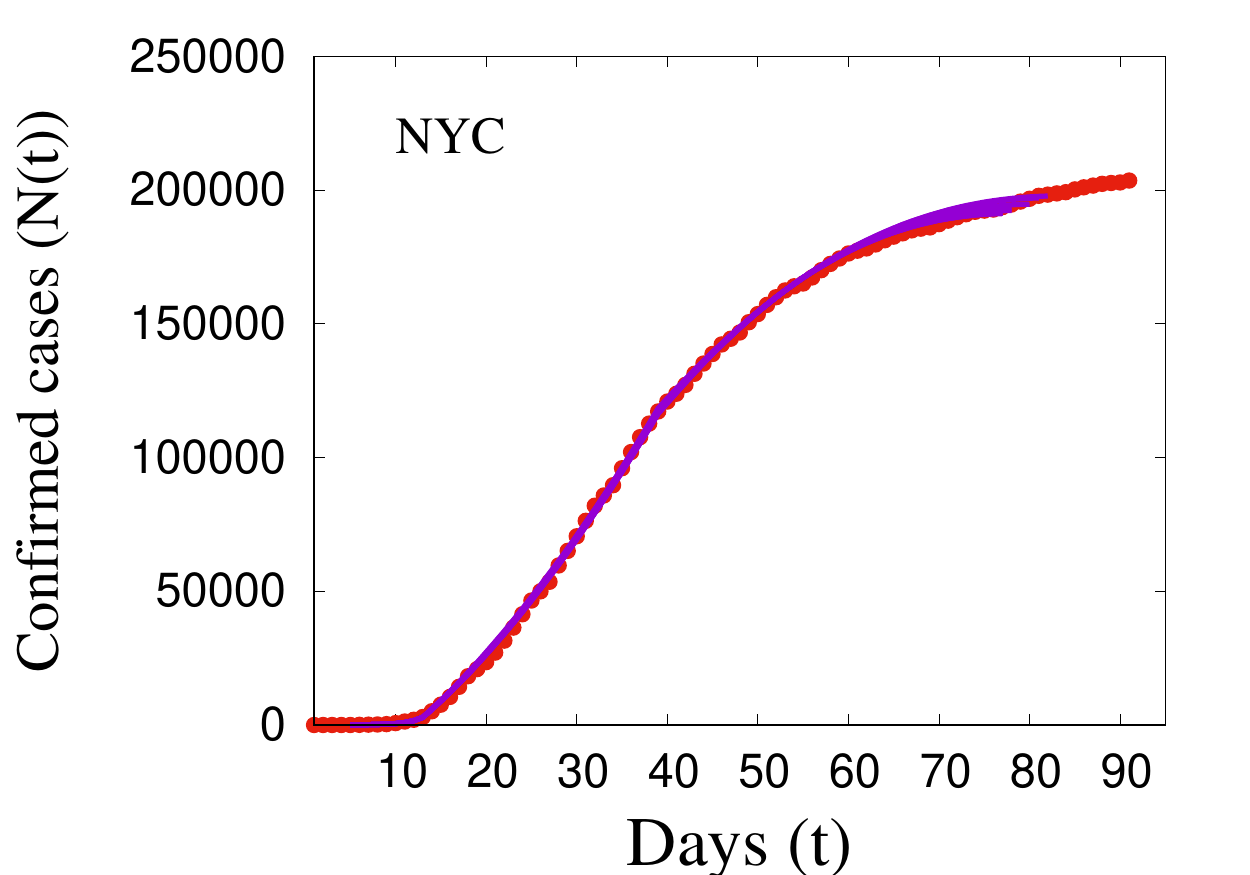}
\includegraphics[scale=0.63]{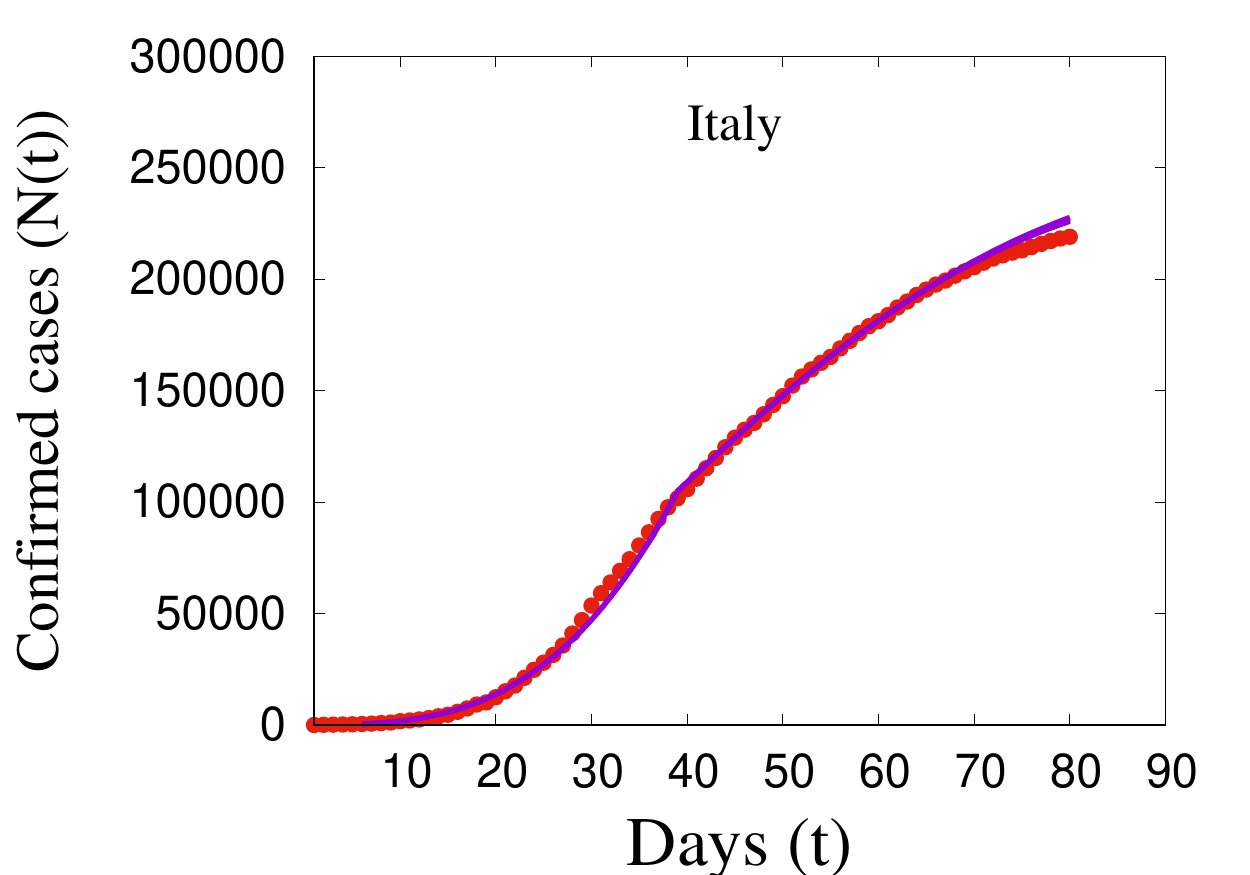}\\
\includegraphics[scale=0.63]{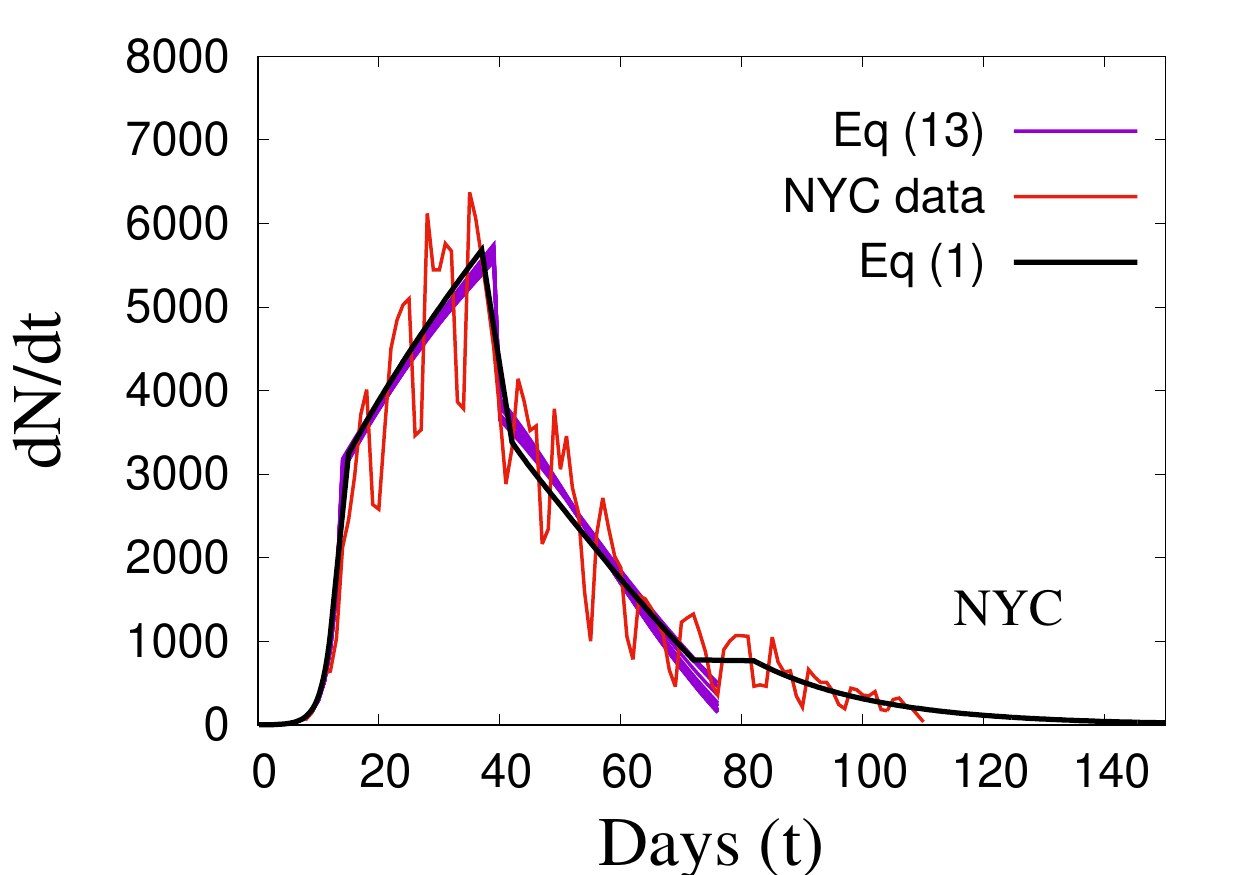}
\includegraphics[scale=0.63]{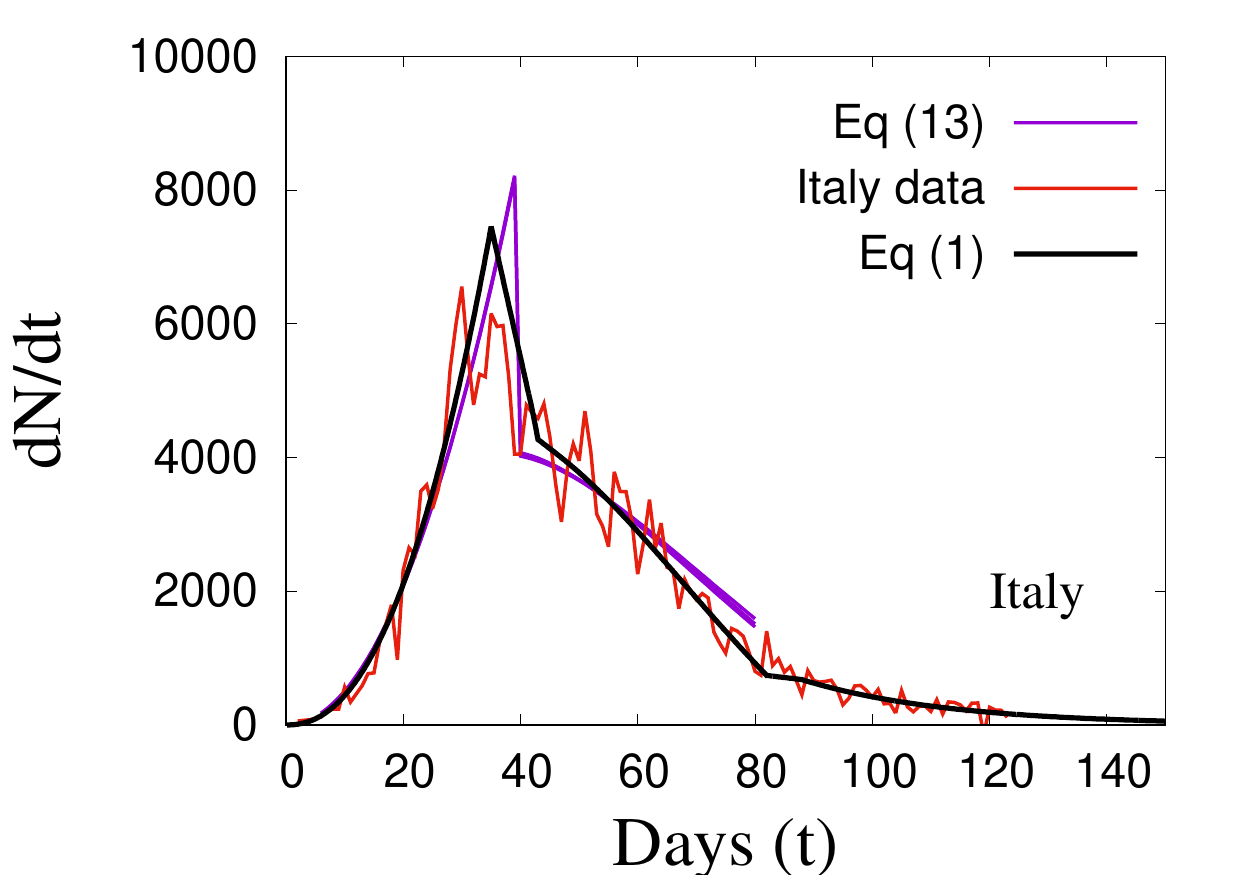}
\caption{\label{fig:projection} The time evolution trajectories as obtained from the proposed differential equation (Eq.(13)). A subset of the data is utilized to constrain the parameter set of the model which are then used to predict future time evolution. The data (cumulative and differential) for NY City and Italy are shown by the red solid lines while the blue solid lines are obtained using data up to time 55th days and 60th days for NY City and Italy, respectively.}
\eef{}

\subsection{Projection for a few other countries}
Having validated the model and then showing its predictive ability, we
proceed to study the data on other countries, Russia, Brazil, India and the USA, where the number of
Covid-positive cases are still rising substantially.
For India, we also choose two of its biggest
cities, Mumbai and Delhi. We first constrain the parameters by fitting
the available data as detailed above, both for
Eq. \eqref{eqn:model_eq} and the differential equations (Eq (13)). We then
proceed to predict the future time-evolution trajectories up to the
onset of saturation.

\subsubsection{Russia}
In Fig. \ref{fig:russia_plot} we plot the cumulative number of Covid-19 positive cases for
Russia \cite{covid_wiki_rus} with the effective starting date ($t_0$) as March 03, 2020. The data of the time-trajectory shows the expected exponential rise followed by a power-rise growth as in Eq. \eqref{eq:model_eq1} and \eqref{eq:model_eq2}.  We fit the data with the combined
equations and find $\alpha_{i}^e = 0.2239$ and $\alpha_i = 4.886$. It
is interesting to note that this is towards the higher value of $\alpha_i$  that shown in Table 1 for other countries in Europe. From the fitted
results we find the transition from the rising to the mitigation
period has happened around time $t_m = 67$, which corresponds to around May 10, 2020. It is interesting to note that the mitigation period of Russia is quite longer compared to other European countries, perhaps due to larger population.
Using the available data up to
August 10 we extract the parameters and use those to predict the time evolution
trajectory till the time $t = 200-215$, which we believe, signals the onset of
saturation period ($t_{s_{i}}$) with less than 300 per day infections. We also use the proposed differential
equation for the time evolution (Eq (13)) and constrain the
parameter set using data till August 10. The corresponding time evolution
trajectories (both cumulative and differential) are shown in
Fig. \ref{fig:russia_plot}. The blue bands are obtained using the
constrained parameter set, while the red solid lines are actual data.
The projected results, using both Eq. \eqref{eqn:model_eq} as well as Eq. (13)
show that, with the same prevailing environmental condition as it is now,
the onset of saturation ($t_{s_{i}}$)  will start most probably 
in between $t = 190-215$. That is, by the late September to the middle of October the number of infection will reduce substantially (below 300 per days) with a total cumulative infected number in between 0.95-1.1 million.
With the sustained conditions against disease spread a gradual slow
increment is thereafter expected. It will be interesting to find if
the true data follows the trajectory as we projected in
Fig. \ref{fig:russia_plot}.
\bef[h]
\centering
\includegraphics[scale=0.63]{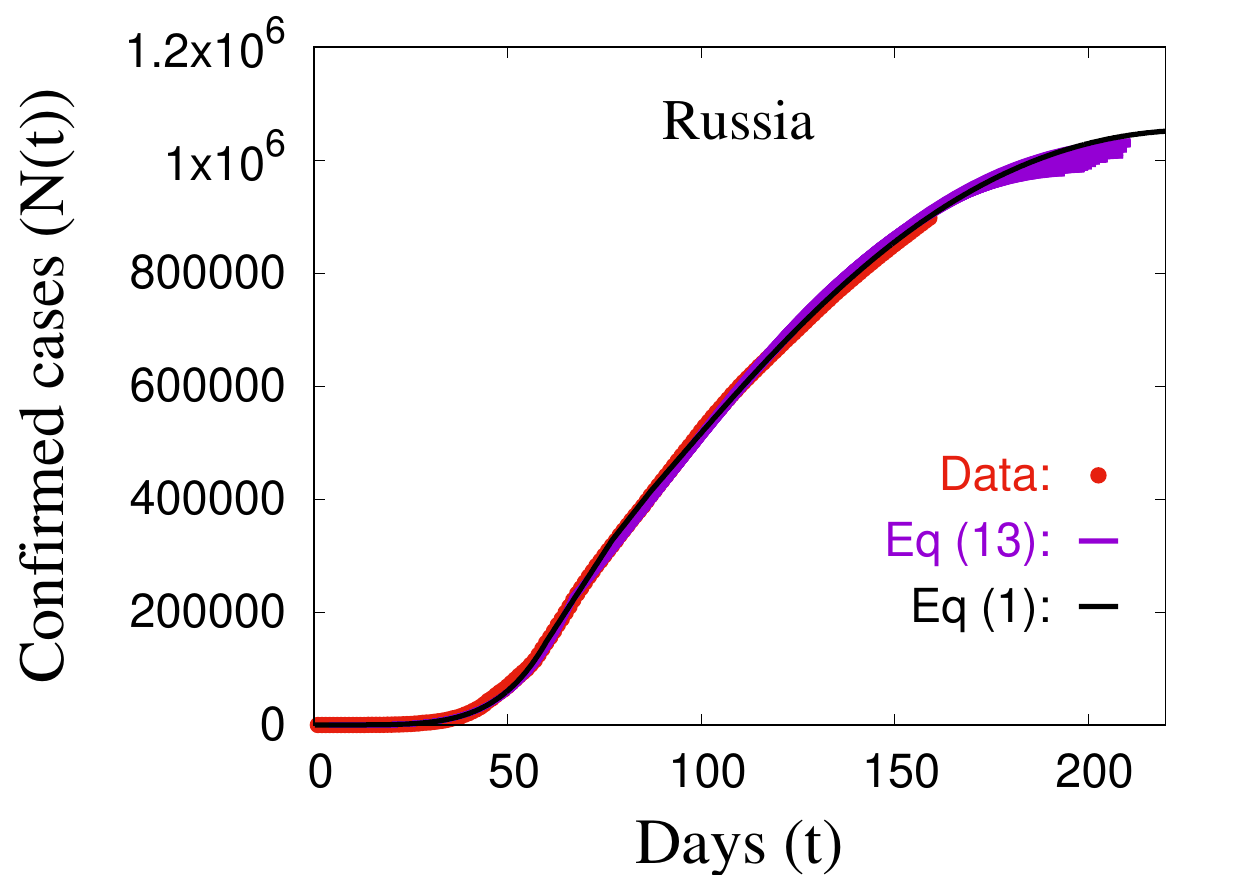}
\includegraphics[scale=0.63]{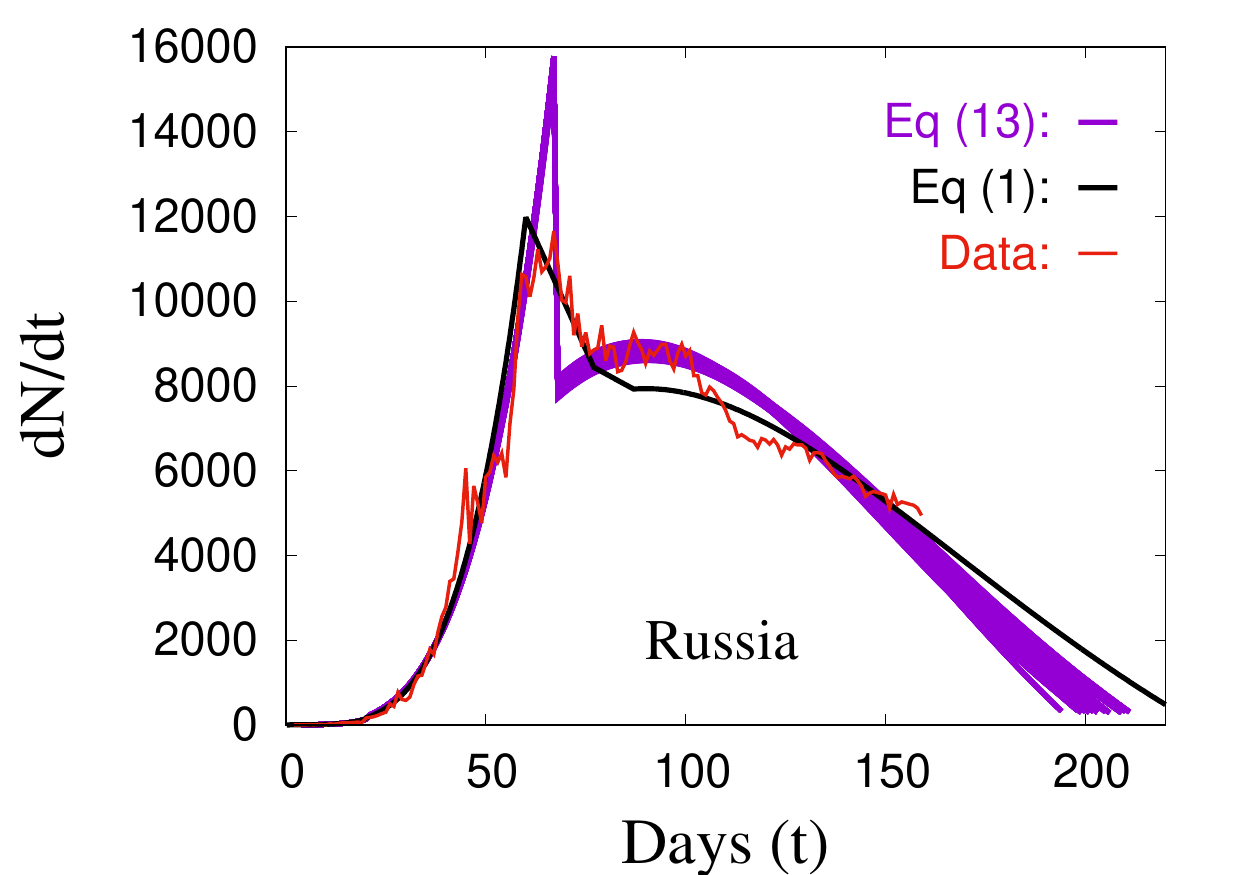}
\caption{\label{fig:russia_plot} The time evolution trajectory of Covid-19 infected population
  for Russia is shown. The red filled circles are actual data from
  Ref. \cite{covid_wiki_rus}. Data up to August 10 are used to obtain
  parameters for Eqs. \eqref{eqn:model_eq} and (13), and projections
  for future evolution are shown by black and blue lines with allowed
  values in the parameter space.}  \eef{}

\subsubsection{Brazil}
The second country whose data we study, and where Covid-19 infection is increasing rapidly, is Brazil.  It has a large population and high population density in a number of big cities. It is natural to expect that without
the strict regulations on human to human contact and proper mechanisms to
tackle a pandemic, a contagious disease like Covid-19 will spread rapidly to a high number,
and both the rising as well as mitigation periods will be prolonged. In
fact, in a short span of time the total number of infection has reached to a
large number and it is now the second most affected country with total
 infections more than 3 million and more than one hundred thousand deaths (Ref. \cite{covid_wiki_brazil} till August 10). The infection is still very much increasing and one would expect that a large population will further be
infected in the foreseeable future. It is thus quite important to project the probable time evolution of the total number of infection.

\bef[h]
\centering
\includegraphics[scale=0.63]{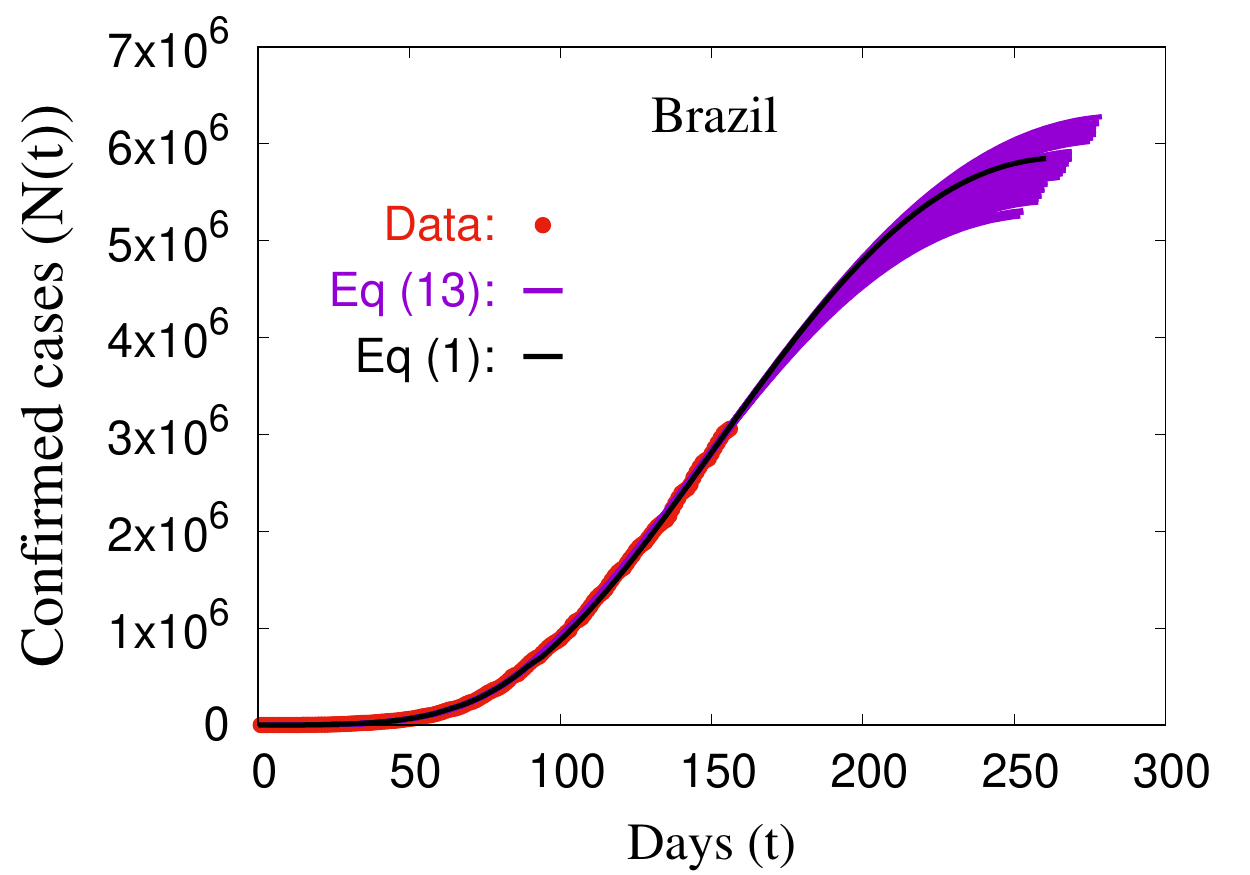}
\includegraphics[scale=0.63]{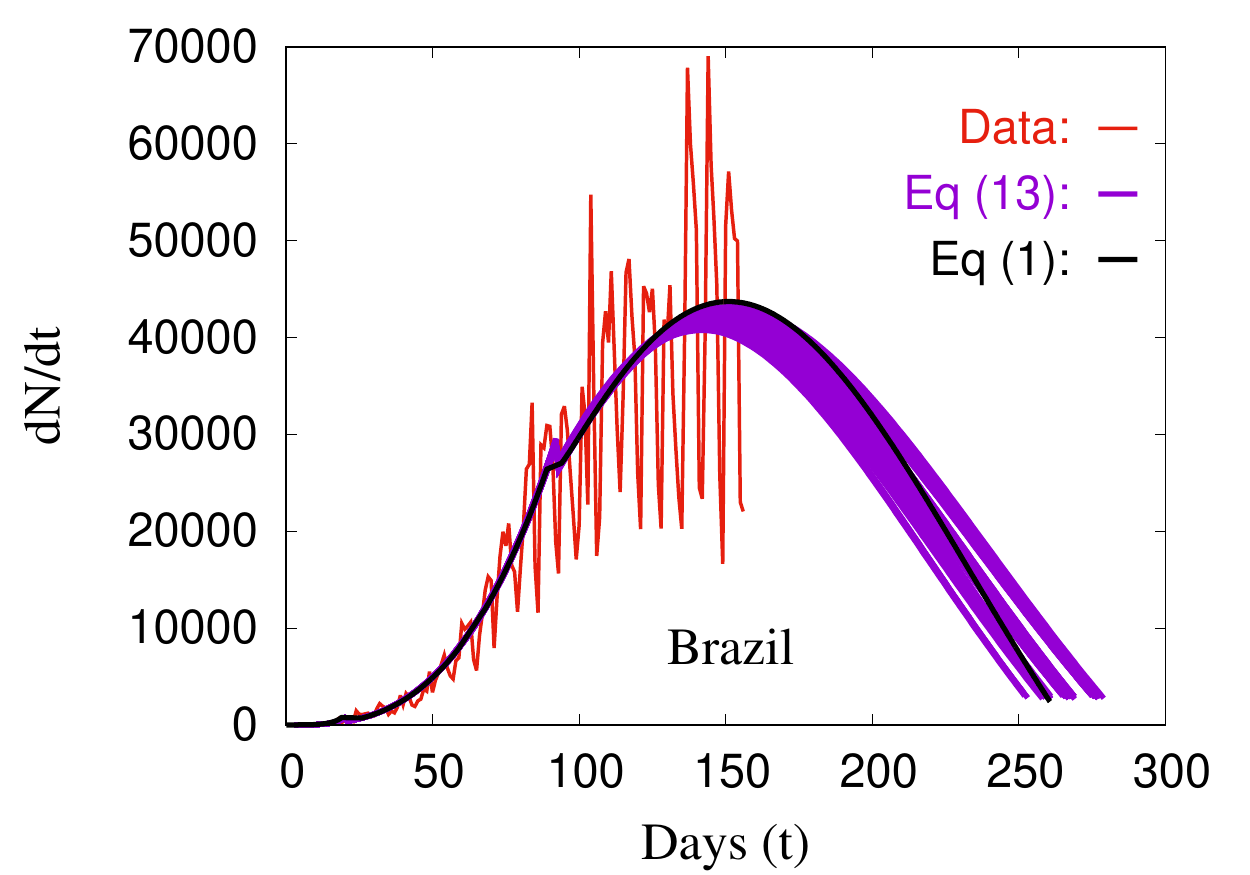}
\caption{\label{fig:brazil_plot} The total number of Covid-19 virus (SARS-CoV-2)-infected
  population (left) and the per day infection (right) are
  shown as a function of time (days) for Brazil. The red filled
  circles are actual data from Ref. \cite{covid_wiki_brazil}. Data up to August  10
  are used to obtain parameters for Eqs. \eqref{eqn:model_eq} and
  (13), and projections for future evolution are shown by black and
  blue lines with allowed values in the parameter space.}
\eef{}

We analyze Brazil's data on the total number of infection till August 10. As in other countries, we find an exponential ($\alpha^{e}_i = 0.3$) followed by a power-law-rise ($\alpha_i = 3.926$) increment at the beginning.
With further analysis our finding is that Brazil is now at the early stage of mitigation at its first wave of infection. The mitigation period has possibly started around day 91 (around the beginning of June)
assuming $t_0$ of infection as March 05, 2020.
Using the available data till August 10 \cite{covid_wiki_brazil}, we first constrain the parameters of
Eqs. \eqref{eqn:model_eq} and Eq. (13). With the
constrained set of parameters, the projection for future trajectories of
the total number of infection, till the possible time for the onset of
saturation, are shown in Fig. \ref{fig:brazil_plot}. The red solid line
is the available data, black solid lines are obtained through
Eq. \eqref{eqn:model_eq} and the blue band represents the projection
through the differential equations in Eq (13). Our results suggest that
the current wave of infection, with the prevailing measures against disease spread, will progress for another 3-4 months
and then it will reach to the beginning of the
saturation period by the end of November with less than 2500 infections per day. The cumulative number of infection by then will reach to 5.5-6.5 million.
However, with more stringent measures to prevent
disease spread or if an effective vaccine becomes available soon,
the mitigation period could be shortened. On the other hand, if the
prevailing conditions of preventive measures become softer and no vaccine is available, then a second wave can start from this mitigation period with more number of infections as is happening now in USA. There is still a large uncertainty in this projection. As mentioned earlier that this projection can be progressively improved with inclusion of more data as the disease progress further. Update results will be posted in the {\it url} mentioned later $^4$.

\subsubsection{India}
The analysis of Covid-19 data on India provides an ideal platform as well as a
challenge to formulate a model for the time evolution trajectory of
the infected population considering the shear numbers for both the
population density (464 per square km \cite{india_density, mumbai_population}) and the total population which is second highest in the world. Moreover, constant
migration of population between cities and countrysides, percentage of young
population, inadequacy in testing facility and several other effective factors (economical, social, religious as well as political) can make the time evolution of the infected population in India very different than that of many other countries.

In India, the first Covid-19 positive case was reported on January 30,
2020. Till March 03, the number of infected patients was limited to 6,
but after that it started to rise as many infected travelers returned
to India from various parts of the world. To reduce the
widespread infection of this virus, the Government of India had
imposed a lockdown starting on March 24, which was subsequently
extended till May 30. After then un-lockdown has started in phases at
different sectors to balance the economic slowdown.  Many of its states and cities have demarcated the containment zones and are periodically imposing further lockdown measures at those places (some time a few days in a week to a more than a week at a stretch) to contain the infection spread. As of August 10, the total number of infected population is more than 2.2 million with a fatality rate of about 1.98\% \cite{covid_wiki_india, covid_india1, covid_india2}.

In Fig. \ref{fig:India_plot}, we plot the cumulative number of confirmed cases \cite{covid_wiki_india, covid_india1} with the
effective starting date ($t_0$) as March 02, 2020. Again we find an exponential  increment followed by a power-law-rise growth as in Eq. \eqref{eq:model_eq1}. We fit the data with the combined equations. It is interesting
to note that $\alpha_{i}^e$ and $\alpha_i$, till the middle of June, was
on the lower side compared to that of 
other countries with a smaller population, which signifies the effect of
 preventive measures at the initial days. However, $\alpha_i$ has started to increase after that (higher percentage of infection is also correlated to the number of tests which has substantially increased during this period). 

\bef[h]
\centering
\includegraphics[scale=0.63]{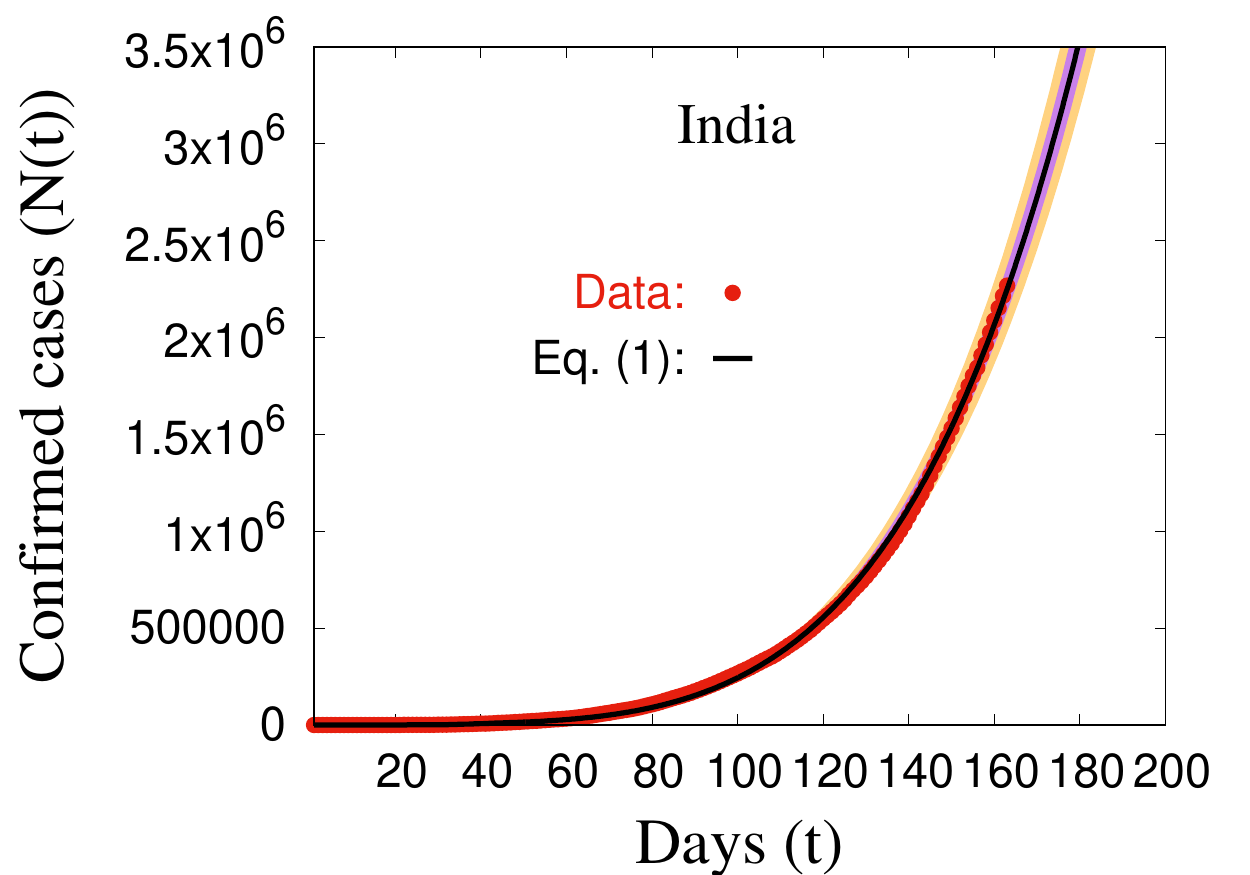}
\includegraphics[scale=0.63]{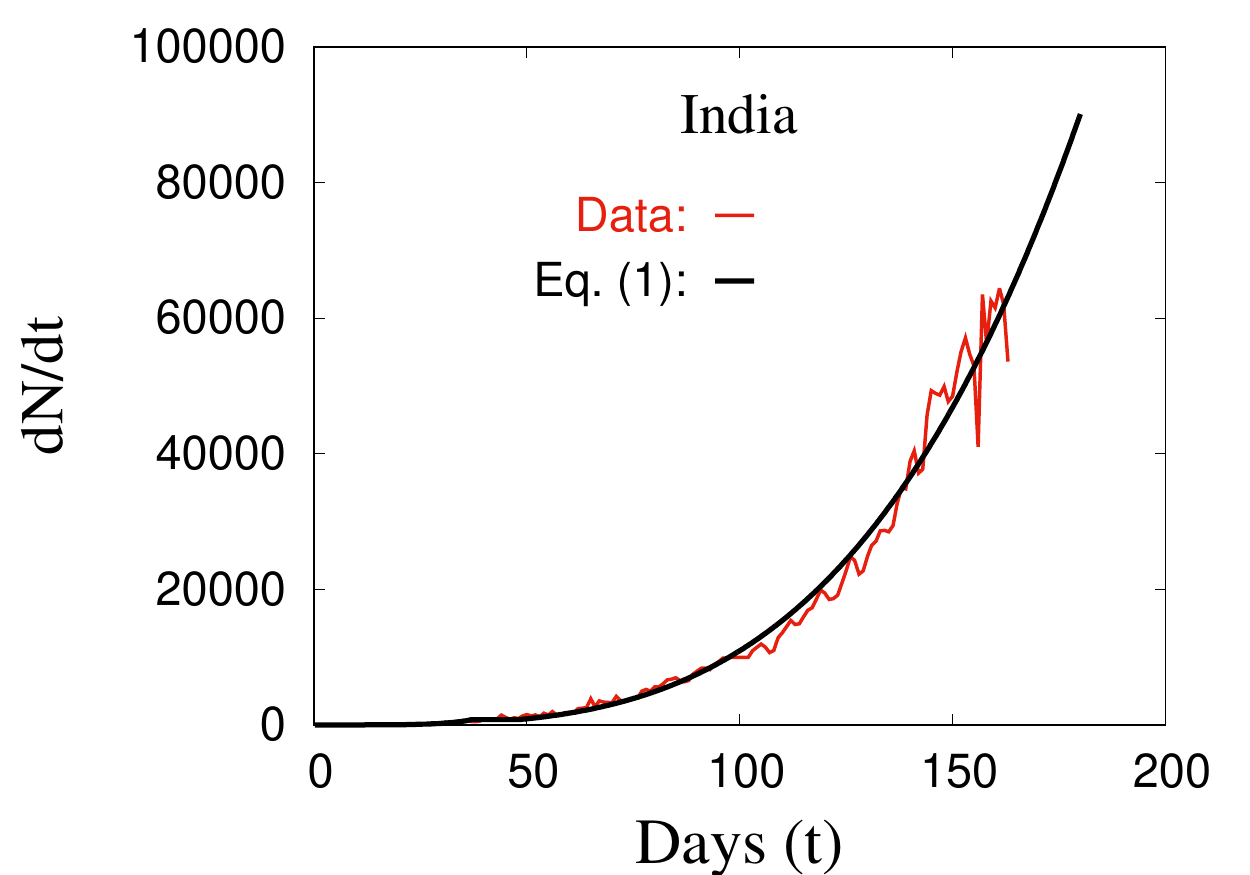}
\caption{\label{fig:India_plot} The time evolution trajectory of the confirmed
  Covid-19 infected population for India is shown. The red filled circles are
  actual data from Ref. \cite{covid_wiki_india}. Data up to July 31
  are employed to obtain parameters for Eq. \eqref{eqn:model_eq} and projections for future evolution are shown by the black line. The two bands show 68\% and 95\% confidence intervals.}
\eef{}

As is evident from the recent data that the disease spread is surging ahead
and there is no signal for slowing down of infection even after passing more than
5 months by when many countries are either in mitigation or saturation
periods in the first wave of infection. This is precisely due to a large number as well as density of
population 
where the strict lockdown measures
cannot be maintained continuously for indefinite periods considering
economical, social and political factors.
Considering these external factors, 
it is expected that the rising, as well as
mitigation periods of a contagious disease, will considerably be longer (and most probably be the longest) compared to those of all other countries.
Given that there is no availability of an effective vaccine in the immediate future and it is not feasible to impose a prolonged lockdown, other possible measures (strict social distancing, contact tracing, mask-wearing, informing all levels of population about the disease transmission etc.) against this contagious disease, which could well be airborne \cite{who}  must be strictly followed to avoid an avalanche of infected population in next few months.

On Indian data, the main observation is that the mitigation period has not started yet. Hence, at this moment, through our model it is not possible to project the time evolution trajectory for a longer period in future.
However, considering the current rate of infection we
project the possible scenario that may happen to the number of total
infection in next three weeks. Fits to the model suggest that, with the current trend, 
the cumulative number of infection will cross 3.5 million at the end of August (Fig. \ref{fig:India_plot}). On a positive note, only very recently the increment coefficient, $\alpha_m$ (of Eq. \eqref{eq:model_eq2}), is found to be reducing, albeit slowly,  with its current value at $4.58$. It will be interesting monitor if $\alpha_m$ continues to reduce further and the infection enters into the mitigation period. We will
continuously monitor the disease spread and will project the time evolution
later more accurately with the availability of more data.

Along with the whole Indian population, we also analyze data for its two
biggest cities, Mumbai and Delhi. Below we elaborate our analysis
for these two cities and subsequently project the most probable trajectory
of infection for the next few months.

\vspace*{0.2in}

\noindent{3.1 {\bf{Mumbai}}}

Mumbai is the largest city in India and also is one of the most
densely populated cities in the world ($>$ 25000 per sq km with a total population more than 20 million \cite{mumbai_population}). Moreover, 
the city has clusters of densely populated areas and a large number of migration happens
very frequently.  Naturally it is intriguing to analyze the time
evolution trajectory of Covid-19 infection of such a city. The first
reported case in the city of Mumbai was on March 11, 2020 and subsequently
the infected number has increased substantially, and as of August 10, 2020,
the total infected number is more than 124 thousand  with a fatality rate of 5.55\% \cite{mumbai_data}. As in other parts of India, Mumbai was also under
lockdown effectively from March 20, and gradual un-lockdown has
started since the beginning of June while placing lockdown measures at containment zones.  

In Fig. \ref{fig:mumbai_plot}, we plot the cumulative number of confirmed Covid-19-infected patients with the effective starting date ($t_0$) as March 14,
2020 \cite{mumbai_data}. This data also shows an exponential increase followed by a power-rise growth in the initial days as in Eq. \eqref{eq:model_eq1}. We fit the data with the combined form and find
$\alpha_{i}^e = 0.176$ and $\alpha_i = 3.436$ which are again smaller compared to many other places. Perhaps this is due to strict lockdown measures at the initial days.
The transition from the
rising to the mitigation period is found to be at time $t_m = 81$,
which corresponds to around June 03, 2020. Due to its huge carrying population
density as well as the total population, it is expected that the mitigation
period will be much longer. With existing data we track the time evolution
with Eq. \eqref{eqn:model_eq} and the fitted result is shown with the
black solid line in Fig. \ref{fig:mumbai_plot}.  We also use Eq. (13)
to find the time evolution trajectory and the solutions with the
restricted parameter space is shown by
the blue band. Results from this work, at this time, project that with the existing measures of
restriction against the infection spread, we expect to see the onset of
saturation period, with about 250 infection per day, around time $t_{s_i} = 195-210$ which corresponds to the end September to early October, with a cumulative number of infection in between 145-160 thousand. However, this will very much depend on the continuation of the prevailing preventive measures against the disease spread, given the large unaffected population with large density in Mumbai.
\bef[h]
\centering
\includegraphics[scale=0.63]{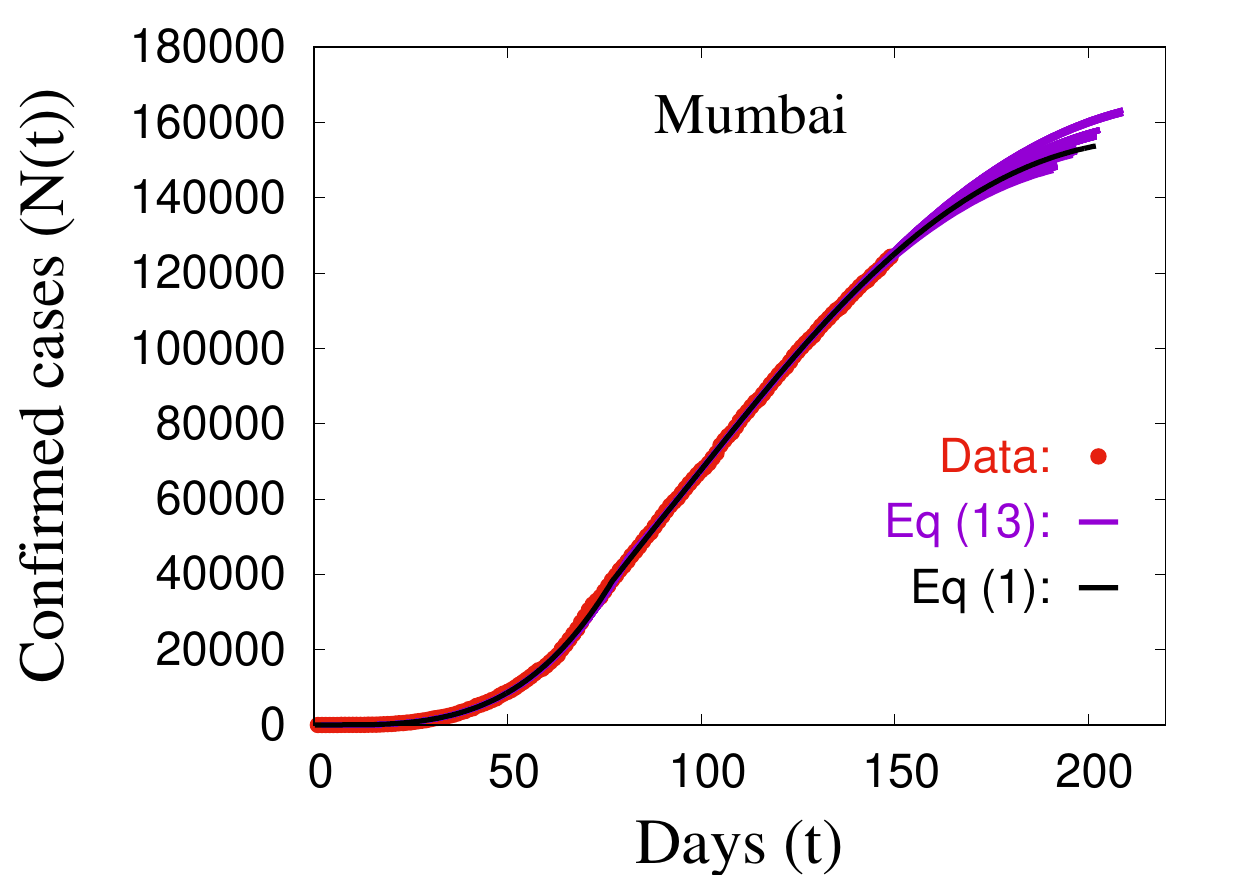}
\includegraphics[scale=0.63]{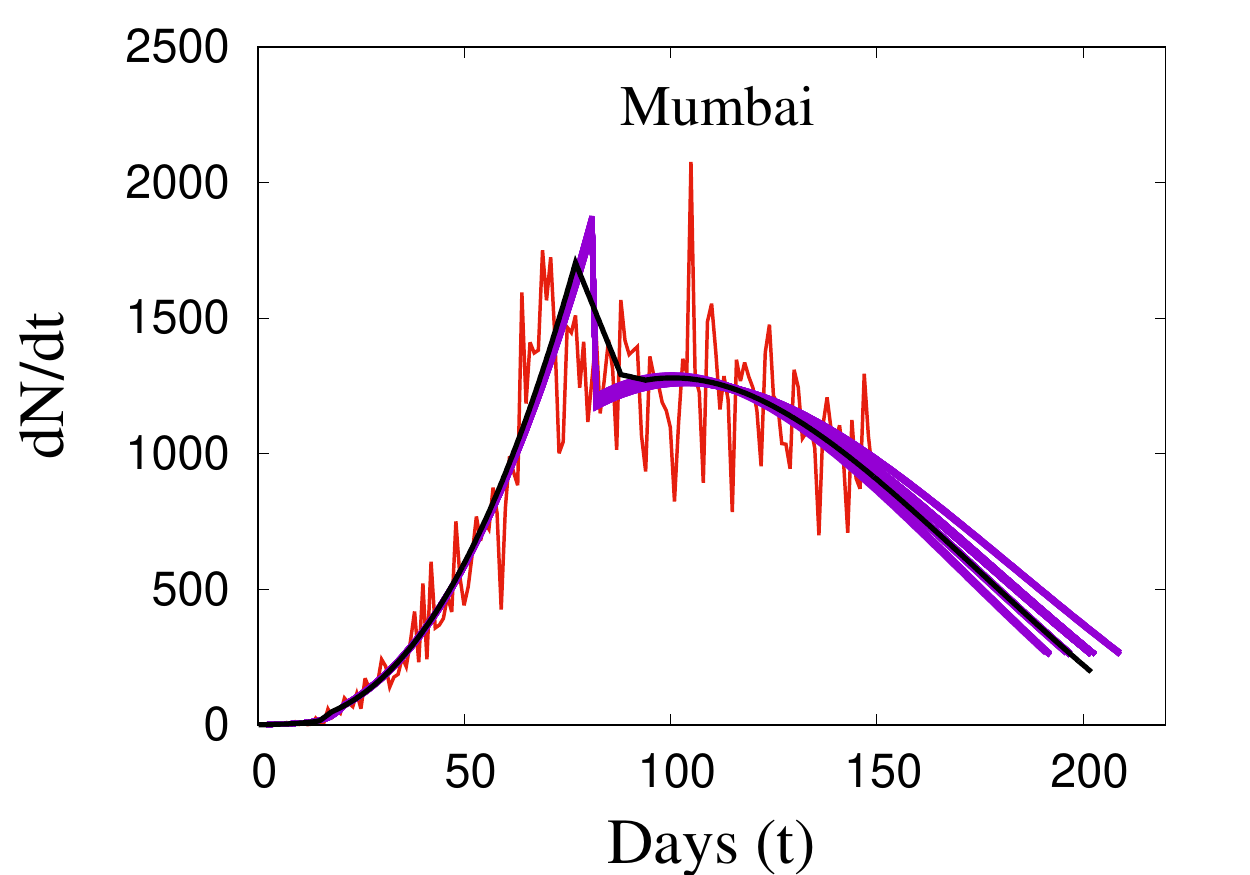}
\caption{\label{fig:mumbai_plot}The time evolution trajectories of the confirmed
  Covid-19 infected population (cumulative and differential) for Mumbai are shown. The red filled circles are
  actual data from Ref. \cite{mumbai_data}. Data up to July 31
  are used to extract parameters for Eqs. \eqref{eqn:model_eq} and
  (13), and projections for future trajectories are shown by black and
  blue lines with allowed values in the parameter space.}
\eef{}

\bef[h]
\centering
\includegraphics[scale=0.63]{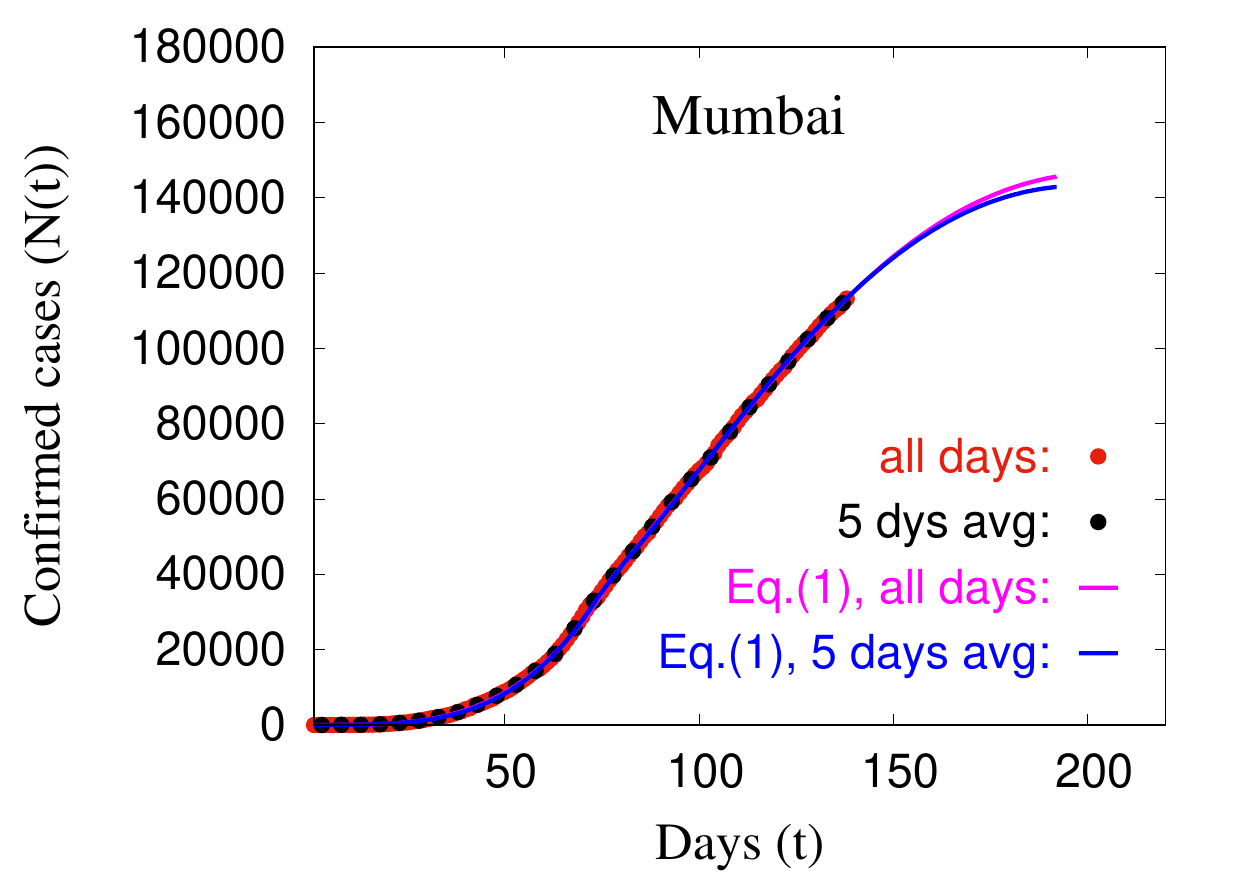}
\includegraphics[scale=0.63]{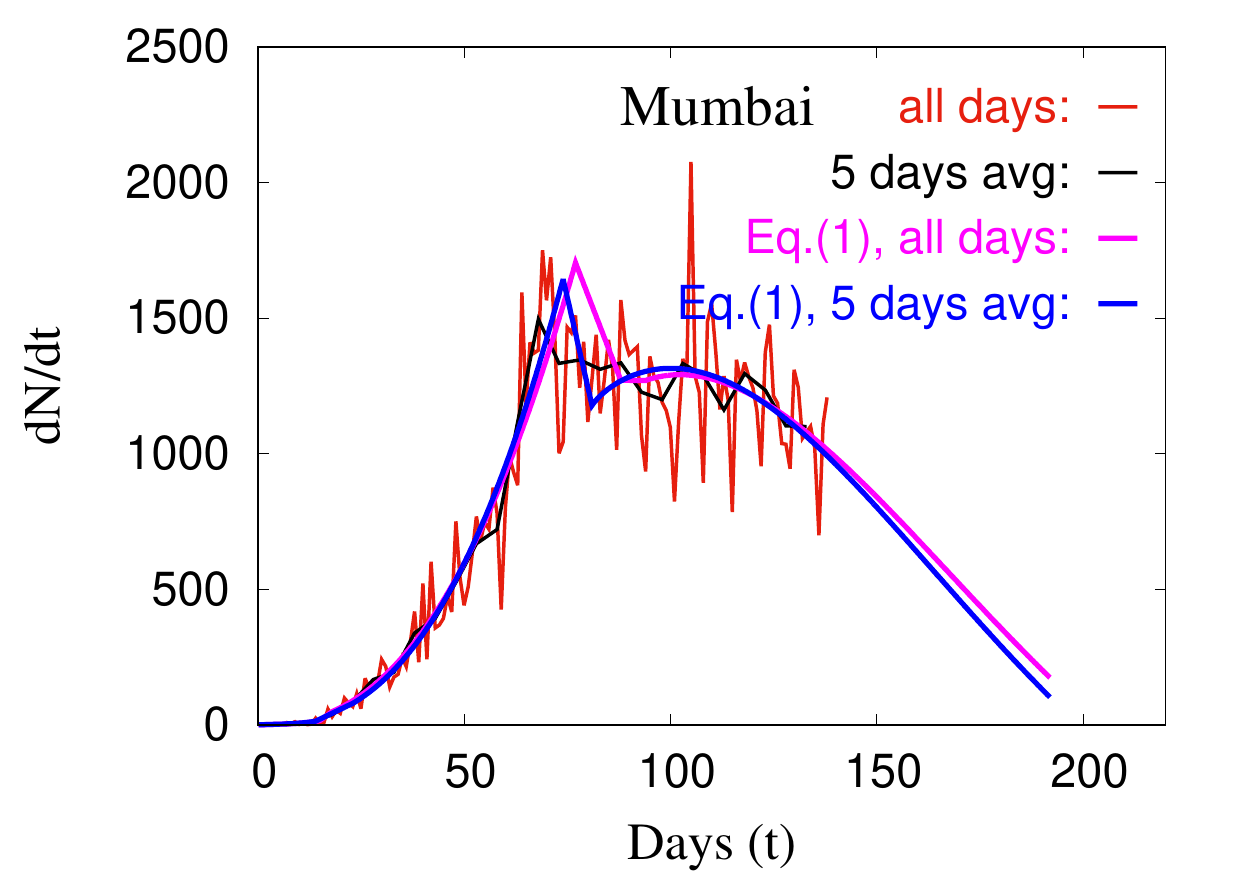}
\caption{\label{fig:mumbai_plot5}The time evolution trajectories of the confirmed
  Covid-19 infected population (cumulative and differential) for Mumbai are shown both for all days data (red point/line) and 5 days average data (black point/line). The results obtained using Eq. \eqref{eqn:model_eq} are shown by magenta and blue lines respectively.}
\eef{}

Similar to our study on England's data we also perform a study on 5 days average data of Mumbai. The results are shown in Fig. \ref{fig:mumbai_plot5}. As one can see that there is effectively no difference between two results (magenta and blue lines), except a very small difference at the saturation period. This shows that our method for obtaining parameters for  Eq. \eqref{eqn:model_eq} is robust.

\vspace*{0.2in}

\noindent{3.2 {\bf{Delhi}}}

Delhi is another highly populated city in the world. Like Mumbai, Delhi
has also a number of densely populated areas and everyday migration
happens to and from Delhi. It is thus also interesting to analyze the
Covid-19 data on Delhi and compared it with that of Mumbai. Below we
elaborate our analysis on Delhi.

The first reported case on Covid-19 infection in the city of Delhi was
on March 2. Subsequently the infected number has increased, and as
of August 10, the total infected number as reported is
more than 146 thousand with a mortality rate of 2.82\% \cite{delhi_data}, which is considerably lower than the same rate of Mumbai.
As in other parts of India, Delhi
was also under lockdown from March 24 and gradual
un-lockdown has initiated from the beginning of June.
However, unlike Mumbai, there was a sudden
jump in the number of infection was observed around fourth week of June, perhaps
due to ease of lockdown and not maintaining other preventive measures.
Since then further measures against the disease spread have been taken and the number of per day infected population has slowed down. In Fig. \ref{fig:delhi_plot}, we plot the cumulative number of infected population \cite{delhi_data} with the effective starting date
($t_0$) as March 2, 2020. This data also shows an exponential
followed by a power-law-rise form as in Eq. \eqref{eqn:model_eq}.  We
fit the data with the combined form and find $\alpha_{i}^e = 0.186$
and $\alpha_i = 4.82$ which are larger than the corresponding exponents
for Mumbai.  The transition from the rising to the mitigation period
is found to be at time $t_m = 103$, which again is larger than Mumbai.
This transition time corresponds to June 25, 2020. Due to its huge carrying
population density as well as the total number of population, we again expect
that the mitigation period will be longer. With the existing data, we track
the time evolution with Eq. \eqref{eqn:model_eq} and the result is
shown with the black solid line in Fig. \ref{fig:delhi_plot}. We also
use Eq (13) and solutions with
the allowed parameters are shown
by the blue band. Our projection on further Covid-19 infection spread
for Delhi is that with the sustained current measures taken against
disease spread, we expect to see the onset of saturation time (about 250 infections per day) around
$t_{s_i} = 180-190$ which corresponds to around the end of September.
However, as mentioned
earlier this will very much depend on whether existing preventive measures
can be maintained over the period. Our model suggests that the infection can be contained substantially with the current measures against disease spread, and by the end of September to the beginning of October it can be reduced to 250 per day with a cumulative infected population within 160-180 thousand.
However, adequate precaution must be taken even
after that considering the large unaffected population and regular migration, particularly as the city is trying to come back to its regular life.
\bef[h]
\centering
\includegraphics[scale=0.63]{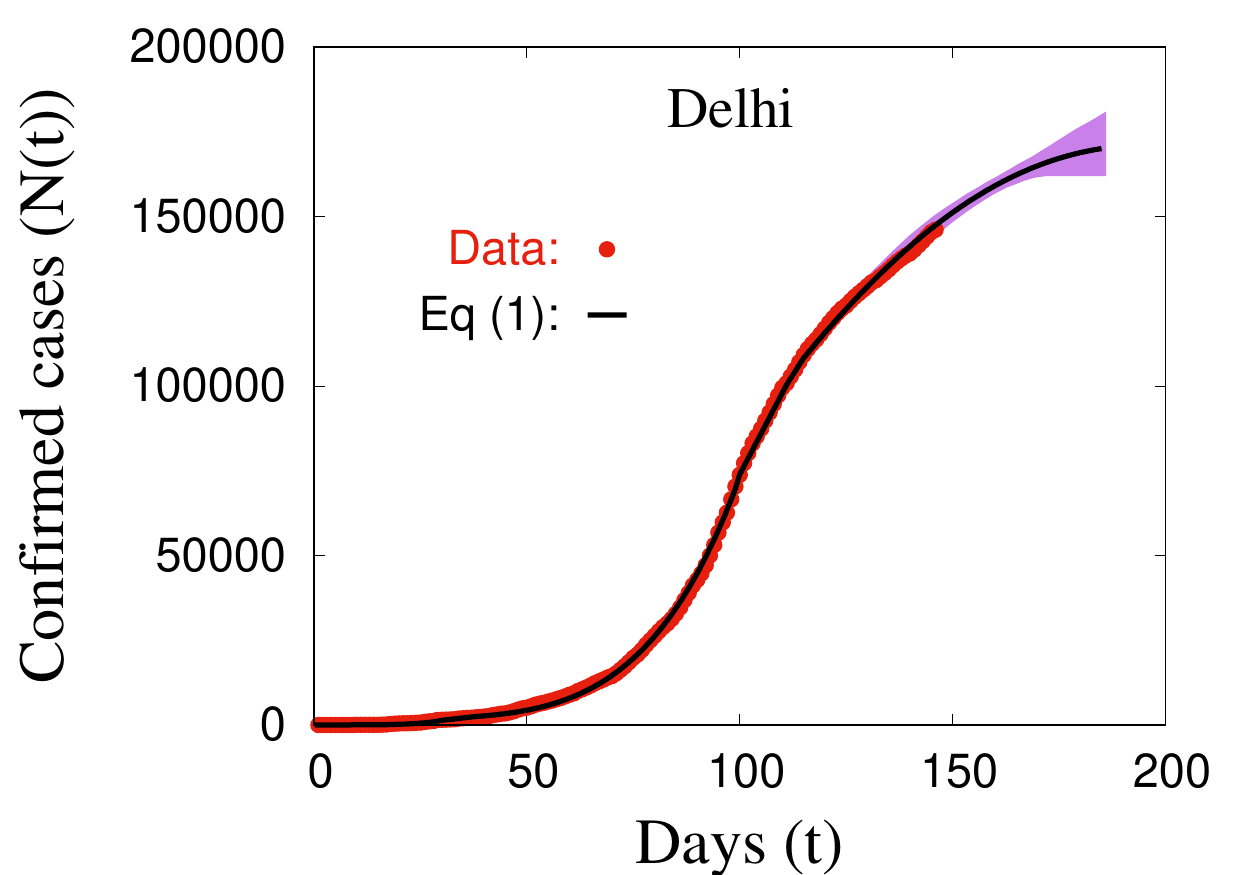}
\includegraphics[scale=0.63]{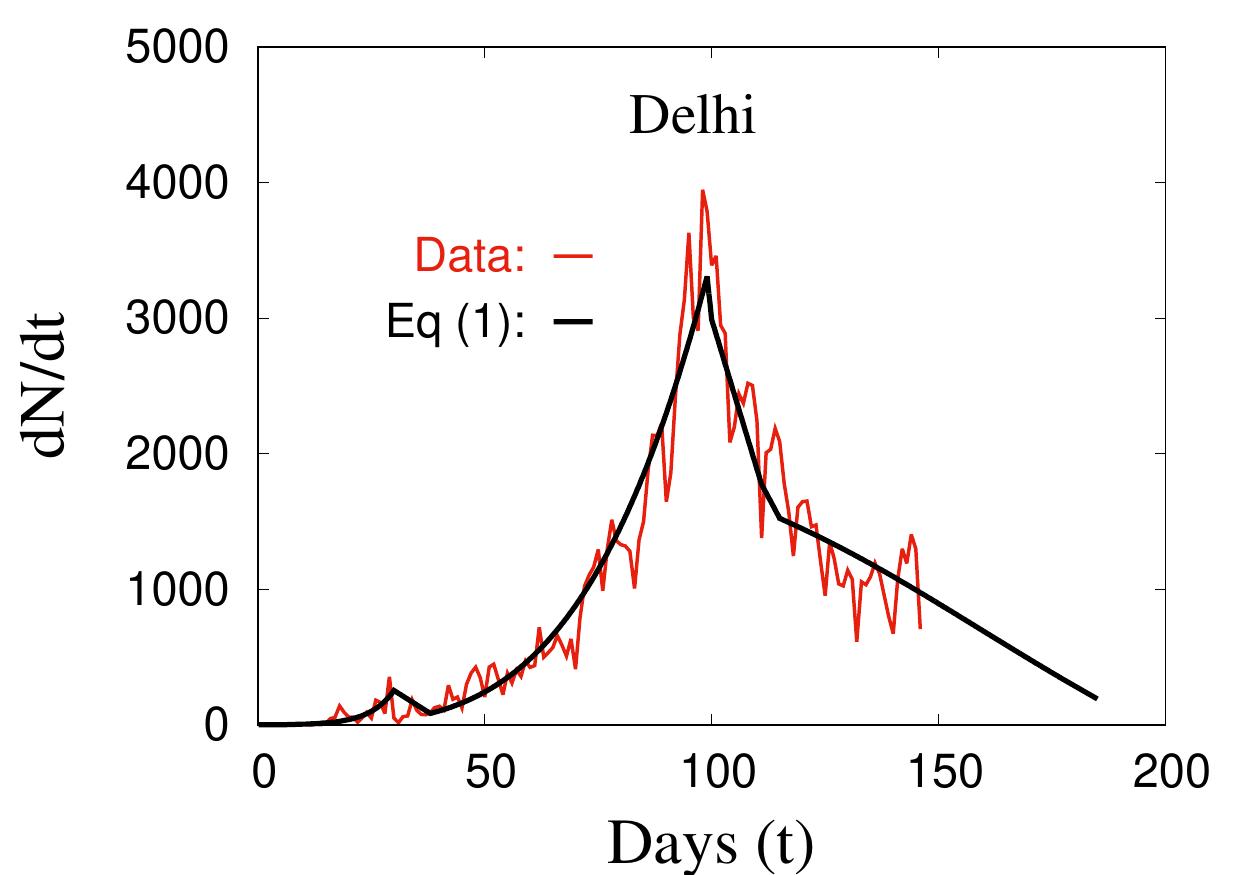}
\caption{\label{fig:delhi_plot}Time evolution trajectories of the
  Covid-19 infected population (cumulative and differential) for Delhi are shown. The red filled circles are
  actual data from Ref. \cite{delhi_data}. Data up to July 26
  are used to obtain parameters for Eqs. \eqref{eqn:model_eq} and
  (13), and projections for future evolution are shown by black and
  blue lines with allowed values in the parameter space. The blue band shows 95\% confidence intervals.}
\eef{}

\subsubsection{USA}
At the end of this work, we analyze the data on Covid-19 infected population of
USA.  It is also a populous country with a large number of big cities with 
large population. We have already analyzed the data on New York City
in the subsections A and B above. To be noted that the infected
population of New York City has already reached the saturation period (stage 3 of Eq. \eqref{eqn:model_eq}, see Figs. \ref{fig:time_traj_valid2}
and \ref{fig:d_time_traj_valid2}). However, unlike New York City, currently
in many places in
USA, the number of infected population are increasing quite rapidly.
In fact, the data on USA is quite interesting and has a unique feature 
which shows a big jump in the number of infected population from a
situation where there was a clear pattern of reduction in the number of
infected cases. Unlike many European countries and New York City, the trajectory of the 
number of infected population in USA does not move towards the 
saturation period. Rather it looks like a new wave of infection has
started with another initial power-law-rise increment and subsequent mitigation.
For a contagious disease this must be related to the change in measures against the disease spread by a large population.


\bef[h]
\centering
\includegraphics[scale=0.63]{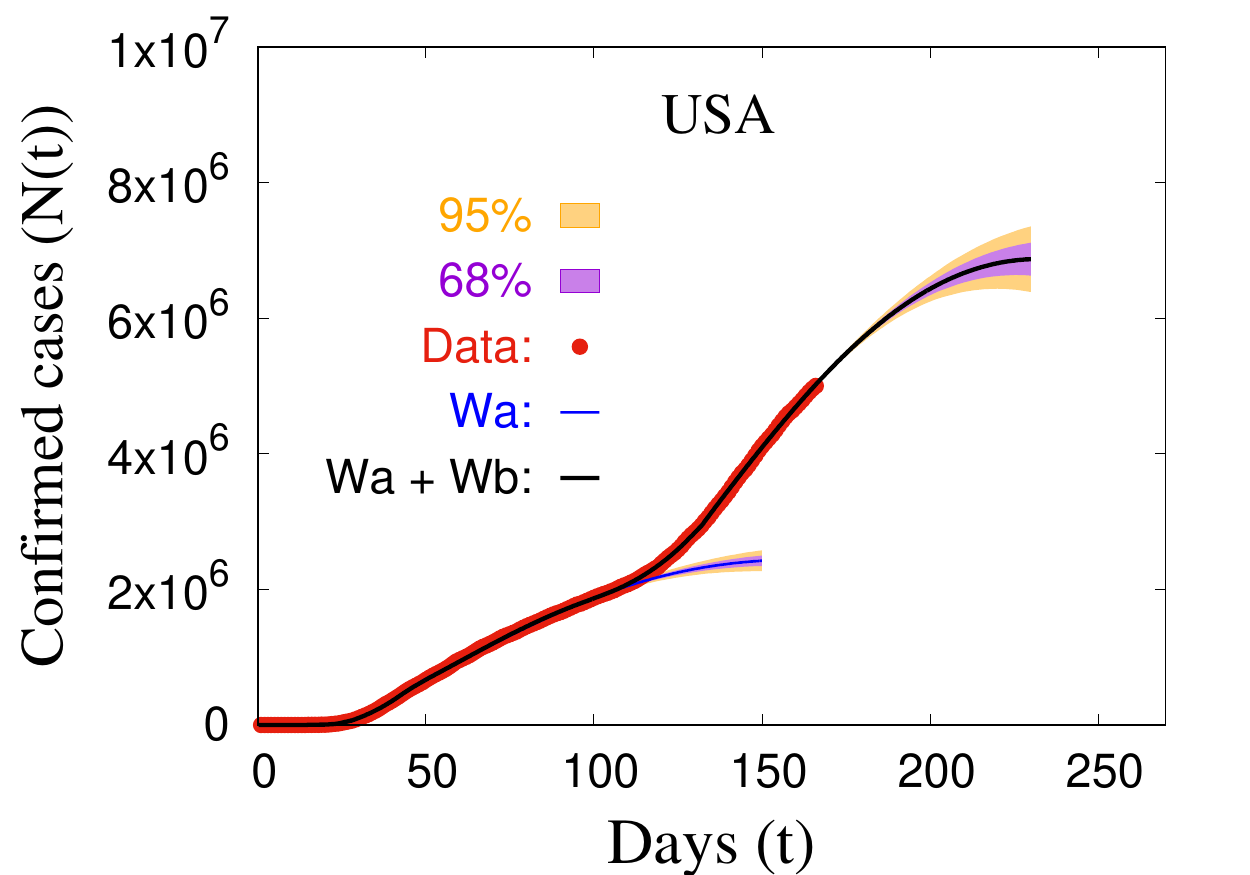}
\includegraphics[scale=0.65]{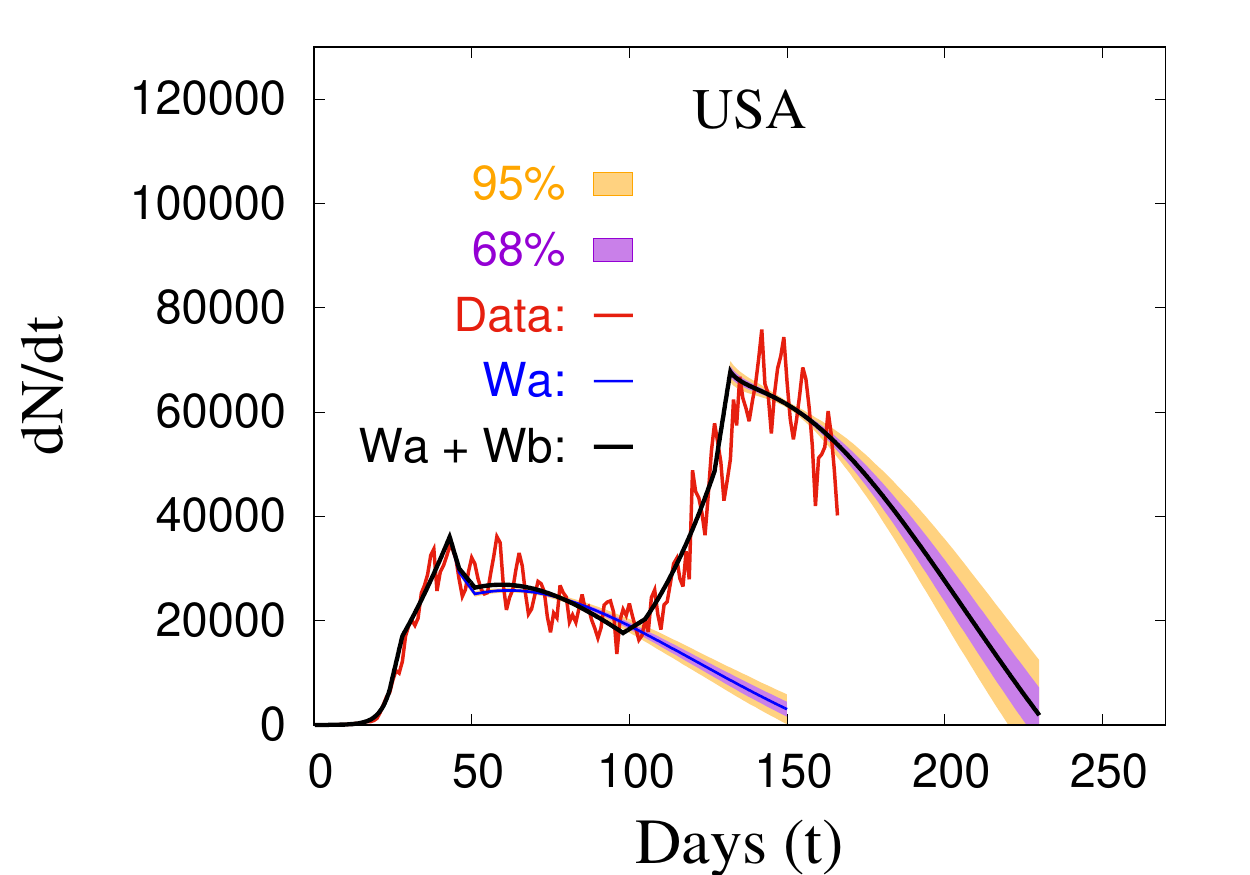}
\caption{\label{fig:USA_plot} The time evolution trajectories of Covid-19 infected population (cumulative and per day) for USA are shown. The red filled circles are
  actual data from Ref. \cite{covid_wiki_usa}. Data up to July 31,
  including both $Wa$ and $Wb$ (see text for details),  are utilized to obtain parameters
  for Eq. \eqref{eqn:model_eq}. Results from our model are shown by the black line.
  The trajectory without red points shows the projection for the total number of infected population in next few months.
  The blue line represents the result for only $Wa$ showing the possible
  trajectory that could have happened had the second increment in $Wb$ not taken place. The two bands represent 68\% and 95\% confidence intervals (see Appendix for details).}
\eef{}

In Fig. \ref{fig:USA_plot}, we plot the cumulative number on infected population \cite{covid_wiki_usa, who} with the
effective starting date ($t_0$) as February 27, 2020. Similar to most of
the previous countries, this data also shows an exponential increment
followed by a power-law-rise, as in Eq. \eqref{eq:model_eq1}.
A simultaneous fit with both forms yields
$\alpha_{i}^e = 0.3075$ and $\alpha_i = 2.744$. This suggests, while
the early rise is much faster than many other countries the power-law rise
was much slower. One needs to remember that a substantial contribution of
this part of the data came from New York City for which
$\alpha_{i}^e = 0.5810$ and $\alpha_i = 1.619$. One can argue from
these two exponents that the lockdown effect of New York City
may have played a key role against the fast rise of US
infection at the earlier months of the infection. As mentioned earlier,
one can observe from the figure that the data after crossing about 100
days (with $t_0$ at Feb 27, 2020) shows
a unique feature: instead of
slowing down with the expected trajectory, as was observed in most other countries,
the number of infection surges once more with a power-law-rise form.
Indeed data after that can be fitted easily again with a power-law-rise form (Eq. \eqref{eq:model_eq1}).

Because of the above-mentioned unique feature of the trajectory
we use a different analysis strategy for this data. First, we try
to find out a maximum time till which the total number of infection
can be analyzed like any other affected regions incorporating an early
increment of exponential-cum-power-law-rise followed by a mitigation
period. We call that part of data as $Wa$. The maximum of $Wa$ is
chosen dynamically with the usual condition of minimum $\chi^2(Wa)
(= \chi_{i}^2(Wa)+\chi_{m}^2(Wa))$.  Then we introduce another
exponential-cum-power-law-rise followed by a mitigation period,
considering as if there is a second wave. We call this part of data as
$Wb$. We then dynamically choose the time of transition from $Wa$ to
$Wb$ so that the total $\chi^2 = \chi^2(Wa) + \chi^2(Wb)$ becomes minimum
within the acceptable limit. We find the minimum $\chi^2$ fit leads to
$t_m(Wa) = 45$, the transition time from $Wa$ to $Wb$ at around
time $t_{tr} = 103$, and $t_m(Wb) = 129-140$.
Interestingly in this way we could track the
time evolution of the total number of infection from the initial period
to date very well. Result using Eqs. \eqref{eq:model_eq1}
and \eqref{eq:model_eq2} is shown in Fig. \ref{fig:USA_plot} with the
black solid line, whereas the red solid line represents the actual data
\cite{covid_wiki_usa, who}.
To be noted that the
power-law-rise exponent $\alpha_i$ in $Wb$ (5.594) is much bigger
than its corresponding value in $Wa$ (2.744). In fact, it is the biggest
among all other power-law-rise exponents for any affected
regions that we study. This is a clear signal that the infection had spread
very rapidly in the early period of $Wb$, and unless adequate measures are
imposed against the disease spread a substantial increase in the
number of infection is expected in foreseeable future. However, on a positive note, it is worthwhile to mention that our model also allows a possibility that another mitigation period has started, albeit slow, at around $t= 129-140$, $i.e.$, around the second week of July. With this slow rate of mitigation, and in no change in existing conditions, this disease can rise another 2.5-3 months infecting about 6.5-7 million people (shown by the black line in Fig. \ref{fig:USA_plot}), before reaching to the saturation period with less than about 3000 infections per day by the end of October to the beginning of November. Thereafter the onset of saturation period can possibly be achieved by maintaining the same restrictions against the infection spread. We project this as the ideal case scenario  with continuation of the existing restrictions and current trend of reduction. Although this is an ideal situation and in reality there will be deviation in restrictions and so in this projection, still this can be taken as a possible scenario.  However, if the current value of the percentage of per day increment sustains and does not reduce further in the next 2 weeks, this situation can change.  We will discuss next about such a possible situation with a set of speculated data.

Before that, here one point is worthwhile to mention that had the infection trajectory follow
the similar pattern as that of most other countries ({\it i.e.}, without
$Wb$, as shown by the blue line in Fig. \ref{fig:USA_plot}), around 150 days ({\it i.e.}, by now) the onset of saturation period would had started with a
much reduced number of infections (about 3000 per day).
This is strongly indicative that by
maintaining the measures against the infection spread for longer time,
as it was in the early days of $Wa$, perhaps the disease spread could
have been substantially contained by now and perhaps the
onset of saturation period could have happened as well!

\vspace*{0.2in}

\noindent{4.1 {\bf{A speculative but possible scenario:}}}

At the end we would discuss about a speculative scenario on the disease spread in next few months based on a speculative set of data for next 2 weeks. We
make following two assumptions: i) in next two weeks the number of per
day infection does not increase more than 1.25\% of previous day's number (current rate is about 1.1-1.2\%) and ii) it also does not decrease less than 0.9\%. With these assumptions
we generate a set of speculative data for next 2 weeks.
We then use this
data, and use our model, with $t_m(Wb) = 129$, to get the future trajectory.
In Fig. \ref{fig:usa_plot1} we show the corresponding results. The speculative data is shown with the blue line and the
projection is shown by the green line. The original data till August 10 is
shown with the red line, and projection with the same $t_m$ is
represented by the black line. This speculative scenario points that
if the percentage of infection, as mentioned above, does not decrease further in next 2 weeks and sustains in between 0.9-1.25\%, then within this model there is high possibility (95\% confidence intervals) that the number of infection by the end of October will be about 7-7.5 million with 10000-15000 infections per day.

\bef[h]
\centering
\includegraphics[scale=0.63]{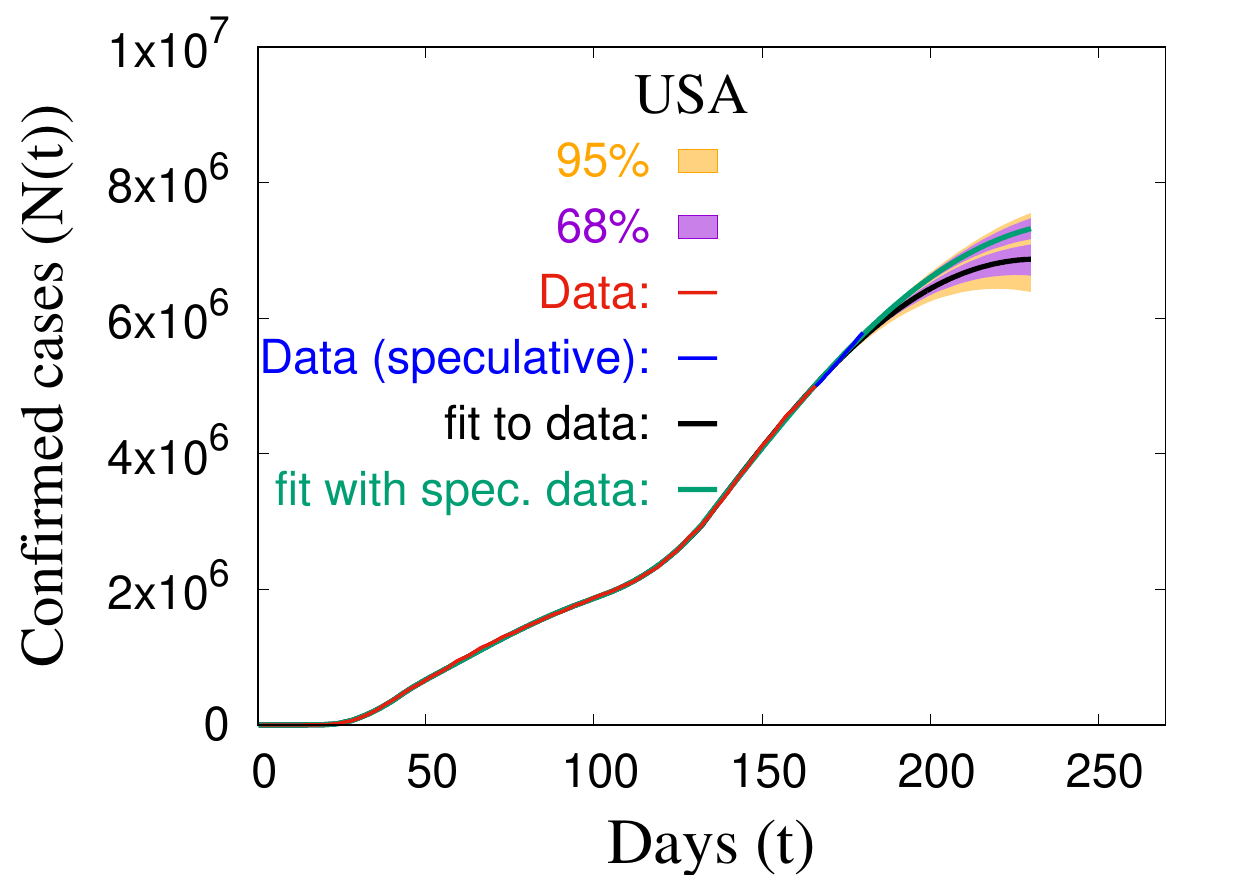}
\includegraphics[scale=0.65]{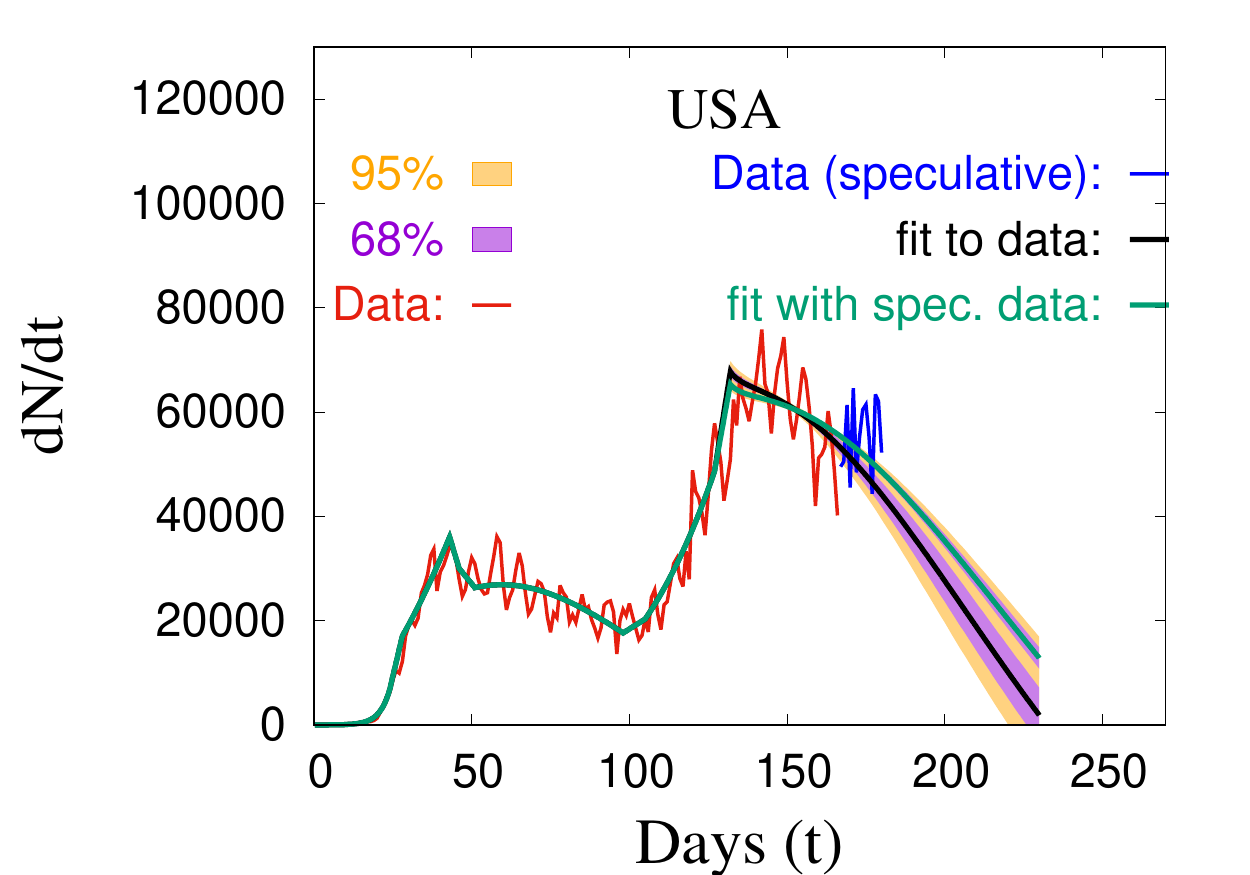}
\caption{\label{fig:usa_plot1} A speculative scenario considering next 2 weeks data as shown by blue lines. The red filled circles are
  actual data from Ref. \cite{covid_wiki_usa}. Results from our model are shown by the black line. The green line shows a possible scenario as projected by our model considering the speculated data (blue line) along with the actual data (red points). The two bands represent 68\% and 95\% confidence intervals. }
\eef{}

Summary of all projections described above is given
in Table \ref{tab:summary}.

\begingroup
\begin{table}[ht]
	\centering
	\begin{tabular}{|c|c|c|c|}
          \hline
        {Countries} & \multicolumn{3}{c|} {Projection}\\
\cline{2-4}
 and Cities   &$t_{s_i}$&N &${dN\over dt}$\\
 \hline
 Russia & End Sept $-$ early Oct & $0.95-1.1\times 10^6$ & $< 500$\\
Brazil & Mid Nov $-$ mid Dec
&$5.5-6.5 \times 10^6$& $< 2500$\\
India &--   & $ \sim\, 3.5 \times 10^6$ (with current rate)   & $ > 75000$\\
& & End August & \\
 Mumbai &End Sept $-$ Early Oct &$145-160 \times 10^3$& $< 250$\\
 Delhi &End Sept $-$ Early Oct &$160-180 \times 10^3$& $< 250$\\
 USA & End Oct (with the current trend)  &  $\sim\, 6.5-7 \times 10^6$   & $ <  5000$\\
 &  Early Nov  (increment within 0.9\% -1.25\%)  &   $ < \, 7-8.5 \times 10^6$   &  $\sim\, 10000-12000$\\
    \hline
        \end{tabular}
        \vspace*{0.1in}
	\caption{\label{tab:summary}{Current projections for several regions where Covid-19 disease is spreading. For USA two possible situations are shown (see text for details). These projections are made within 95\% confidence intervals of the proposed model assuming there will be no change in the existing conditions at these places.}}
\end{table} 
\endgroup


\section{Summary and Conclusions}~\label{sec:conclusions}
In this work, using a data-driven statistical model, we present an
analysis of the time-evolution of the Covid-19 infected human population for
a number of countries and cities. By analyzing the time-dependent data
of the infected population, we
find that there is a common pattern of infection over a long
time-period irrespective of different prevailing conditions at the
affected regions. The total infection time can broadly be divided into
three periods: {\it i}) initial increment with an exponential rate
followed by a power-law rise or simply a power-law rise from the
beginning, {\it ii}) a mitigation period and {\it iii}) a saturation
period. Of course, the transitions from one period to the next one is
a smooth cross-over without any discontinuity.  We propose a
mathematical model for the time evolution of the number of infected
population and show that it can very well track the number
of infected population at various affected regions, from the initial
days to saturation period. Based on the proposed mathematical
trajectory we then formulate a set of differential equations which
 can equally well describe the same time evolution. The parameters of the
model (for both the mathematical trajectory (Eq. \eqref{eqn:model_eq})
and differential equations (Eq.(13)) are extracted by dynamically
performing a simultaneous $\chi^2$-minimization against the available
data.  Our proposed model is
then employed to project the probable number of infected population in
future till the saturation period is reached.
To demonstrate the predictive ability of the model we use a
subset of the data and show that using the proposed model we can
project the future time evolution of the number of total infected
population from a week to a month and even more. Projection from this model
can be progressively
improved by including more and more data. However, it is
based on the assumption that the existing measures against the disease
spread remain to be the same throughout this period and the affected region is isolated (i.e., new infections are not coming from outside). After
validation and showing the predictive ability of our model, we analyze
the similar data on a number of other countries where Covid-19
infection is currently increasing rapidly. We observe that for Russia
the per day infection has already peaked around May 10, and it is
heading towards the onset of saturation period. With the continuation
of the existing measures against the disease spread, we expect to see
the onset of saturation period by the end of September to the beginning of 
 October with less than 500 infections per day and the cumulative
number of infected population around $0.95-1.1$ million.  For Brazil, our
finding is that it is now crossing the peak of disease spread (that is
per day maximum number of infection). However, due to large unaffected
population and the unavailability of an effective vaccine, its
mitigation period will be prolonged. With the existing measures
against the disease spread we expect to see the onset of the saturation period
by the middle of November to the middle of December, with $5.5-6.5$ million cumulative infection and less
than 2500 infection per day.
Our analysis on India shows that it is still in the power-law
rise period even after passing more than 5 months of the disease
spread. With the current trend of infection we expect that there would
be more than 3 million infection by the end of August.  We also expect
that even with the continuation of the existing measures against the
diseases spread, the mitigation period will be quite long considering
its huge unaffected population. Hence it is even more important to
maintain the existing measures against the disease spread, at least
for the next few months, given that no effective vaccine would be
available before that.  We will continue to monitor the disease
propagation and progressively improve our projection on the future
trajectory with the availability of more data{\footnote{The updated
    figures on projection will be available at the following url:
    \url{https://theory.tifr.res.in/~nilmani/covid19}.}}. We also
analyze data on two of its biggest cities: Mumbai and Delhi. For Mumbai, our
finding is that it has entered into the mitigation period. If the 
current conditions against the disease spread can be continued throughout, by
the beginning of October we expect to see the onset of saturation
period, with the cumulative infected population about 145-160 thousand
and with less than 250 infections per day. For Delhi, we project that
 the
onset of saturation period, with a cumulative number of infection of
about 160-180 thousand and per day infections below 250, can be
achieved by the end of September to the beginning of October.
Finally on USA's data, we find
that there is a distinct feature in the time evolution trajectory.
 Our model clearly shows
that this data can be explained very well by allowing two waves of
infection: first one with the usual exponential-cum-power-rise
followed by a mitigation period up to a time, and then the second one started
 with that mitigated number with a
power-rise followed by a very slow mitigation.  The exponent in the
second power-rise is bigger than that of the first one suggesting a
very large number of population can get infected if adequate preventive
measures against it are not followed. Our result also shows that had
the mitigation period continued for the first wave, by now, a saturation period with much
less number of infection could have been achieved.
With the current rate of slow mitigation, but with existing restrictions, we project that this disease can rise another 3-4 months infecting about $6.5-7$ million people, before reaching to the saturation period with less than 5000 infections per day. However, we project this as an ideal case scenario. 
If the rate of per day increment remains within 0.9-1.25\% and no further reduction happens in next 2 weeks, within a
speculative scenario, we show that the total number of infection can be around 7-7.5 million
by the early November, with more than 10000-12000 infections per day. All the projections are made with 95\% confidence intervals in the proposed model given the reported available data. These projections can be progressively improved with the availability of more data in next few weeks$^3$. We summarize these projections in Table \ref{tab:summary}.

In the proposed model the parameters are assumed to be mean-field-type
effective parameters with no time and other
dependences. However, in reality, there is a number of external
parameters, which could be environmental, medical, economical, social
as well as political. It is not clear how the model parameters are
correlated with those external factors, and it would be worthwhile to
study their inter-relations. Such a study is necessary to understand
the significance of the model parameters.  It may be possible to
constrain the parameters more stringently using some other data, for
example, the number of active cases  and the number of fatalities. We will
pursue such a study in future. The predictive ability of the model can be improved further if it is possible to find a robust method to quantify the errors in the reported data.



\section{Acknowledgments}
We thank TIFR for support. GS also acknowledges WOS-A grant from the Department of  Science  and  Technology  (SR/WOS-A/PM-9/2017), India. NM would like to thank  Giorgio Sonnino for discussions, Abhishek Dhar for comments, and Girish Kulkarni for discussions on error analysis. 
  



\bibliography{refs}{}

\section{Appendix}
Here we show some of the numerical details, particularly describing the method employed for obtaining the confidence intervals.

For a given data set \{days, number of infection and error $\equiv$ $t$, $y(t)$ and $\sigma(t),\, t = 1,..., N$\} and the model function ($f(t)$), where $f(t)$ is the time evolution trajectory as in Eqs. \eqref{eqn:model_eq} and (13), we calculate the minimum of $\chi^2$, defined as
\begin{equation}
\chi^2(f) = \sum_{t = 1}^{N} {{(y(t) - f(t))^2}\over {\sigma^2(t)}} .
\end{equation}
A non-linear fitting method is utilized for the minimization. The total minimum $\chi^2$ is obtained by fitting all periods simultaneously (by varying $t_e, t_m, t_{s_i}$ together). For Eq. (13) this minimization is performed for each trajectory with a given set of parameters and we choose only those which are within the acceptable minimum $\chi^2$. The non-linear fitting method employed here has been extensively used previously for lattice gauge theory calculations and data analysis \cite{fitting_ref1, fitting_ref2, fitting_ref3}.

To maximize the probability
of the observed data given the model, we follow Ref. \cite{data_ana_ref1} and calculate the {\it likelihood}
of the parameters, defined as
\begin{equation}
\mathcal{L} = \prod_{i = 1}^N p\left(y_i|x_i,\sigma_{yi},\{\alpha\}\right).
\end{equation}
Here we use the generic notation $x$ for time ($t$) and $\{\alpha\}$ represents all parameters of the model, and the frequency distribution $p\left(y_i|x_i,\sigma_{yi},\{\alpha\}\right)$ is taken to be Gaussian
\begin{equation}
\prod_{i = 1}^N p\left(y_i|x_i,\sigma_{yi},\{\alpha\}\right) = {1\over \sqrt{2\pi\sigma^2_{yi}}} exp \left(- {{(y_i - f(i,\{\alpha\}))^2}\over {2 \sigma^2_{yi}}}\right).
\end{equation}
We then adopt the Bayesian way of data-analysis \cite{data_ana_ref2} using
\begin{equation}
p(\mathrm{model|data}) = {{p(\mathrm{data|model})p(\mathrm{model})}\over p(\mathrm{data})},
\end{equation}
to generalize the above frequency distribution as
\begin{equation}
  p\left(\{\alpha\}|\{y_i\}_{i=1}^N, I\right)
 = {{p\left(\{y_i\}_{i=1}^N|\{\alpha\}, I\right) {p\left(\{\alpha\}| I\right)}} \over{p\left({\{y_i\}_{i=1}^N} |I \right)}},
\end{equation}
where
\begin{itemize}
\item $I$ : all the {\it prior} knowledge of the data and the problem.
\item $p\left(\{\alpha\}| I\right)$: the {\it prior} probability distribution for the
parameters ($\{\alpha\}$)  that represents all knowledge except the data
\item  $p\left(\{\alpha\}|\{y_i\}_{i=1}^N, I\right)$: the {\it posterior} probability distribution for the parameters ($\{\alpha\}$) with the given data and the prior knowledge
  \item $p\left({\{y_i\}_{i=1}^N} |I \right)$: A normalization constant
  \end{itemize}
One gets the peak of the {\it posterior} probability at the best-fit values of the
parameters ($\{\alpha\}$) while its moments provide the uncertainties of these parameters.

Since the above {\it posterior} probability distribution with the given number of parameters of our model is quite complicated, we adopt a Markov-Chain-Monte-Carlo (MCMC) method to generate this distribution with the given data set and model. The {\it priors} for the parameters are chosen uniformly within the relevant ranges. We use more than 50000 MCMC steps in most cases after suitably adjusting the autocorrelation times. Each MCMC steps are initialized with a {\it Gaussian ball} around the maximum likelihood result.

In Fig. \ref{fig:usa_err_ana}  we show the projections of the {\it posterior} probability distributions of model parameters for the mitigation period. Since this period is most complicated as well as important  and more number of parameters are involved, we choose this period to show results from the above-mentioned analysis method. Fig. \ref{fig:usa_err_ana} is for the data on USA and the extracted parameters correspond to Fig. \ref{fig:USA_plot}. The bands represent the one, two and three sigma confidence intervals.
\bef[h]
\centering
\includegraphics[scale=0.4]{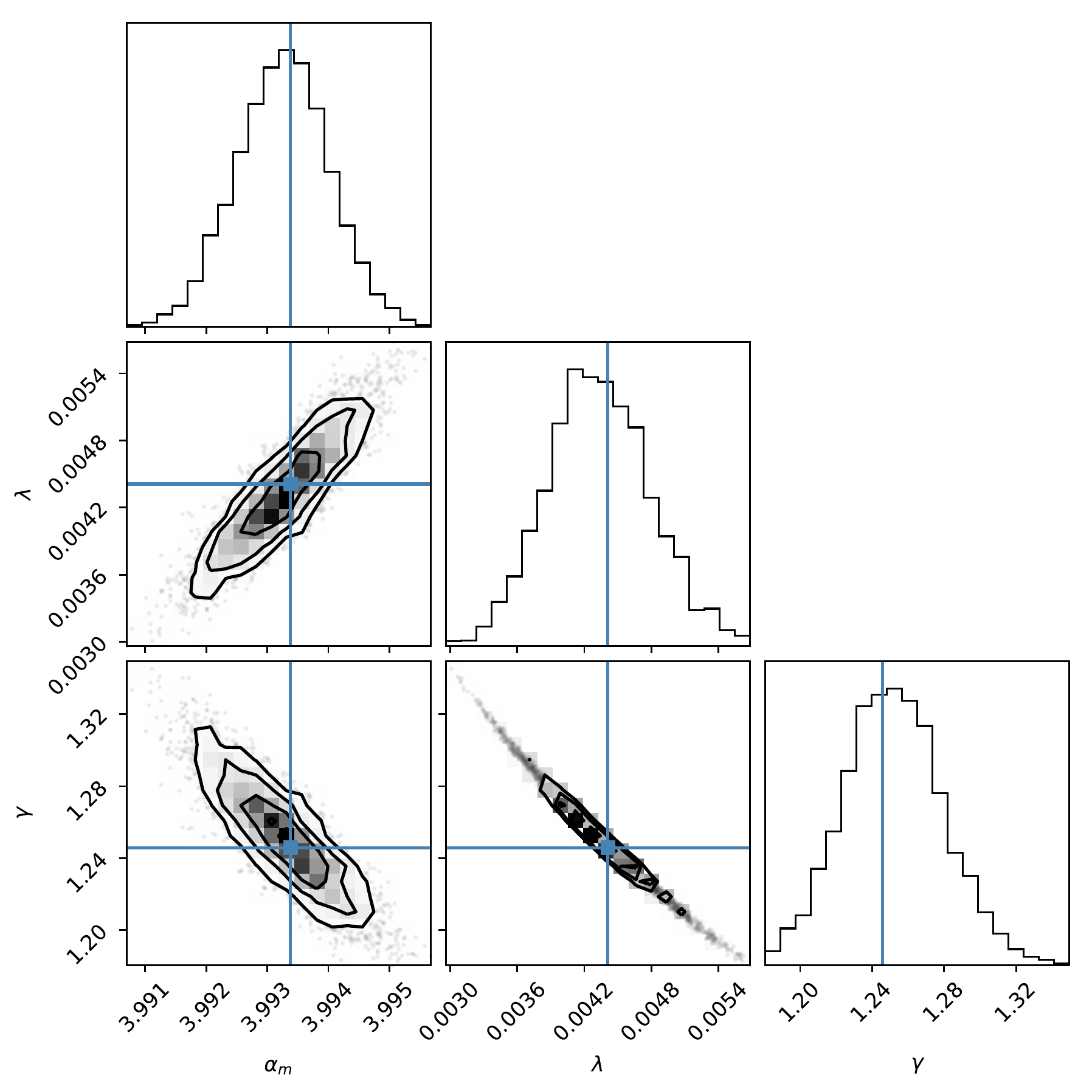}
\caption{\label{fig:usa_err_ana}Projections of the {\it posterior} probability distributions of model parameters, $\alpha_m, \lambda$ and $\gamma$, corresponding to the mitigation period (Eq. \eqref{eq:model_eq2}) of USA. Data represents the cumulative number of infection for USA \cite{covid_wiki_usa}.}
\eef{}
In Fig. \ref{fig:nyc_err_ana} we show similar results for the NY city in its saturation period.
\bef[h]
\centering
\includegraphics[scale=0.5]{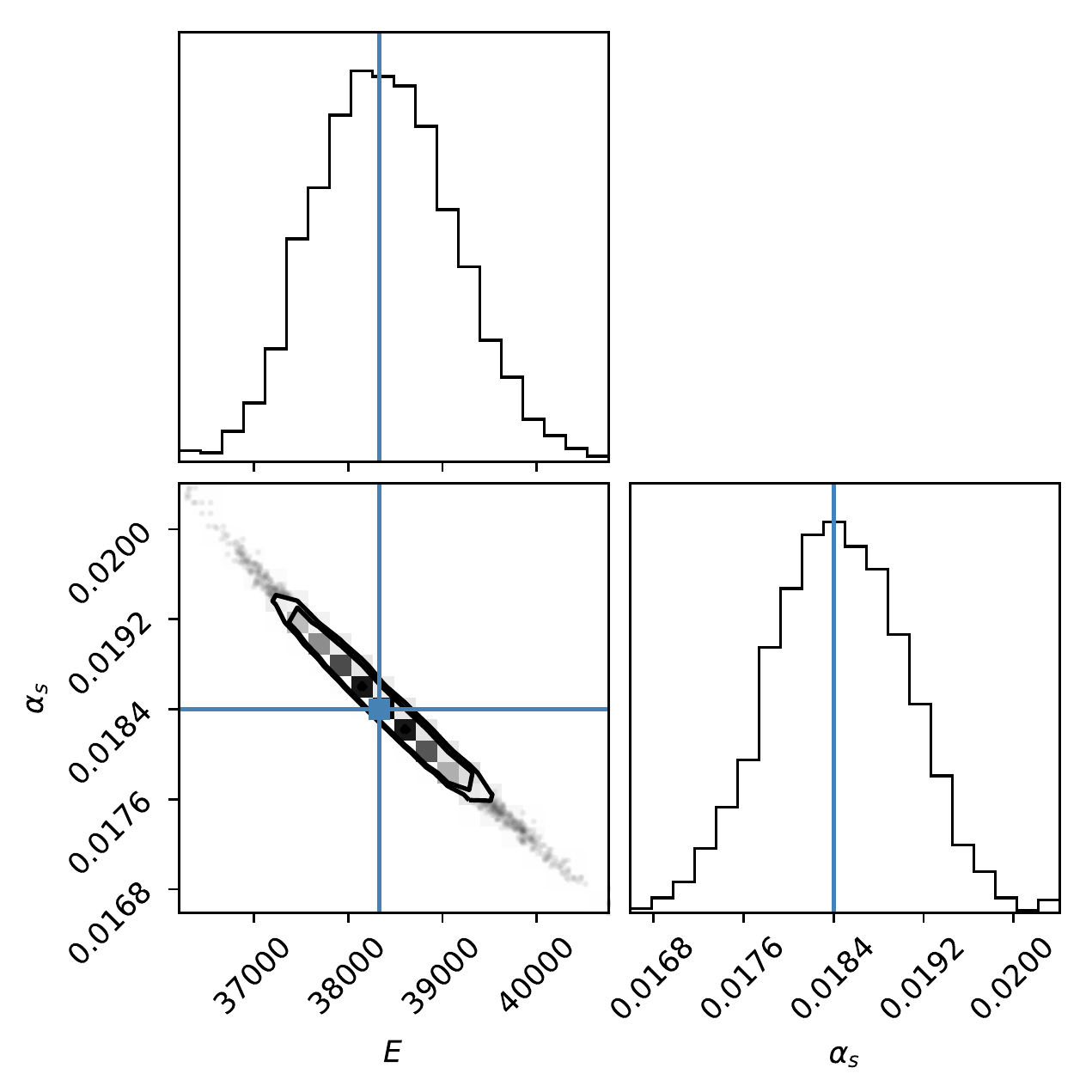}
\caption{\label{fig:nyc_err_ana}Projections of the {\it posterior} probability distributions of model parameters, $E$ and $\alpha_s$, corresponding to the saturation period (Eq. \eqref{eq:model_eq3}) of NY city. Data represents the cumulative number of infection for NY city~\cite{covid_wiki_nyc}.}
\eef{}

Finally the fitted error in the cumulative number of infection, particularly for projection purpose, was calculated by bootstrapping the MCMC samples. In Figs. \ref{fig:USA_plot} and \ref{fig:usa_plot1} we show such errors by 68\% and 95\%  confidence intervals.
\end{document}